\documentclass[a4paper,11pt]{article}
\RequirePackage{snapshot}  

\usepackage{jheppub} 

\usepackage{bm,amsmath,amssymb,slashed,graphicx,%
            enumerate,alltt,xspace,multirow,xcolor,mathrsfs}
\usepackage{fancyvrb}
\usepackage{booktabs,tabularx}
\usepackage{graphicx} \usepackage{subcaption} \usepackage{xspace}
\usepackage[utf8]{inputenc} \usepackage[compat=1.0.0]{tikz-feynman}
\usepackage{scalerel} \usepackage{layouts}

\usepackage{lineno}
\usepackage{printlen}
\usepackage{wasysym}
\usepackage{grffile}
\usepackage{soul}

\makeatletter
\g@addto@macro\bfseries{\boldmath}
\makeatother

\definecolor{labelkey}{rgb}{0,0.5,0.0}

\usepackage{listings}
\definecolor{darkgreen}{rgb}{0,0.4,0}
\definecolor{lightblue}{rgb}{0.0,0.5,1.0}
\definecolor{grey}{rgb}{0.5,0.5,0.5}

\definecolor{darkgreen}{rgb}{0,0.4,0}
\definecolor{grey}{rgb}{0.5,0.5,0.5}
\definecolor{orange}{rgb}{0.9,0.5,0.0}
\definecolor{lightblue}{rgb}{0.0,0.5,1.0}
\allowdisplaybreaks 

\definecolor{semiblue}{rgb}{0.3,0.3,0.8}

\newcommand{\as}{\alpha_s}

\newcommand{\dd}{\;\mathrm{d}}

\newcommand{\nc}{N_\text{\textsc{c}}} 
\newcommand{\nf}{n_{\!\:\!f}}

\newcommand{\itilde}{{\tilde \imath}}
\newcommand{\jtilde}{{\tilde \jmath}}

\newcommand\sss{\mathchoice%
{\displaystyle}%
{\scriptstyle}%
{\scriptscriptstyle}%
{\scriptscriptstyle}%
}
\newcommand{\dis}{\rm \sss DIS}

\newcommand{\mathd}{{\rm d}}

\newcommand{\muF}{\mu_\text{\textsc{f}}}
\newcommand{\etaq}{\bar{\eta}_{\sss Q}}
\newcommand{\betaps}{\beta_{\sss \rm PS}}
\newcommand{\betaobs}{\beta_{\sss \rm obs}}

\tikzstyle{block} = [rectangle, minimum width=1.0cm, minimum height=0.75cm, thin, draw=black]
\tikzstyle{blob} = [circle, minimum width=0.5cm, thin, draw=black]
\tikzset{blackarrow/.style={-stealth, semithick, draw=black}}
\tikzset{connection/.style={inner sep=0,outer sep=0}}

\newcolumntype{C}{>{\centering\arraybackslash}X}

\title{Next-to-leading-logarithmic PanScales showers for Deep Inelastic Scattering and Vector Boson Fusion}

\preprint{CERN-TH-2023-079, OUTP-23-05P}

\newcommand{\OXaff}{Rudolf Peierls Centre for Theoretical Physics, Clarendon Laboratory, Parks Road,
  University of Oxford, Oxford OX1 3PU, UK}
\newcommand{\CERNaff}{Theoretical Physics Department, CERN, CH-1211 Geneva 23, Switzerland}

\author[a]{Melissa van Beekveld,}%
\author[b]{Silvia Ferrario Ravasio,}%

\emailAdd{melissa.vanbeekveld@physics.ox.ac.uk}
\emailAdd{silvia.ferrario.ravasio@cern.ch}

\affiliation[a]{\OXaff}
\affiliation[b]{\CERNaff}

\date{Received: date / Accepted: \today}

\abstract{
  We introduce the first family of parton showers that achieve
  next-to-leading logarithmic~(NLL) accuracy for processes involving a
  $t$-channel exchange of a colour-singlet, and embed them in the
  PanScales framework.
  These showers are applicable to processes such as deep inelastic
  scattering~(DIS), vector boson fusion~(VBF), and vector boson
  scattering~(VBS).
  We extensively test and verify the NLL accuracy of the new showers 
  at both fixed order and all orders across a wide range of observables. 
  We also introduce a generalisation of the Cambridge-Aachen jet algorithm 
  and formulate new DIS observables that exhibit a simple resummation structure. 
  The NLL showers are compared to a standard transverse-momentum ordered dipole shower, serving as a proxy
  for the current state-of-the-art leading-logarithmic showers available in public codes.
  Depending on the observable, we find discrepancies at NLL of the order of $15\%$.
  We also present some exploratory phenomenological
  results for Higgs production in VBF.
  This work enables, for the first time, to resum simultaneously global
  and non-global observables for the VBF process at NLL accuracy. 
}


\begin{document}

\maketitle

\section{Introduction}
\label{sec:intro}

Particle-physics collider experiments
provide us with a unique opportunity to test our
knowledge of the fundamental interactions of elementary particles.
Accurately predicting signatures originating from Standard Model (SM) physics is crucial to fully harness the potential of the data, and have sensitivity to possible signals originating from beyond-the-Standard-Model scenarios.
Deciphering the nature of the Higgs boson and its interaction with other
SM particles is indeed one of the main objectives of the physics programme
of the Large Hadron Collider (LHC).
General purpose Monte Carlo~(GPMC) event generators are
fundamental tools in this context.
They play a crucial role in our understanding of the phenomenology of
colliders, thanks to their ability to describe much of the data from
the LHC  and its predecessors.

Parton showers lie at the core of GPMCs, and describe the energy
degradation of highly-energetic partons that are produced in the hard
scattering process, through radiation of soft and/or collinear
partons.
They enable us to simulate arbitrarily complex collider events,
characterised by a large multitude of particles, whose modelling
involves physics across a broad range of scales.
Despite their fundamental role in collider phenomenology, only in the
recent years more attention has been dedicated to
understanding and improving the formal accuracy of parton
showers.
Parton showers resum logarithmic terms in the perturbative series.
These terms arise from soft and collinear divergences, 
and can become large when exploring physics across a wide range of scales.
Consequentially, terms of the form $\alpha_s^n\sum_{i=1}^{n+1} c_i
L^{i}$, where $L$ is a large logarithm of a ratio of two disparate
scales, will occur at each perturbative order $n$, with $c_i$ a
coefficient that depends on the observable.
For example, with the strong coupling $\alpha_s(m_Z) = 0.118$, a hard scale of
$m_Z = 91.18$~GeV, and a hadronic scale
of $\Lambda = 1$~GeV, 
the next-to-leading logarithmic (NLL) corrections of
the form $\alpha_s \ln(m_Z/\Lambda)$ become of the order of $0.5$.
This constitutes an $\mathcal{O}(1)$ correction, which needs
to be resummed to all orders in the perturbative coupling
to obtain an accurate prediction. 

The Herwig7~\cite{Bahr:2008pv,Bellm:2019zci} angular-ordered shower,
which is based on the coherent branching
formalism~\cite{Gieseke:2003rz}, correctly resums NLL
terms (provided a careful interpretation of the ordering
variable is performed~\cite{Bewick:2019rbu,Bewick:2021nhc}) for global
observables, but not for non-global ones~\cite{Banfi:2006gy}.
The NLL terms of the latter cannot be captured by the coherent branching formalism, but
require a dipole approach~\cite{Gustafson:1987rq}, as they are
sensitive to the full angular distribution of soft emissions.
Another practical advantage of a dipole shower, is that matching with
fixed-order matrix elements is much simpler.
This has led to
numerous techniques to perform multi-jet merging at
next-to-leading order~(NLO)~\cite{Catani:2001cc,
Krauss:2002up,Lavesson:2008ah,Hoeche:2009rj,
Giele:2011cb,Platzer:2012bs,Lonnblad:2012ix,Frederix:2012ps,Lonnblad:2012ng,Bellm:2017ktr,Brooks:2020mab}
, and matching to fixed-order accuracies, i.e.~(next-to-)next-to-next-to-leading order ((N)NNLO)~\cite{Hamilton:2012rf,Alioli:2013hqa,Hoche:2014dla,Monni:2019whf,Campbell:2021svd,Prestel:2021vww}.%
\footnote{LO multi-jet merging techniques that can be applied to an
angular-ordered showers are discussed in
Refs.~\cite{Mangano:2001xp,Hamilton:2009ne,Martinez:2021chk}.}
Despite the large body of work on improving the fixed-order accuracy
of showers, until recently, relatively little has been done on
improving their logarithmic accuracy.  Indeed, all standard
dipole showers currently embedded in public
GPMC~\cite{Schumann:2007mg,Platzer:2009jq,Hoche:2015sya,Cabouat:2017rzi,Brooks:2020upa,Sjostrand:2006za,Sjostrand:2014zea,Gleisberg:2008ta,Sherpa:2019gpd,Bierlich:2022pfr}
are only leading logarithmic~(LL), i.e.~they resum terms that are
proportional to $\as^n L^{n+1}$.

The focus of this paper is to design the first dipole showers for processes
characterised by the $t$-channel exchange of a colour-singlet
that reach NLL (here defined to be single-logarithmic, $\as^n L^n$) accuracy 
for both global and non-global observables. 
This work follows on earlier developments of parton showers with a controlled
logarithmic accuracy by the PanScales collaboration, like those
for dijet production in $e^+e^-$ collisions~\cite{Dasgupta:2020fwr}. 
Key components in the design of showers is a careful construction
of the recoil distribution after a new emission has been generated, 
and its interplay
with the ordering variable of the shower~\cite{Dasgupta:2018nvj}.
Several other groups have also been investigating NLL-accurate showers
in the context of $e^+e^-$ collisions.
Ref.~\cite{Forshaw:2020wrq} introduced an algorithm that is shown to analytically  reproduce
NLL accuracy for the thrust distribution and subject multiplicity. 
The $\Lambda$-ordered Deductor shower is shown to be NLL accurate
for the thrust distribution in Ref.~\cite{Nagy:2020dvz}. 
Ref.~\cite{Herren:2022jej} presents Alaric, 
a shower that is proven to be NLL accurate for a wide range of global observables.
Designing NLL showers for hadronic colliders bring additional
complications with respect to that for $e^+e^-$ collisions, as the
treatment of the recoil from the initial-state is more subtle.
In Refs.~\cite{vanBeekveld:2022zhl,vanBeekveld:2022ukn}, we presented
NLL-accurate dipole showers for the production of a colour
singlet at the LHC.
At present, no other showers exists with demonstrated NLL
accuracy (for both global and non-global observables) for such
processes.%
\footnote{A discussion on the treatment of the transverse-momentum
recoil for initial-state radiation in an angular-ordered shower can be
found in Ref.~\cite{Bewick:2021nhc}, as well as in the Herwig++
manual~\cite{Bahr:2008pv}.
This topic has also been addressed in Ref.~\cite{Nagy:2009vg} in the context of
the transverse-momentum distribution
of the $Z$ boson in Drell Yan production, albeit
without a claim on the logarithmic accuracy that is achieved.
}
The present work extends the set of NLL accurate showers to
processes characterised by a $t$-channel exchange of a colour singlet, such
that now all processes with at most two partons at the first contributing order
can be described at single-logarithmic accuracy for generic observables. 

\begin{figure}[t]
  \centering
  \begin{subfigure}{0.32\textwidth}
    \centering
    \includegraphics[width=0.8\textwidth]{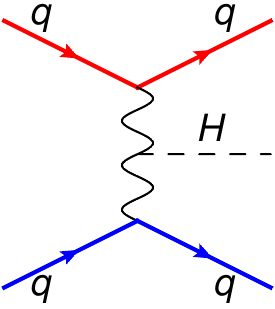}
    \caption{}
    \label{fig:VBF}
    \end{subfigure}
  \begin{subfigure}{0.32\textwidth}
    \centering
    \includegraphics[width=0.8\textwidth]{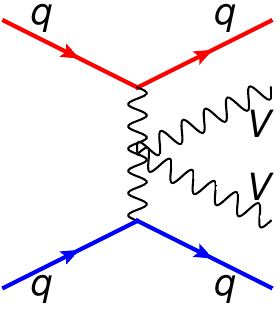}
    \caption{}
    \label{fig:VBS}
    \end{subfigure}
  \begin{subfigure}{0.32\textwidth}
    \centering    
    \includegraphics[width=0.8\textwidth]{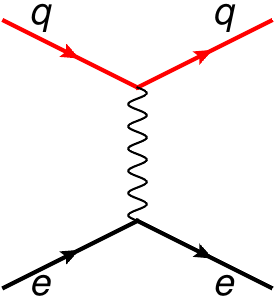}
    \caption{}
    \label{fig:DIS}
    \end{subfigure}
    \caption{Feynman diagrams contributing to (a) Higgs production in VBF, (b) VBS and (c) DIS. Colour-connected quark lines are highlighted in red and blue. Leptons are represented with black solid lines.}
\end{figure}

Examples of processes with a $t$-channel exchange of a colour-singlet,
are vector-boson fusion~(VBF, Fig.~\ref{fig:VBF}) and vector-boson
scattering~(VBS, Fig.~\ref{fig:VBS}).
Measurements of VBF and VBS processes can provide valuable insight
into the electroweak (EW) and Higgs
sectors~\cite{Covarelli:2021gyz,BuarqueFranzosi:2021wrv}.
Higgs production in VBF is the second most abundant production mode
for the Higgs boson at the LHC.
It has a clear experimental
signature, given by the presence of two back-to-back hard jets in the
forward and backward regions of the detectors, 
and hence it is the preferred
channel to for example measure the Higgs coupling to
muons~\cite{ATLAS:2020fzp,CMS:2020xwi}, tauons~\cite{CMS:2022kdi, ATLAS:2022yrq}, and to study Higgs to invisible
decays~\cite{ATLAS:2022yvh, CMS:2022qva}.
Run III data will enable us to perform precise determinations of the
size of gauge-bosons self-interactions. Modifications to VBS
processes are predicted in models of physics beyond the SM, through
changes to the Higgs boson couplings to gauge bosons and the resonant
production of new particles.
Differences in the parton-shower modelling turn out to be one of the leading uncertainty in these
kind of processes~\cite{Ballestrero:2018anz,Jager:2020hkz,Hoche:2021mkv, CMS:2020xwi, ATLAS:2022tnm, ATLAS:2022ooq, ATLAS:2022yrq,ATLAS:2020rej, CMS:2022kdi, CMS:2021kom,CMS:2022uhn,Buckley:2021gfw}.
Having access to several NLL-accurate showers, such as the ones we present in this article,
will help to assess more realistic shower uncertainties, and possibly reduce them. 

Higher-order QCD contributions to VBF and VBS processes are computed
using a factorised
approach~\cite{Han:1992hr,Bolzoni:2010xr,Bolzoni:2011cu},
i.e.~neglecting non-factorisable corrections stemming from the
exchange of partons between the two hadronic sectors, which are
coloured in red and blue in Figs.~\ref{fig:VBF} and~\ref{fig:VBS}.
Non-factorisable corrections appear only from NNLO and are typically
colour-suppressed with respect to the factorisable
ones~\cite{Liu:2019tuy,Dreyer:2020urf}.
In the factorised approximation, radiative corrections to VBF
and VBS are closely related to those for deep inelastic
scattering~(DIS, Fig.~\ref{fig:DIS}).%
\footnote{
VBF can be treated using the structure function approach,
i.e.~treating the two quark lines as two seperate copies of DIS.
Instead, for VBS, interferences between diagrams with a $t$-channel
exchange of a boson and the tagged bosons being emitted from the quark
legs directly spoil this picture.
From the QCD point of view,
one can still use the factorised approximation
and neglect gluon exchanges between the two quark lines,
also see Ref.~\cite{BuarqueFranzosi:2021wrv} and references therein.
However, for the VBS topology, 
one needs to decide whether a boson
needs to be boosted together with all partons belonging to 
one hadronic sector, to preserve its angle with respect to the
original-final state quark, which is sensible when that
boson was emitted directly from the quark line.
Otherwise, the shower may reshuffle the post-branching momenta in such
a way that the boson becomes collinear to the jet, or that a collinear
boson is suddenly emitted at a wide angle: in both situations, the
matrix element used to generate the process would then no longer
describe the post-branching situation with the new four-momenta.
}
For this reason, the first part of this article focuses on the
formulation of the
PanScales showers for DIS processes. 
We then extend our showers to handle two copies of DIS, so
that they can also be used to describe VBF and VBS events
in the factorised approach.
After including mass effects, which we leave for future work,
this framework will also enable us
to handle $t$-channel single-top production.
These showers will have full colour accuracy at LL, and leading colour
accuracy at NLL, i.e.~single-logarithmic in our definition.
This is the first time this logarithmic accuracy is reached, either
analytically or numerically, for the VBF and VBS
processes. 

In addition to its relevance to hadron-collider processes involving a colourless $t$-channel exchange, 
an improved theoretical framework for deep inelastic scattering (DIS) can be directly applied to interpret data gathered from electron-proton ($ep$) colliders like the Hadron Electron Ring Accelerator (HERA). 
These colliders provide an ideal environment for investigating the internal structure of hadrons and conducting accurate studies of quantum chromodynamics (QCD).
The new-generation of lepton-hadron colliders, such as the
Electron-Ion Collider~(EIC), whose construction is planned to start
next year, would also benefit from more accurate MC generators.

The paper is structured as follows.
In Sec.~\ref{sec:breit} we review the standard kinematic
variables used in DIS.
In Sec.~\ref{sec:psgen} we present the common building-blocks of a
generic dipole-shower for DIS.
In Sec.~\ref{sec:dipolekt}, we summarise the main features of a
standard dipole shower, while in
Sec.~\ref{sec:panshowers} we present the new PanScales showers for
DIS, whose extension to VBF/VBS is discussed in
Sec.~\ref{sec:VBF-shower}.
We then perform fixed-order tests of such showers in
Sec.~\ref{sec:fo-test}.
The all-orders validations of the PanScales showers are instead
performed in Sec.~\ref{sec:ao-test}.  We test numerous observables,
designed to probe soft and/or collinear emissions
that an NLL shower should accurately describe:
the DGLAP evolution of the parton distribution
functions (Sec.~\ref{sec:dglap}), average particle
multiplicity (Sec.~\ref{sec:multiplicity}), several continuously-global
observables (Sec.~\ref{sec:global-obs}), and the amount of radiation in a
rapidity slice (Sec.~\ref{sec:non-global-obs}), which we consider as a proxy
for a generic non-global observable.  
In Sec.~\ref{sec:pheno} we show some phenomenological results
for Higgs production in VBF, and in Sec.~\ref{sec:conc} we 
present our conclusions.

\section{DIS definition and kinematics}
\label{sec:breit}
We consider the DIS process $\gamma^* p\to X$, where $p$ is the
incoming proton, $\gamma^*$ is a space-like photon (or more
generically a colourless boson) and $X$ is a generic hadronic
final-state.
We denote with $q_{\dis}^{\mu}$ the photon momentum, and we define
\begin{equation}
Q^2 = -q_{\dis}^2 >0\,.
\label{eq:Q2toqdis}
\end{equation}
The (massless) proton momentum is denoted by $P^{\mu}$.
We introduce two light-like reference vectors $n_1$ and $n_2$, such
that $n_1$ is parallel to the incoming proton i.e.
\begin{equation}
  n_1^{\mu} = x_{\dis} P^{\mu}\,,
  \label{eq:n1_breit}
\end{equation}
and
\begin{equation}
  n_2^{\mu} =   q_{\dis}^{\mu} + n_1^{\mu}\,.
   \label{eq:n2_breit}
\end{equation}
Requiring that $n_2^2=0$ leads to
\begin{equation}
x_{\dis} = \frac{Q^2}{2 \, q_{\dis} \cdot P}\,.
\end{equation}
Furthermore, we see that $n_1 \cdot n_2 = Q^2/2$. 
At the partonic level, the lowest-order contribution reads $
\gamma^*\, q \to q$, where the incoming (outgoing) quark momentum is
precisely $n_1^{\mu}$ ($n_2^{\mu}$).
More generally, if the sum of the momenta of the final-state partons 
has an invariant mass $M_{\sss X}$, one may always
parameterise their collective momentum $p_{\sss X}^{\mu}$ as
\begin{equation}
p_{\sss X}^{\mu} = n_2^{\mu} + \frac{M_{\sss X}^2}{Q^2} n_1^{\mu}\,.
\end{equation}
Momentum conservation then implies that the incoming parton has momentum
\begin{equation}
p_1^{\mu} = \left( 1 +    \frac{M_{\sss X}^2}{Q^2}\right) n_1^{\mu} = x_1 P^{\mu}\,, \quad \mbox{with } \quad x_1= x_{\dis}  \left( 1 +    \frac{M^2_{\sss X}}{Q^2}\right).
\end{equation}
Indeed, with these definitions we have
\begin{equation}
	\label{eq:mom-consv-dis}
	p_1^{\mu} + q_{{\rm dis}}^{\mu} = p_{\sss X}^{\mu}\,.
\end{equation}
We work in the Breit frame~\cite{Webber:1993bm}, which is defined 
as the frame where
$n_1$ and $n_2$ are back-to-back and aligned along the $z$-axis, i.e.
\begin{subequations}
\begin{align}
  n_1^{\mu} = \frac{Q}{2}\left(0,0,-1; 1\right), &\qquad n_2^{\mu} = \frac{Q}{2}\left(0,0,+1; 1\right),\\
  q_{\dis}^{\mu} & =  Q\left(0,0,+1; 0\right),
\end{align}
\end{subequations}
with $Q \equiv \sqrt{Q^2}$. 
The final-state momenta can then
always be decomposed in terms of Sudakov variables, i.e.~
\begin{align}
	k_i^{\mu} = \alpha_i n_1^{\mu} +  \beta_i n_2^{\mu} + k_{\perp i}^{\mu}\,,
\end{align}
where $k_{\perp i}^{\mu}$ is a generic space-like vector orthogonal to $n_{1,2}^\mu$.
Partons are said to reside in the current hemisphere when $\beta_i > \alpha_i$.
The remnant hemisphere instead contains partons with $\alpha_i > \beta_i$. 
When constructing the shower, it will turn
out to be convenient to introduce the reference vector
\begin{equation}
Q^{\mu} = n_1^{\mu} + n_2^{\mu}\,,\label{eq:panscalesQ}
\end{equation}
which has norm  $Q^2=-q_{\dis}^2$.
We note that the energy
component in the Breit frame of a given momentum $p^\mu$
 is obtained through
\begin{equation}
E_{\sss \rm Breit} = \frac{p \cdot Q}{Q}\,.
\end{equation}
We will use the reference vector $Q^{\mu}$ in the formulation of the
PanScales showers to measure angular distances in the Breit frame
instead of the dipole frame.

\section{Dipole showers for DIS and VBF/VBS}
\label{sec:psgen}
The fundamental building block for a dipole shower is a $2\to 3$
branching kernel~\cite{Gustafson:1987rq}.  In these types of showers
each emitter $\itilde$ is colour-connected (understood in the limit of
a large number of colours $\nc$) to a spectator $\jtilde$, such that
the branching is $\itilde \jtilde \to i jk$, with $k$ the radiated
parton.
Dipole showers for DIS (and VBF or VBS) need to handle emissions that
come from dipoles that contain two final-state partons (FF dipoles),
or one parton in the initial state and one in the final state (IF
dipoles).

First, we consider an FF dipole with pre-splitting momenta
$\tilde{p}_i$, $\tilde{p}_j$, and post-splitting momenta $p_i$, $p_j$,
$p_k$.  The momentum of the radiated parton $k$ may be decomposed as
\begin{equation}
p_k^\mu = z_i \tilde{p}_i^\mu + z_j \tilde{p}_j^\mu + k_{\perp}^\mu\,.
\end{equation}
where $k_\perp^\mu$ is a space-like vector orthogonal to
$\tilde{p}_{i,j}^\mu$.  In this notation with $\tilde{p}_i$ identified
as emitter, $z_i$ corresponds to the collinear momentum fraction
 carried away by the emission $p_k$, defined relative to the pre-branching
momentum. 
 The emission probability that describes
correctly radiation in the soft and collinear limit can be written as
\begin{multline}
\label{eq:dipole-prob-FF}
     \mathd \mathcal{P}_{\itilde \jtilde \to ijk} =
    \frac{\as(\mu_\text{\textsc{r}}^2)}{2\pi}\left(1 +\frac{\as(\mu_\text{\textsc{r}}^2) K}{2 \pi}\right) \frac{\mathd v^2}{v^2} \mathd
    \bar{\eta} \frac{\mathd\phi}{2\pi}
 \left[ g(\bar{\eta}) z_i P^{\rm {FS}}_{ik} (z_i) + g(-\bar{\eta}) z_j P^{\rm {FS}}_{jk} (z_j) \right],
\end{multline}
where $\mu_\text{\textsc{r}}^2$ is the renormalisation scale.
The phase space of the emission is parameterised by the shower
variables $v$ and $\bar{\eta}$, where $v$ can be identified with the
shower ordering scale, and $\bar{\eta}$ is a rapidity-like auxiliary
variable.
The exact relation between $z_i, (z_j), k_{\perp}$ and the shower variables $v,
\bar{\eta}$ is shower dependent and will be detailed in the following sections.
The azimuthal angle of the radiation defined with respect to the plane
spanned by the two pre-branching dipole momenta is denoted by $\phi$.
To achieve NLL accuracy one needs to evaluate the running of the
coupling at two loops, with $\mu_{{\rm R}}$ equal to the
transverse-momentum of the emission in the soft-collinear limit.
In addition, the soft-collinear gluon emission probability must
include an $\alpha_s K/ (2\pi)$ correction term with
$K=(67/18-\pi^2/6)C_A-5\nf/9$~\cite{Catani:1990rr}.
We take $C_A=3$, $T_R = 1/2$ and we work with $\nf=5$ light flavours
(unless otherwise stated).
The definition of the DGLAP final-state splitting functions $P^{\rm
  {FS}}(z)$ that we use is given in appendix A of
Ref.~\cite{vanBeekveld:2022zhl}, with their soft limit obtained as $z
\to 0$.
Following Ref.~\cite{Dasgupta:2020fwr}, we use two definitions of
$g$:
\begin{equation}
    g(\bar{\eta}) = g^\text{dip.}(\bar{\eta}) \equiv 
    \begin{cases} 
        0 &\text{ if } \bar\eta < -1 \,,\\ 
        \displaystyle \frac{15}{16} \left(\frac{\bar{\eta}^5}{5} - \frac{2\bar{\eta}^3}{3} + \bar{\eta} + \frac{8}{15} \right) & \text{ if } -1 < \bar{\eta} < 1 \,, \\
        1 &\text{ if } \bar{\eta} > 1\,,
    \end{cases}
    \label{eq:geta}
\end{equation}
or
\begin{equation}
  g(\bar{\eta}) = g^\text{ant.}(\bar{\eta}) \equiv \frac{e^{\bar{\eta}}}{e^{\bar{\eta}}+e^{-\bar{\eta}}}=\frac{e^{2\bar{\eta}}}{e^{2\bar{\eta}}+1}\,.
  \label{eq:fantenna}
\end{equation}
We use $g^\text{ant.}$ for antenna showers, where one makes no
distinction between an emitter/spectator to distribute the
transverse-momentum recoil, while we use $g^\text{dip.}$ when only one
of the parent partons takes the transverse recoil.

Instead, for IF dipoles where we define the parton $\itilde $ as
belonging to the initial-state, we write the momentum of the radiated
parton as
\begin{equation}
p_k^{\mu} = \frac{z_i}{1-z_i} \tilde{p}^{\mu}_i + z_j \tilde{p}_j^{\mu} + k^{\mu}_{\perp}\,.
\end{equation}
The emission probability then takes the form
\begin{multline}
\label{eq:dipole-prob-IF}
     \mathd \mathcal{P}_{\itilde \jtilde \to ijk} =
    \frac{\as(\mu_\text{\textsc{r}}^2)}{2\pi}\left(1 +\frac{\as(\mu_\text{\textsc{r}}^2) K}{2 \pi}\right) \frac{\mathd v^2}{v^2} \mathd
    \bar{\eta} \frac{\mathd\phi}{2\pi}
    \\
    \times
 \left[   \frac{x_i f_i(x_i, \muF^2)}{\tilde{x}_i f_{\itilde}(\tilde{x}_i,\muF^2)} g(\bar{\eta}) z_i P^{\rm {IS}}_{ik} (z_i) + g(-\bar{\eta}) z_j P^{\rm {FS}}_{jk} (z_j) \right],
\end{multline}
where $f_i$ is the PDF of the incoming parton $i$, and $x_i = \frac{\tilde{x}_i}{1-z_i}$.
The choice of the factorisation scale $\muF$ depends on the ordering
variable $v$, and it must be chosen such that for a hard-collinear
initial-state branching, $\muF$ is equal to the transverse momentum of
the emission.
The definition of the DGLAP initial-state splitting functions is also
given in Appendix A of Ref.~\cite{vanBeekveld:2022zhl}.

The DIS invariants $q^2_{\dis}$ and $x_{\dis}$,
introduced in the previous section, determine the
structure of the event.
At LO these two invariants are the only two quantities needed to
describe the interaction between the incoming lepton and nucleon.
Although unitarity of the shower is preserved, the real-virtual
cancelation on more differential LO observables is spoiled
when modifying the DIS invariants through the generation of an emission. 
Preserving the LO DIS invariants is therefore crucial to ensure that the
shower does not alter the description of any inclusive observable.
Furthermore, this also simplifies the inclusion of higher-order
corrections in future works, especially in the context of VBF if the
two hadronic sectors evolve completely independently.
For these reasons, the conservation of the DIS invariants $q^2_{\dis}$
and $x_{\dis}$ is a fundamental property of all the showers we
consider/develop here.%
%

\subsection{A standard transverse-momentum-ordered dipole shower}
\label{sec:dipolekt}

In this section, we briefly summarise the fundamental features of a
standard dipole shower, which we use as a proxy to illustrate the
behaviour of current
standard leading-logarithmic showers that are used for DIS
and VBF/VBS phenomenological studies.
It is based on a Catani-Seymour dipole-local map, 
as first explored in Refs.~\cite{Gustafson:1987rq, Catani:1996vz}. 
We refer to this shower as ``Dipole-$k_t$'', and its kinematic maps
(with local momentum conservation in the IF and FF dipoles) are
presented in Refs.~\cite{vanBeekveld:2022zhl,vanBeekveld:2022ukn}.%
\footnote{In particular, the kinematic maps coincide with 
	those of the Dire-v1 shower~\cite{Hoche:2015sya}.}
All publicly-available dipole showers for DIS collisions, 
such as Pythia8 with dipole-local recoil~\cite{Cabouat:2017rzi},
Sherpa~\cite{Schumann:2007mg}, and Herwig's dipole shower~\cite{Platzer:2009jq}
share a great degree of similarity with Dipole-$k_t$ in the small transverse-momentum limit.
Differences can be large away from the small transverse-momentum limit.
However, the degrees-of-freedom relevant for logarithmic accuracy, 
as shown in~\cite{Dasgupta:2018nvj},
namely the frame in which the emitter and spectator are chosen (the dipole centre-of-mass frame), 
the ordering variable of the shower (transverse-momentum ordered), and the transverse-momentum recoil scheme,
are the same for the showers mentioned above.%
\footnote{Albeit an antenna shower, this also makes us believe that
Vincia~\cite{Ritzmann:2012ca} has the same logarithmic order of 
accuracy as Dipole-$k_t$.}
Here we review the basic characteristics of the Dipole-$k_t$ shower. 

The ordering variable $v$ of the Dipole-$k_t$ shower algorithm is
transverse-momentum-like, i.e.~it corresponds to the transverse
momentum of the emitted particle in the limit that this emission is
soft-collinear.  The rapidity-like variable $\bar{\eta}$, used to partition the
dipole in two halves, can be related to the collinear momentum
fraction $z$.  This relation reads
\begin{equation} \label{eq:eta-dipolekt}
	\bar{\eta}  = 
	\begin{cases}
		\frac{1}{2} \ln\frac{z^2 \, \tilde{s}_{ij}}{v^2} \,  & \text{(final state),} \\
		\frac{1}{2}\ln\frac{z^2 \tilde{s}_{ij}}{(1-z)^2v^2} & \text{(initial state)},
	\end{cases}
\end{equation}
with $\tilde{s}_{ij} = 2 \tilde{p}_i \cdot \tilde{p}_j$ the dipole invariant mass. 
From this it is clear that $\bar{\eta}$ partitions the dipole in its rest frame. 
In the Dipole-$k_t$ shower, momentum conservation is fully local. 
This implies that when an emission occurs from a FF dipole, the transverse momentum recoil is entirely taken up by the emitter, which corresponds to the original dipole leg closer in angle to the emission. 
Conversely, in an IF dipole, the transverse momentum recoil is always absorbed by the final-state leg.
The factorisation and renormalisation scales are set equal to
$v$. Further details on the kinematic maps used in the Dipole-$k_t$
shower are given in Appendix~B.1 of
Ref.~\cite{vanBeekveld:2022zhl}.%
\footnote{Conversely to
Refs.~\cite{vanBeekveld:2022zhl,vanBeekveld:2022ukn}, here we use only
the local variant of Dipole-$k_t$ and we do not consider the global
variant.  The latter algorithm does not preserve the DIS invariants
(specifically $x_{\dis}$ and $y_{\dis}= \frac{Q^2}{x_{\dis} s}$) and for
this reason it has never been used in phenomenological applications to
DIS or VBF/VBS.}
%

\subsection{PanScales showers for DIS}
\label{sec:panshowers}
The PanScales showers need a reference momentum $Q^\mu$ to define a
common frame where to measure angular distances for all the emissions,
which here we take to be the Breit frame.
We set this equal to $Q^\mu$ as introduced in eq.~\eqref{eq:panscalesQ},
which in the Breit frame reads $Q^\mu = (0,0,0;Q)$.
The ordering variable $v$ is defined in such a way that for a
soft-collinear emission of transverse momentum $k_\perp$ and rapidity
$\eta$, we have
\begin{equation}
  v \approx k_\perp e^{-\betaps|\eta|}\,.
\end{equation}
We also introduce the shower variable $\etaq$, that corresponds to
the rapidity of a soft-collinear emission in the Breit frame.
Given the shower variables $v$ and $\etaq$, we can define a
transverse-momentum auxiliary variable
\begin{align}
  \kappa_\perp \equiv  \rho v e^{\betaps |\etaq|}\,,
  \label{eq:kappaperp}
\end{align}
where we have used
\begin{align}
  \tilde{s}_i    = 2 \tilde{p}_i \cdot Q\,, \quad  \tilde{s}_j = 2 \tilde{p}_j \cdot Q\,, \quad
  \tilde{s}_{ij} = 2 \tilde{p}_i \cdot \tilde{p}_i\,, \quad \rho = \left(\frac{\tilde{s}_i\tilde{s}_j}{\tilde{s}_{ij}Q^2} \right)^{\betaps/2}\,.
\end{align}
To achieve NLL accuracy, the renormalisation scale at which the
coupling constant is evaluated is set to $\kappa_\perp$, while for the
factorisation scale we choose
\begin{equation}
\muF = Q \left( \frac{v}{Q} \right)^{\frac{1}{1+\betaps}}\,.
\end{equation}
We also introduce the variables
\begin{align}
\alpha_k    \equiv  \sqrt{\frac{\tilde{s}_i}{\tilde{s}_j \tilde{s}_{ij}}} \kappa_\perp e^{\bar{\eta}_Q}\,, \qquad
\beta_k    \equiv  \sqrt{\frac{\tilde{s}_j}{\tilde{s}_i \tilde{s}_{ij}}} \kappa_\perp e^{-\bar{\eta}_Q},
\end{align}
which enable us to write the light-cone momentum fraction at which we
need to evaluate the DGLAP splitting probabilities appearing in
eqs.~\eqref{eq:dipole-prob-FF},~\eqref{eq:dipole-prob-IF}.
If the original-dipole legs $\itilde$ and $\jtilde$ are final-state
partons, we define
\begin{equation}
  z_i = \alpha_k\,, \qquad z_j = \beta_k\,,
\end{equation}
while if $\itilde$ is an incoming parton, the definition of $z_i$ is
modified and we instead use
\begin{equation}
  z_i = \frac{\alpha_k}{1-\alpha_k}\,.
\end{equation}

\subsubsection{PanGlobal}
\label{sec:panglobal}
This section details the kinematic mapping of an antenna shower with global
transverse-momentum recoil.
We refer to this shower as PanGlobal.
The treatment of the longitudinal recoil is similar to the proposal for the PanGlobal variant for hadron collisions of
Refs.~\cite{vanBeekveld:2022zhl,vanBeekveld:2022ukn}.
The main difference is represented by the boost that is performed to
achieve momentum conservation in the perpendicular component.
This choice is motivated by the fact that in our case we want to
preserve $q_{\dis}^{\mu}$, i.e.~the momentum of the
$t$-channel exchanged boson, 
while for colour-singlet production the most natural
variables to preserve are the invariant mass and the rapidity of the colour
singlet system.
One begins by introducing some intermediate post-branching dipole
momenta, which read
\begin{subequations}
	\label{eq:mapping-PG}
\begin{align}
	 \bar p_i^{\mu} & = r_{\rm \sss L} (1\pm a_k) \tilde{p}_i^{\mu}\,, \\ 
	\bar p_j^{\mu} &=   r_{\rm \sss L} (1\pm b_k) \tilde{p}_j^{\mu}\,, \\
  \bar p_k^{\mu} & =  r_{\rm \sss L}(a_k \tilde{p}_i^{\mu}  + b_k \tilde{p}_k^{\mu}  + k_{\perp}^{\mu}) \,,
\end{align}
\end{subequations}
with $a_k=\alpha_k$, $b_k=\beta_k$ and
\begin{equation}
  k_\perp^{\mu}  = \sqrt{a_k b_k \tilde{s}_{ij}} \left( \hat{k}_{\perp1}^{\mu}  \sin \phi +  \hat{k}_{\perp2}^{\mu}  \cos \phi \right),
\end{equation}
with $\hat{k}_{\perp 1,2}$ two vectors with norm $-1$ orthogonal to
$\tilde{p}_{i,j}$ such that $\hat{k}_{\perp1}\cdot\hat{k}_{\perp2}=0$.
The signs in eq.~\eqref{eq:mapping-PG} depend on whether the $\itilde$
or $\jtilde$ is incoming ($+$) or outgoing ($-$).
The momentum mapping of eq.~\eqref{eq:mapping-PG} clearly does not
conserve momentum.
Indeed we have
\begin{equation} 
	\bar{p}_{\sss X}^{\mu} - \bar{p}_1^\mu =  q_{\rm dis}^{\mu} + r_{\rm \sss L} k_{\perp}^\mu + (r_{\rm \sss L} -1)(\tilde{p}_j^\mu \pm \tilde{p}_i^\mu)\,.
\end{equation}
where we have the $-$ sign for $i=1$ (an IF dipole) and $+$ otherwise (FF dipoles), and $\bar{p}_X$ is the sum of the momenta of all the post-branching final-state partons.
At variance with the original proposal Refs.~\cite{vanBeekveld:2022zhl,vanBeekveld:2022ukn}, we have introduced a local rescaling factor $r_{\rm \sss L}$, whose value depends on the type of the dipole. In particular, we have
\begin{equation}
  r_{\rm \sss L} =
  \begin{cases}
      \displaystyle
      \frac{\tilde{s}_i+\tilde{s}_j}{\tilde{s}_i+\tilde{s}_j + 2 k_\perp \cdot Q} \qquad &\mbox{FF dipoles} \\
        \displaystyle
       \hspace{1.5cm} 1 &\mbox{IF dipoles}.
  \end{cases}
  \label{eq:rLdef}
\end{equation} 
As explained in Appendix~\ref{sec:resc} and in
Ref.~\cite{DOUBLESOFT}, the local rescaling factors $r_{\rm \sss L}$
ensure that triple-collinear FF configurations (where
$k_t^2$ is small, but $k_\perp \cdot Q$ can potentially be large) do not result in a
large boost, or a large rescaling for $\tilde{p}_1$ in Eq.~\eqref{eq:rescale-p1} below.
The form of $r_{\rm \sss L}$ originates from imposing $\tilde{p}_{\sss X}\cdot Q = \bar{p}_{\sss X}\cdot Q$,
which acts as to preserve the energy of final state before and after the emission. As further detailed in Appendix~\ref{sec:resc}, no such factor is necessary for IF dipoles, hence we set $r_{\rm \sss L}=1$.
We stress that for small values of $ 2 k_\perp \cdot Q$, i.e.~when either $a_k$ or $b_k$ are small, we have $ r_{\rm \sss L} = 1$.

There is a considerable amount of freedom in how to 
implement the momentum reshuffling to restore momentum conservation.
However, at NLL accuracy, it is important to ensure that partons in
the remnant hemisphere are only marginally affected by the recoil
from emissions widely separated in angle.
For instance, one could choose to only boost the partons in the current
hemisphere.
However, we can find a configuration where the current hemipshere
is populated by only soft wide-angle  emissions.
Constructing the boost such that all the recoil is given to soft partons would then lead to infrared unsafe results.~\cite{Antonelli:1999kx}.
For this reason, we devised a smooth Lorentz transformation that acts
on all partons, but primarily modifies those with a substantial
component along $n_2^\mu$.
First, we first adjust the
momentum of the incoming parton $p_1^\mu$ so that
\begin{equation}
(p_1 + q_{\rm dis})^2=\bar{p}_{\sss X}^2.
\end{equation}
Requiring that $p_1^{\mu}$ only has a component in the direction of
$n_1^{\mu}$, gives us
\begin{align}
\label{eq:rescale-p1}
  p_1^{\mu} & = \left( 1 + \frac{\bar{p}_{\sss X}^2}{Q^2} \right) n_1^{\mu} 
 = \left(\frac{Q^2 + \bar{p}_{\sss X}^2}{Q^2+\tilde{p}_{\sss X}^2}\right) \tilde{p}_1^{\mu}\,,
\end{align}
where $\tilde{p}_{\sss X}^{\mu}= \tilde{p}_1^{\mu} + q_{\dis}^{\mu}$ denotes the pre-branching partonic final-state.
Finally, we boost the post-branching partonic final-state momenta from
$\bar{p}_{\sss X}^{\mu}$ to
\begin{equation}
  p_{\sss X}^{\mu} = n_2^{\mu} + \frac{\bar{p}_{\sss X}^2}{Q^2} n_1^{\mu}\,.
\end{equation}
The boost is constructed to ensure that partons parallel to
the incoming proton do not acquire any transverse-momentum. 
Instead, this component is absorbed by partons carrying a
substantial fraction of the original final-state quark momentum
$n_2^\mu$.
The boost that achieves this is derived in Appendix~\ref{sec:disboost}
and reads
\begin{align}
  \Lambda^{\mu \nu} =& g^{\mu \nu} + \frac{2 n_1^\mu}{Q^2} \left[ (\beta-1) n_2^\nu + \frac{p_t^2}{\beta Q^2} n_1^\nu + p_\perp^\nu\right] + \frac{2 n_2^\mu n_1^\nu}{Q^2} \frac{1-\beta}{\beta}-\frac{2 p_{\perp}^\mu n_1^\nu}{\beta Q^2}\,
  \label{eq:disboost}.
\end{align}
The components $\beta$ and $p_{\perp}$ are defined through a Sudakov decomposition
of the final-state momentum sum
\begin{equation}
  \bar p_{\sss X}^{\mu} = \frac{\bar{p}_{\sss X}^2 + p_t^2}{\beta Q^2} n_1^{\mu} + \beta n_2^{\mu} + p_\perp^{\mu}\,,
  \label{eq:px}
\end{equation}
with $p_\perp$ a space-like vector orthogonal to $n_{1}^{\mu}$,
$n_2^\mu$ with norm $-p_t^2$.  Notice that for $a_k$ and/or $b_k$ very
small, we have $1-\beta \sim p_t \sim \kappa_t$, so the boost
minimally alter the final-state momenta. Similarly, when $a_k$ or
$b_k$ are small, for FF emissions we have that $p_{1} \approx
\bar{p}_1$, so also the additional rescaling we apply to the incoming
parton.

\subsubsection{PanLocal}
\label{sec:panlocal}
In this section, we describe how to implement a dipole shower with
local recoil for DIS, which we refer to as PanLocal.%
\footnote{We omit the description of an antenna PanLocal version, as
future applications such as NLO matching or multi-jet merging would be
substantially more cumbersome, as found in Ref.~\cite{Hamilton:2023dwb}.}
For all three dipole types (IF, FI and FF), we parameterise the
momentum of the radiated parton as
\begin{equation}
  \bar p_k^{\mu} = a_k \tilde{p}_i^{\mu} + b_k \tilde{p}_k^{\mu} + k_{\perp}^{\mu}\,.
\end{equation}
The emitter can either be an initial-state parton (for IF dipoles), or
a final-state one (for FI and FF dipoles).
In the former case we use
\begin{equation}
  a_k = \alpha_k, \qquad b_k = \beta_k(1+\alpha_k)^{\frac{2}{1+\betaps}}.
  \label{eq:bkpanlocal}
\end{equation}
This choice was introduced in Ref.~\cite{vanBeekveld:2022zhl} to
restore transverse-momentum ordering for very hard-collinear
emissions, and serves to avoid unphysical correlations between emissions
in opposite hemispheres.
Like in Refs.~\cite{vanBeekveld:2022zhl,vanBeekveld:2022ukn}, the new
momenta of the emitter and the spectator become
\begin{align}
  \bar p_i^{\mu}  =& (1+a_k) \tilde{p}_i^{\mu}  + \frac{a_k b_k}{1+a_k} \tilde{p}_j^{\mu} + k_\perp^{\mu} \,, \\
  \bar p_j^{\mu}  =& \left(1 -\frac{b_k}{1+a_k} \right) \tilde{p}_j^{\mu} \,.
\end{align}
Although the momentum is locally conserved (i.e. $\bar{p}_{\sss
  X}^\mu-\bar{p}_1^\mu = q_{\dis}^\mu$), we now end up in a situation
where the incoming parton is no longer aligned with the beam direction
$n_1^{\mu}$.
The Lorentz transformation that we apply to realign the incoming parton
with the beam differs from the one applied for colour-singlet
production, since in this case we want to preserve the DIS
invariants.  
After this transformation, the momentum of the incoming
parton becomes
\begin{equation}
  p_i^\mu = p_1^{\mu} = \left( 1 + \frac{\bar{p}_{\sss X}^2}{Q^2} \right) n_1^{\mu}\,,
\end{equation}
and the sum of final-state partons
\begin{equation}
  p_{\sss X}^{\mu} = n_2^{\mu} + \frac{\bar{p}_{\sss X}^2}{Q^2} n_1^{\mu}\,.
\end{equation}
To achieve this, we first rotate all momenta except the photon
momentum $q_{\rm dis}^{\mu}$ with a rotation matrix $R(\theta)$. This
matrix is defined to align the post-branching incoming parton
momentum $\bar{p}_1$ along the direction of $n_1^{\mu}$.
This operation introduces a momentum imbalance since we do not change
the photon momentum $q_{\rm dis}^{\mu}$.  To restore momentum 
conservation we
proceed in the same way as for PanGlobal, i.e.~we Sudakov decompose
the sum of (rotated) final-state momenta as
\begin{align}
(R(\theta) \cdot \bar{p}_{\sss X})^\mu = \frac{\bar{p}_{\sss X}^2 + p_t^2}{\beta Q^2} n_1^\mu + \beta n_2^\mu + p_\perp^\mu\,,
\end{align}
and apply the boost of eq.~\eqref{eq:disboost} to all partons
including the initial-state one.

In the case where the emitter is instead a final-state parton we set
\begin{equation}
	a_k = \alpha_k, \qquad b_k = \beta_k\,.
\end{equation}
The momenta of the emitter and spectator in the case of an FF/FI
dipole reads
\begin{align}
	\bar p_i^{\mu}  =& (1-a_k) \tilde{p}_i^{\mu}  + \frac{a_k b_k}{1-a_k} \tilde{p}_j^{\mu}  - k_\perp^{\mu} \,, \\
	\bar p_j^{\mu}  =& \left(1 \pm \frac{b_k}{1-a_k} \right) \tilde{p}_j^{\mu} \,,
\end{align}
where the $+$($-$) sign is required for FI(FF) dipoles.
Note that no boost/rotation needs to be performed in these cases since
the initial-state parton does not acquire a transverse momentum
component.
Hence, $\bar{p}_{i,j,k}^{\mu} = p_{i,j,k}^{\mu} $ for FI and FF
dipoles.

\subsection{Extension to VBF and VBS}
\label{sec:VBF-shower}
From the point of view of QCD radiative corrections, VBF
(Fig.~\ref{fig:VBF}) and VBS (Fig.~\ref{fig:VBS}) can be seen as a
double copy of DIS (Fig.~\ref{fig:DIS}).
Non-factorisable corrections arising from the exchange of partons
between the two hadronic sectors are colour suppressed and only
contribute from order $\alpha_s^2$.
They were computed for the first time in Ref.~\cite{Liu:2019tuy} for
Higgs production in VBF, using the eikonal approximation, and found to
be typically ten times smaller than the NNLO factorisable
corrections~\cite{Cruz-Martinez:2018rod,Dreyer:2018rfu,Asteriadis:2021gpd},
as confirmed by the phenomenological study of
Ref.~\cite{Dreyer:2020urf}, which also addresses double Higgs
production.

For this reason, to shower a VBF or VBS process, we treat
the two hadronic sectors as two separate and independent DIS
processes.
This is done by labelling the two sectors $a$ and $b$,
and using for each of them a separate reference vector $Q^\mu_{a}$
and $Q^\mu_b$
that reads
\begin{align}
Q^\mu_{a} = n_{1, a}^{\mu} + n_{2, a}^{\mu}\,,
\end{align}
where $n_{1,a}^\mu, n_{2,a}^\mu$ are the
light-like reference vectors introduced in Sec.~\ref{sec:breit}
such that $q^{\mu}_{\dis, a} = n_{2, a}^{\mu}- n_{1, a}^{\mu}$.
Here $q^\mu_{\dis, a}$ describes the four-momentum of
the $t$-channel colour-singlet boson that is exchanged
between the initial- and final-state partons. 
A similar definition is employed for sector $b$, with different
light-like reference vectors $n_{1,b}^\mu$ and $n_{2,b}^\mu$.
Since the showering of the two hadronic sectors is completely
factorised, we have the freedom to choose different values for the
shower starting scale for the two different sectors, as well as the factorisation and renormalisation
scales employed in the underlying fixed-order calculation.

Once we have established SL accuracy 
for the showers in the DIS process, 
extending this accuracy to VBF topologies is conceptually straightforward (especially in the case of VBF), 
with the exception of non-factorisable corrections, which are not included.%
\footnote{We remind the reader that interferences between the
  diagrams where the tagged bosons in the VBS process are connected to
  the $t$-channel propagator (Fig.~\ref{fig:VBS}), and those where the
  tagged bosons in the VBS process are emitted directly from the quark
  legs spoil the structure-function picture, so higher-order
  corrections to the VBS process are not simply those for DIS squared. }
The terms that we are neglecting are
colour-suppressed NLL corrections.
Describing correctly the factorisable contributions is however
sufficient to
obtain full-colour accuracy at LL, and NLL accuracy in the large $\nc$
limit. 
Additionally, non-factorisable contributions are typically suppressed
after applying VBF cuts~\cite{Dreyer:2020urf}, and also unknown in their complete form
(only approximated results have been obtained at fixed order in Ref.~\cite{Liu:2019tuy}).
%

\section{Fixed-order tests}
\label{sec:fo-test}
One fundamental property that NLL-accurate parton showers must satisfy
is that emissions widely separated in rapidity or transverse momentum
must be independent~\cite{Dasgupta:2018nvj}.
This requirement follows from the factorisation
properties of the underlying QCD matrix element 
in the limit of soft and/or collinear emissions.
It can be directly translated into a ``fixed-order'' criterion: ``a
soft emission can alter the momentum of a previously radiated parton
only if they are very close in rapidity and emitted with a
commensurate transverse momentum''.

To better understand this requirement, consider an emission with
momentum $\tilde{k}_1^{\mu}$ from a dipole.
After this, the shower will try to generate a second emission, with
momentum $k_2^{\mu}$.
The redistribution of momenta (either directly in the kinematic map or
via a momentum-conserving boost) may alter the kinematics of the first
emission, i.e.~$\tilde{k}_1^{\mu} \to k_1^{\mu}$.
The result is that the event now has two emissions with momenta
$k_1^{\mu}$ and $k_2^{\mu}$, but the first emission was generated with
a matrix element corresponding to $\tilde{k}_1^{\mu}$.
The matrix element corresponding to $k_1^{\mu}$ and $\tilde{k}_1^{\mu}$
are not the same when these momenta are vastly different, hence this
would destroy the factorisation property it must obey when $k_2^{\mu}$
is either collinear to any of the partons in the pre-branching event
or very soft.
This process gets repeated for every emission that follows, leading to
a wrong logarithmic exponentiation starting at NLL (at leading colour
accuracy).
To verify the condition above, in Sec.~\ref{sec:fo-test-contours} we
investigate the behaviour of the showers introduced in the previous
section in the presence of two gluons emitted at commensurate values of
the ordering variable $v$.%
\footnote{Note that this is a necessary, but not sufficient
requirement to reach NLL accuracy for global observables. 
For the latter, it is important that any pair of emissions are described
accurately in the shower, while in the fixed-order test of Sec.~\ref{sec:fo-test-contours}
only the action of the second emission on the first emission's momentum
is probed. }

Another fixed-order criterion that we test is that
subleading-colour corrections are correctly implemented in case of
strongly-ordered emissions.
Indeed, dipole showers are implemented in the large-$\nc$ limit, 
but subleading (i.e.~$\propto 1/\nc^2$) corrections in the
LL contribution have the same numerical size of NLL terms, thus from a
phenomenological point of view they have the same relevance and must
be included.
For this reason, in Sec.~\ref{sec:colour} we considered two
strongly-ordered emissions and we assess whether subleading-colour
corrections have been correctly incorporated in the shower at fixed
order through a comparison to the exact analytic matrix element.
Note that this algorithm is expected to
not only yield the correct result at LL full-colour, but also at NLL for many
observables, such as average particle multiplicity (Sec.~\ref{sec:multiplicity}) 
and global
observables (Sec.~\ref{sec:global-obs}), as well as for the DGLAP
evolution~(Sec.~\ref{sec:dglap}).

The final ingredient necessary to reach NLL is the implementation of
spin correlations to correctly reproduce the azimuthal distributions
of the radiation.
To this aim, in appendix~\ref{sec:spin} we show that the
algorithm introduced in Ref.~\cite{vanBeekveld:2022zhl} for PanScales
showers for hadron collisions can be applied also to DIS-type
processes.
However, since spin correlations do not impact the NLL resummation of
the observables we use to validate the NLL accuracy of our showers at all
orders (detailed in Sec.~\ref{sec:ao-test}), we do not include them in
our all-order tests.

\subsection{Lund-plane contours}
\label{sec:fo-test-contours}
In this section, we investigate if emissions of commensurate hardness
are independent when widely separated in angle, as required to achieve
NLL accuracy.
We consider the DIS process $q\, \gamma^* \to q$, with $x_{\dis} =
0.01$, and we emit a gluon $g_1$ from the $qq$ dipole at a fixed
shower evolution scale of $\ln v_1/Q=-20$, azimuthal angle $\phi_1 =
0$, and a few fixed values of $\bar{\eta}_{Q,1}$.
We then examine the impact of a second gluon $g_2$, emitted at a scale
$\ln v_2/Q = \ln v_1/Q -2$, on the kinematics of $g_1$, varying
$\bar{\eta}_{Q,2}$.  Its azimuthal angle is fixed at $\phi_2 = 0$.
We parameterise the available phase space of the emissions in terms of
the Lund variables $\ln k_{t,i}/Q$ and $\eta_i$~\cite{Andersson:1988gp}. 
Some care has to 
be taken in defining these variables, in particular for emissions
off the final-state quark.
If only one gluon emission is present, we calculate two angular
distances, one that parameterises the distance from the beam ($d_{\sss
  kB}$), and one to the final-state quark ($d_{\sss kF}$)
\begin{equation}
	\label{eq:angular-dist}
d_{\sss kB} = 1-\cos \theta_k \,, \qquad d_{\sss kF} = 1-\cos \theta_{kq}\,.
\end{equation}
where $\theta_k$ is the angle between $k$ and the beam, while
$\theta_{kq}$ is the angle between $k$ and the final-state quark $q$,
both defined in the Breit frame.
We then define the angle
\begin{equation}
  \cos \theta \equiv 
    \begin{cases}
     +\cos\theta_{kq}  \qquad \mbox{ if }  d_{\sss kF}<d_{\sss kB}\,,\\
     -\cos\theta_k     \qquad \mbox{ if }  d_{\sss kF}>d_{\sss kB}\,,
    \end{cases}
    \label{eq:costheta}
\end{equation}
so that the Lund variables read
\begin{equation}
	\label{eq:lundvar}
  \eta    = \frac{1}{2} \ln \frac{1+\cos\theta}{1-\cos\theta}\,, \qquad
  \ln \frac{k_t}{Q}     = \frac{1}{2} \ln\frac{ E_k^2(1-\cos^2\theta)}{Q^2}\,.
\end{equation}
The sign in eq.~\eqref{eq:costheta} is chosen such that partons in the
current hemisphere have positive rapidity, while partons in the
remnant hemisphere have negative rapidity.
Now let us move to the case where two gluons are present.
In this case, we first calculate the angular distances of
eq.~\eqref{eq:angular-dist} for both gluons.
Focusing on emissions on the primary Lund plane, there are two
possible scenarios: the smallest angular distance can either be to the
beam, or the final-state quark. 
Note that the third scenario, where the second gluon is
closest to the first emitted gluon, is instead described by a
secondary Lund plane, which is not considered here. 
If the smallest angular distance is $d_{k_iB}$, we simply use the
definition of the Lund variables of eq.~\eqref{eq:lundvar} for both
gluon emissions.
If instead the smallest angular distance is between any of the two
gluons and the final-state quark, we merge them into a
new `final-state' momentum $p_F^{\mu}$.
We then recalculate $d_{\sss kF}$ using that momentum, and calculate
the new Lund variables for the last recombined emission.

Notice that this definition of the Lund variables shares some
similarities with the variables associated with the generalisation of
the Cambridge/Aachen algorithm for DIS, detailed in
Appendix~\ref{sec:decl}.
The main differences are first that for the current case, we are only
interested in describing the `primary' Lund plane, ignoring the case
where the two emissions would first cluster together, as we want to
focus on the case in which the emissions are widely separated in
angle.
Secondly, to define the transverse momentum of the splitting we always
use the energy of the radiated gluon, while in the flavour-blind
algorithm of Appendix~\ref{sec:decl} we take the energy of the softest
parton.

\begin{figure}[t]
    \centering
    \includegraphics[page=4, width=0.45\textwidth]{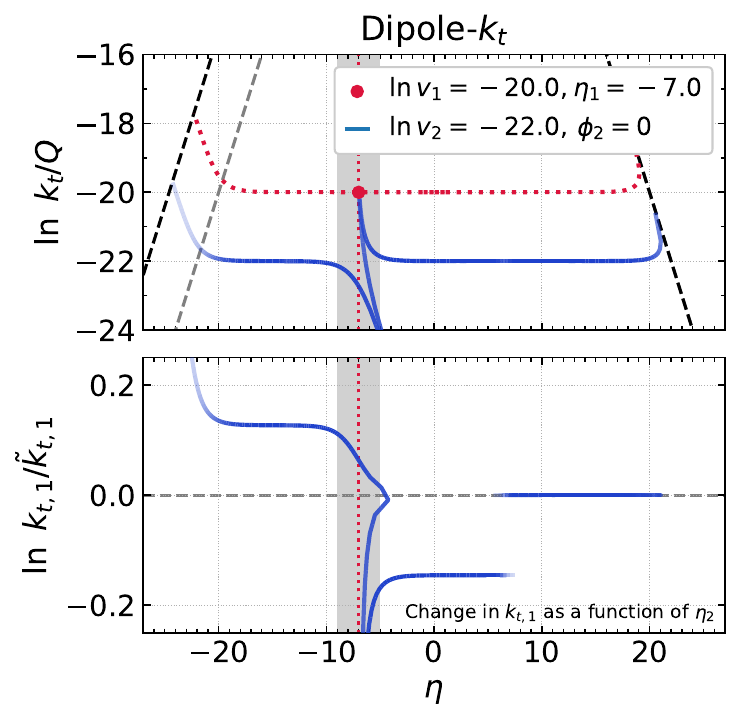}
    \includegraphics[page=5, width=0.45\textwidth]{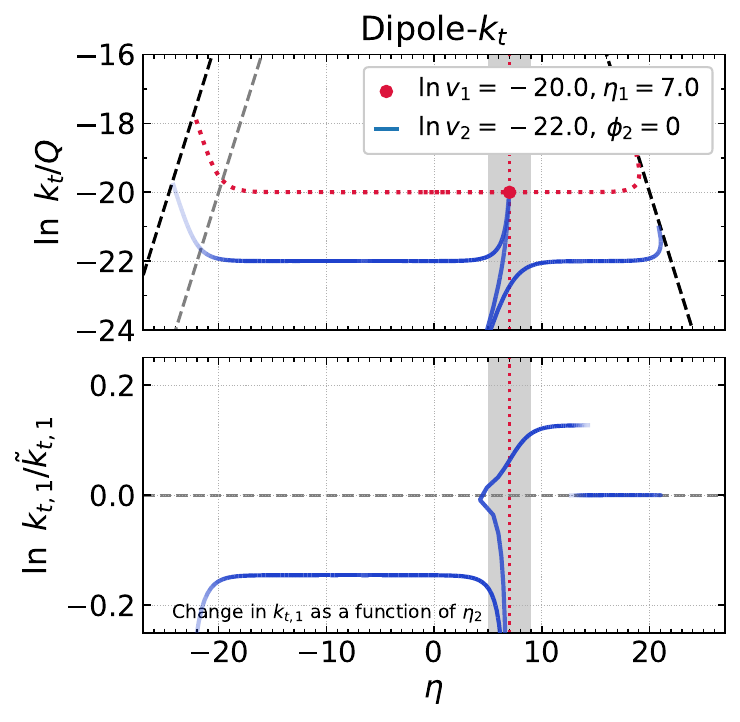}
    \includegraphics[page=12, width=0.45\textwidth]{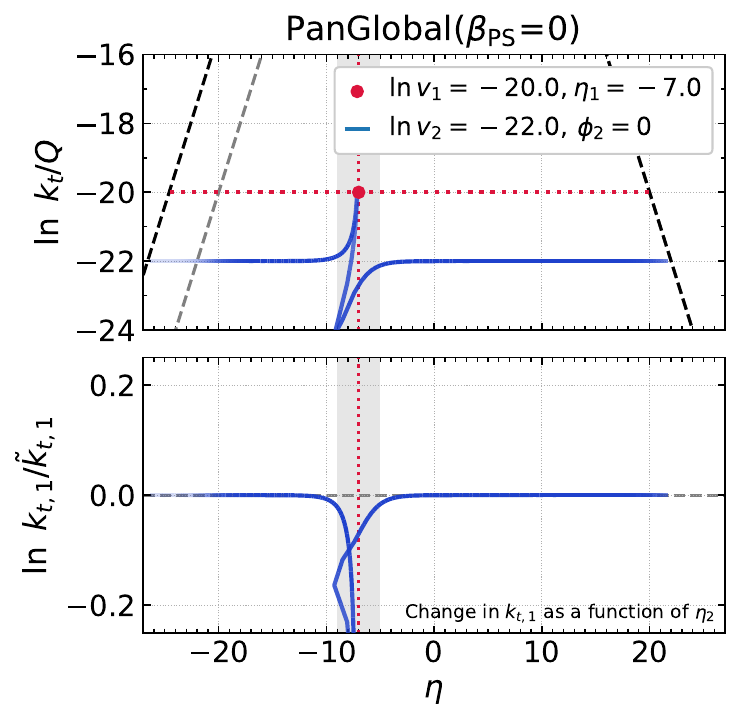}
    \includegraphics[page=13, width=0.45\textwidth]{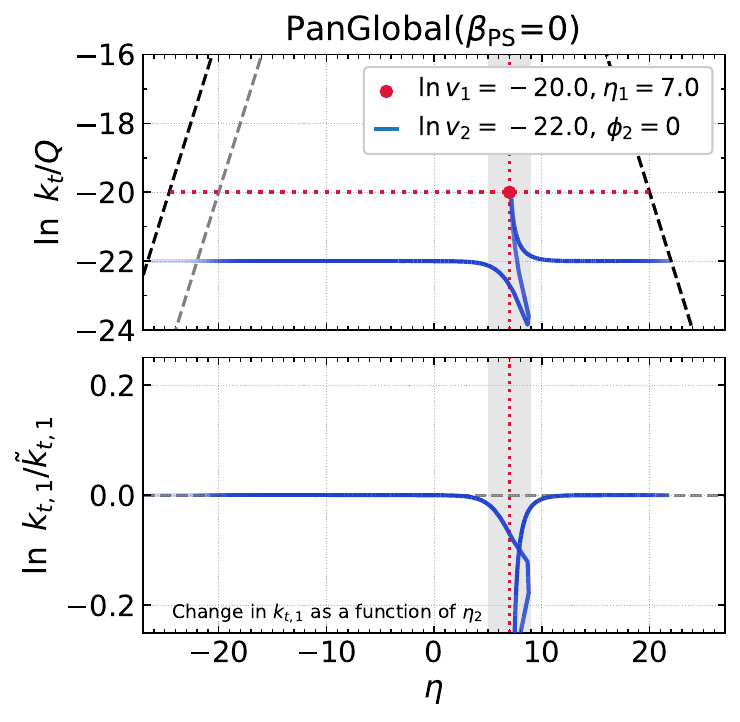}
    \caption{Double-emission contours for Dipole-$k_t$ (top) and
      PanGlobal($\betaps = 0$) (bottom) for a first soft-collinear
      emission in the remnant hemisphere~(left) and in the current
      hemisphere~(right) for a DIS process with $x_{\rm DIS}=0.01$ and
      an incoming proton with negative rapidity in the Breit frame.
      The phase-space contours are shown as a function of $\ln
      k_{t}/Q$ and $\eta$. A red dot indicates the kinematics of the
      first emission which is fixed at $\eta_1 = -7$ (left) or $\eta_1
      = 7$ (right) and $\ln v_1/Q=-20$ (in the plot labels, values of
      $v_i$ are always expressed in units of $Q$).
      The contour of the first emission, obtained by fixing $\ln
      v_1/Q$ but varying $\eta_1$, is shown with the mostly horizontal
      red dotted line.  That of the second emission at $\ln v_2/Q=-22$
      is drawn as a blue solid line.
      The colour shading of the lines indicates the branching
      probability.
      In the bottom panels, we show the logarithm of the ratio between
      the transverse momentum of the first emission after ($k_{t,1}$)
      and before the second emission took place
      ($\tilde{k}_{t,1}$). This is expected to be zero (dashed grey
      line) except when the two emissions are close in rapidity (grey
      vertical band, where the rapidity of the first emission is
      indicated with a vertical red dotted line).
    }
    \label{fig:contour-dipole-kt-soft}
\end{figure}
On the top two panels of Fig.~\ref{fig:contour-dipole-kt-soft} we illustrate
the two-emission contours for Dipole-$k_t$, a
transverse-momentum ordered shower. 
We notice that the first emission erroneously takes the
transverse-momentum recoil if
\begin{equation}
\eta_2 < \frac{1}{2} \left(\eta_1 -\ln \frac{v_1}{Q} \right).
\end{equation}
This is because this shower uses a fully local map, such that the
first gluon always absorbs the recoil whenever the second emission
comes from the new initial-final $(q_I g_1)$ dipole in the region
$\eta_2<\eta_1$, and because the midpoint of the final-final dipole is
assigned in the dipole frame in the region where $\eta_2>\eta_1$.

\begin{figure}[tb]
    \centering
    \includegraphics[page=35, width=0.49\textwidth]{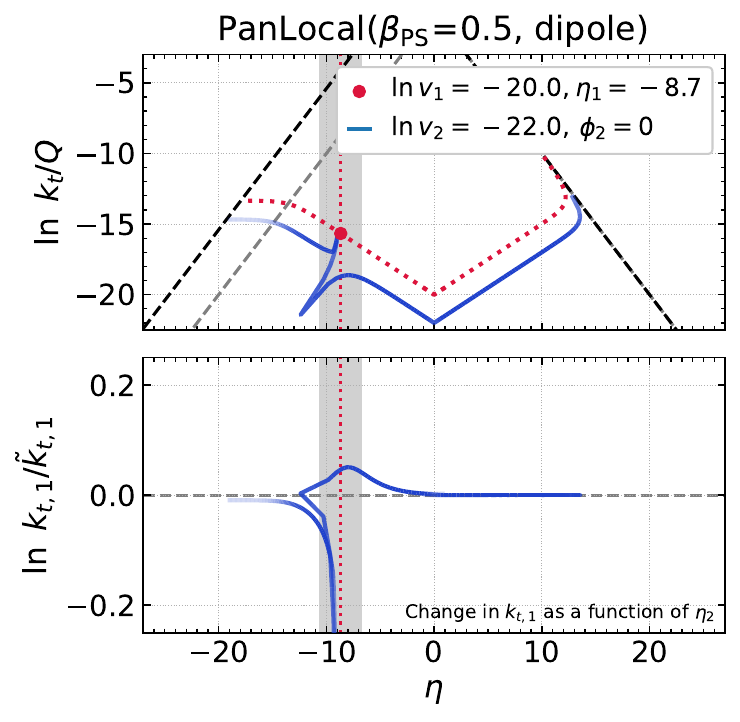}
    \includegraphics[page=19, width=0.49\textwidth]{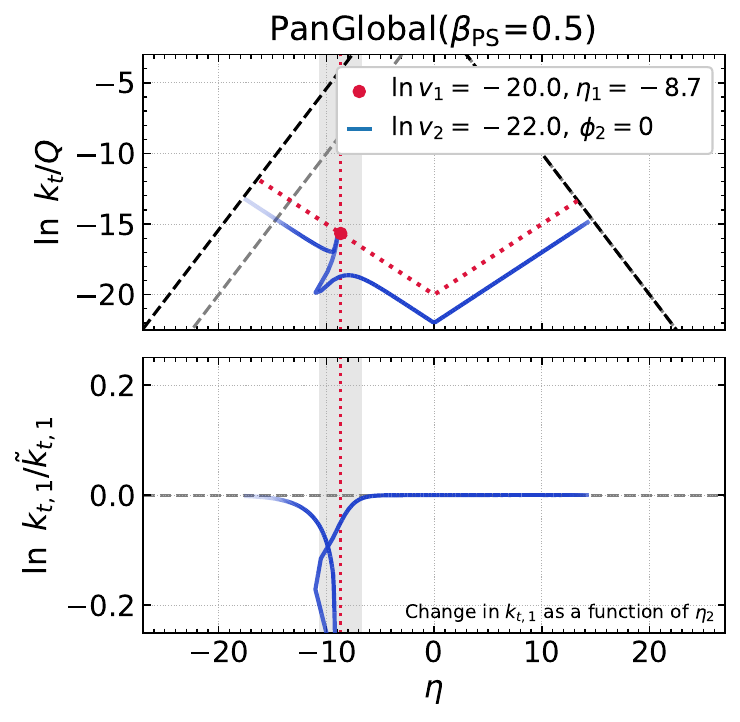}
    \caption{
      Same as Fig.~\ref{fig:contour-dipole-kt-soft}, but for PanLocal($\betaps = 0.5$, left) and 
      PanGlobal($\betaps = 0.5$, right) for $\eta_1 = -8.7$.
    }
    \label{fig:contour-PG-PL-beta05}
\end{figure}

This behaviour can be corrected by either choosing a different
evolution variable, or by conserving the transverse momentum globally,
as done by the PanGlobal($\betaps=0$) shower.
As discussed in Sec.~\ref{sec:panglobal}, the global boost that takes
care of redistributing the transverse momentum imbalance has the
property of affecting mostly those partons at very large (and positive)
rapidity, while leaving the ones living in the remnant hemisphere
unchanged.
Indeed, the bottom two panels of Fig.~\ref{fig:contour-dipole-kt-soft}
show that subsequent emissions widely separated in rapidity leave the
Lund variables associated with the first emission unaffected.
The results obtained after taking a different ordering variable are
shown in Fig.~\ref{fig:contour-PG-PL-beta05} for PanGlobal($\betaps =
0.5$) and PanLocal.  Here the ordering variable is set to $v \approx
k_t e^{-\betaps|\eta|}$, with $\betaps = 0.5$.
As already discussed in Ref.~\cite{Dasgupta:2020fwr} local
transverse-momentum conservation requires one to choose $\betaps>0$.

\begin{figure}[tb]
	\centering
	\includegraphics[page=40, width=0.49\textwidth]{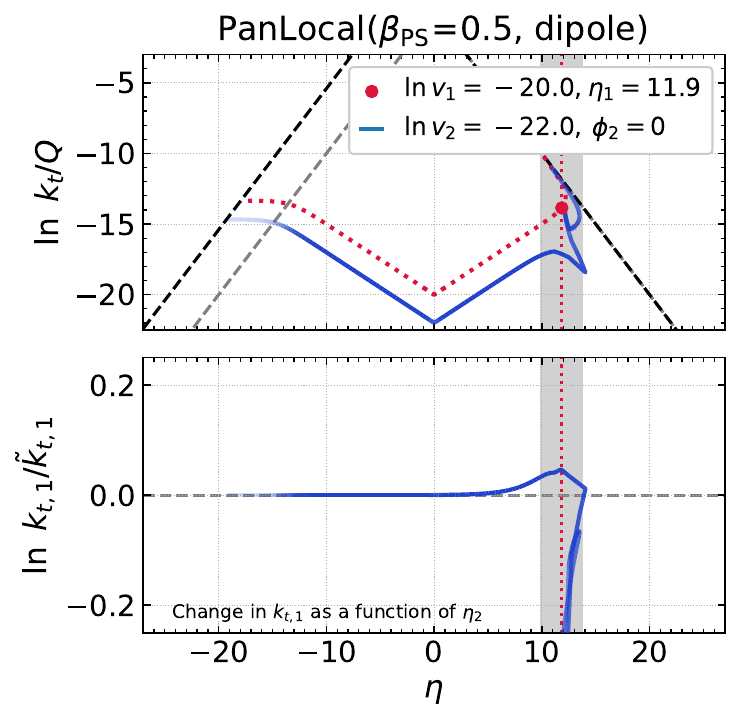}
	\includegraphics[page=24, width=0.49\textwidth]{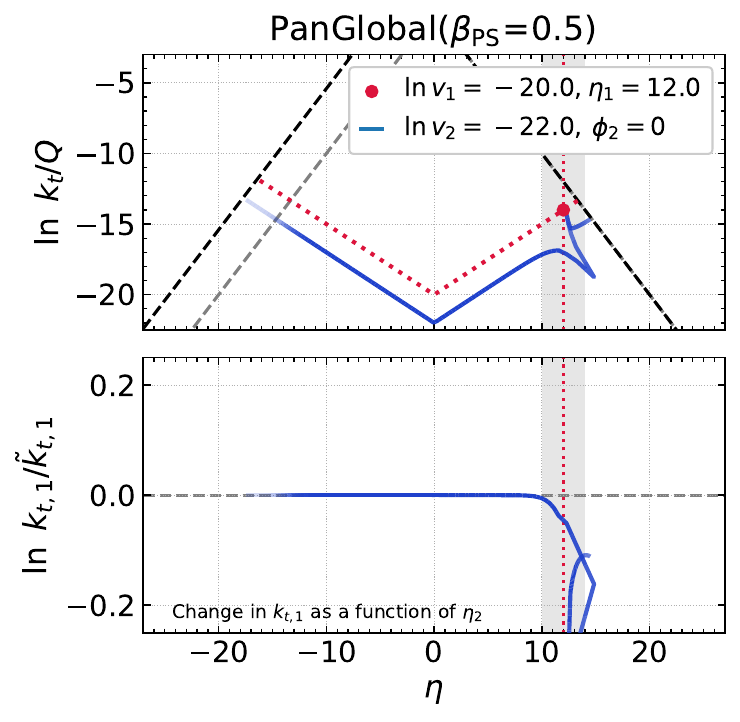}
	\caption{
	  Same as Fig.~\ref{fig:contour-PG-PL-beta05}, but considering a hard-collinear first emission with $\eta_1 = 12$.
        }
	\label{fig:contour-coll}
\end{figure}

Finally in Fig.~\ref{fig:contour-coll} we show results when choosing a
different value for $\eta_1$ resulting in a hard-collinear first
emission.
They are analogous to the case where the first emission is
soft-collinear, but note that care has to be taken for the PanLocal
shower.
At variance with the PanGlobal($\betaps =0.5$) case, for very
collinear initial-state radiation the ordering variable for PanLocal
behaves like a transverse momentum.
As discussed in Ref.~\cite{vanBeekveld:2022zhl} and in
Sec.~\ref{sec:panlocal} around eq.~\eqref{eq:bkpanlocal}, this
modification in the local shower is necessary to prevent a very
hard-collinear emission from significantly impacting the momentum of a
softer previous emission.

We conclude that in the DIS PanScales showers, conversely to
standard dipole showers, emissions widely separated in angle
leave the kinematics of the previous emissions untouched.
The novelty of our showers is that the transverse-momentum recoil due
to initial-state-radiation is effectively absorbed by hard partons in
the current hemisphere, while soft partons or partons collinear to the
initial state remain unaffected, as required from colour coherence.

\subsection{Subleading-colour corrections}
\label{sec:colour}

Dipole showers are developed using the large-$\nc$ approximation of
QCD, and designed to correctly describe not only collinear emissions,
but also soft wide-angle gluon emissions in the large-$\nc$ limit. 
In standard dipole showers, subleading-colour corrections
are included by replacing $C_A/2 = \nc /2$ with $C_F = (\nc^2 -1)/(2\nc)$
when a quark leg is identified as emitter (we refer to this as
the colour-factor-from-emitter scheme, CFFE).
It has been known for quite some time~\cite{Gustafson:1987rq} that
this choice is inconsistent with colour coherence and more
recently~\cite{Dasgupta:2018nvj} it was observed that it leads to
wrong (subleading-colour) contributions already at LL, due to the
incorrect assignment of the emitter.
Since $1/\nc^2 \sim \alpha_s$, these LL mistakes have the same size of
NLL terms, so it stands to reason that these subleading colour terms
should be included in an NLL-accurate shower.

Two schemes that do result in full-colour accuracy for LL terms were
introduced in Ref.~\cite{Hamilton:2020rcu} for showers applicable to
$e^+e^-$ collisions, which also have been generalised for
colour-singlet production at hadron colliders in
Ref.~\cite{vanBeekveld:2022zhl}: the so-called segment and NODS
schemes.
These schemes furthermore result in correct full-colour accuracy at
NLL in the case of global observables, and next-to-double-logarithmic
(NDL) accuracy for jet multiplicity, the latter of which is
sensitive to the integrated rate of double-soft energy-ordered
emissions at commensurate angles.
The segment colour scheme divides each dipole into an arbitrary number
of distinct segments, identified through the generation variable
$\bar{\eta}_Q$.  These segments either have a $C_F$ or $C_A/2$ colour
factor, and are assigned respecting colour coherence.
The nested-ordered-double-soft (NODS) scheme was designed to not only
get the correct integrated rate of two soft and energy-ordered
emissions that occur at commensurate angles, but also describe them
correctly at the differential level.
This is achieved by applying a matrix element correction that describes
a pair of energy-ordered commensurate-angle emissions, well separated
in rapidity from all other emissions. 
For more details on these schemes, see Refs.~\cite{Hamilton:2020rcu,vanBeekveld:2022zhl}.
These algorithms are straightforwardly extended to the DIS case.

\begin{figure}[t]
    \centering
    \begin{subfigure}{0.49\textwidth}
        \includegraphics[page=1, width=\textwidth]{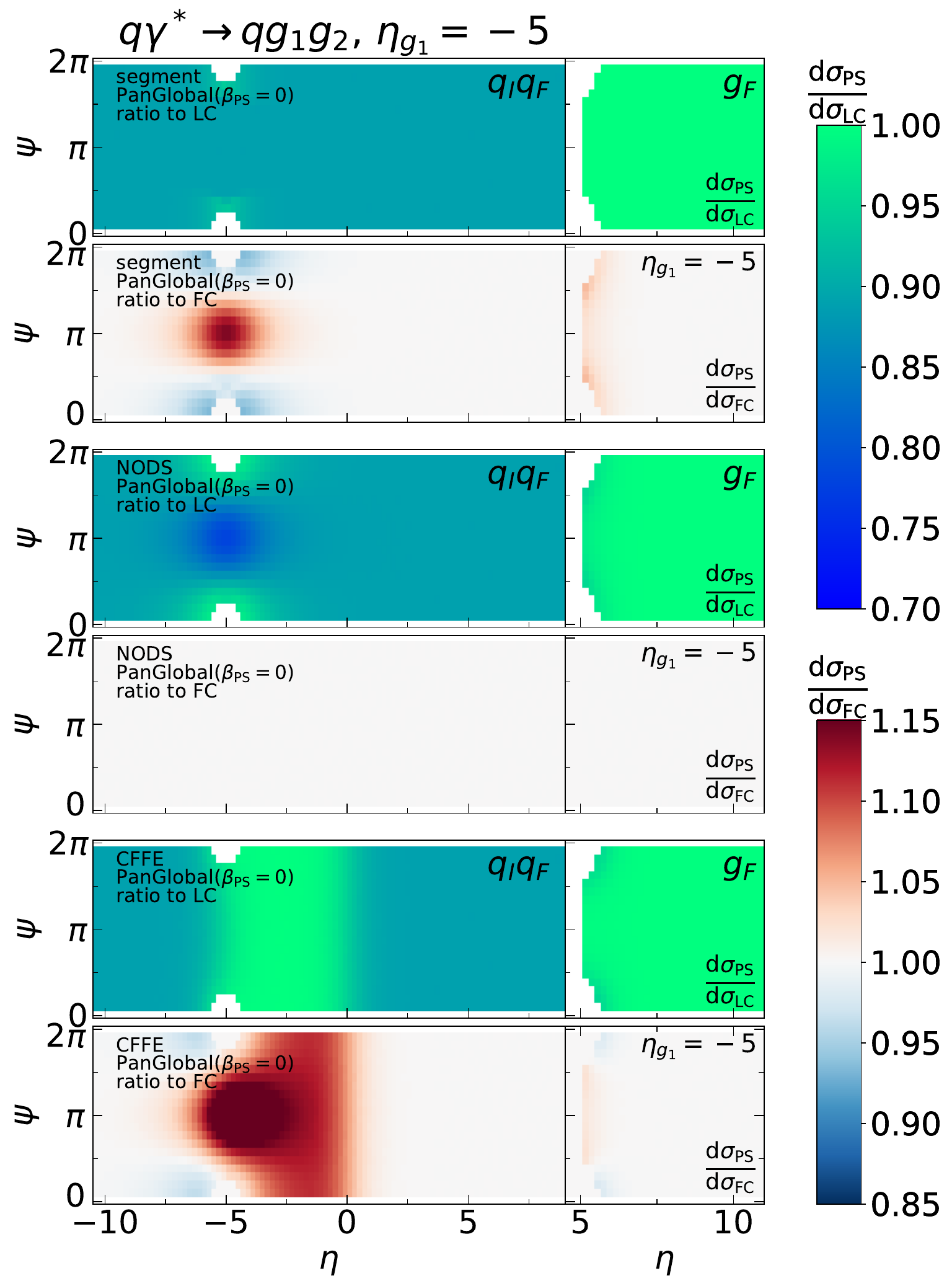}
    \end{subfigure}\hfill
    \begin{subfigure}{0.49\textwidth}
      \includegraphics[page=2, width=\textwidth]{plots/plot-2dcolour.pdf}
    \end{subfigure}
    \caption{Density for the emission of an ultra-soft gluon from a $q
      \gamma^* \rightarrow q g_1$ configuration (left), or from a $g
      \gamma^* \rightarrow q \bar{q} $ system (right).  In both cases,
      the incoming parton has a negative rapidity in the Breit frame,
      the final-state quark a positive one, and 
      the first-emitted final-state parton ($g_1$ in
      the left pane, $\bar{q}$ in the right one) has a rapidity
      $\eta_1=-5$.
      The emission density is illustrated as a function of the Lund
      variables $\eta$ and $\psi$ (defined in the main text). 
      For each configuration, the left panel corresponds
      to the primary Lund plane, while the right panel represents the
      secondary Lund plane (populated by emissions from the parton
      located at $\eta_{1}=-5$, see Appendix~\ref{sec:decl}).
      From top to bottom, the three 
      rows show the results for the PanGlobal($\betaps = 0$)
      shower
      result with the segment, NODS and CFFE colour scheme implemented
      For each row, the upper panel illustrates the ratio
      between the parton shower differential cross section and the
      leading colour~(LC) result, $\dd\sigma_{\rm
        PS}/\dd\sigma_{\rm LC}$,  obtained setting
      $C_A = 2C_F = 3$,
      while the lower panels show the deviation from the full colour~(FC)
      differential matrix element,  $\dd\sigma_{\rm
        PS}/\dd\sigma_{\rm FC}$.
    }
    \label{fig:colour-2dcolour-qq}
\end{figure}
To validate these schemes for DIS, we perform tests of the
differential matrix element produced by the shower after two 
strongly-ordered emissions.
The kinematics of the first emission (which can either be a quark or a
gluon) is fixed at $\eta_{1} = -5$, $\phi_1 = 0$ and $\ln v_1 / Q =
-10$.
The second emission is then emitted at a fixed value for $\ln v_2 / Q
= -60$, and sampled over $\bar{\eta}_{Q,2}$ and $\phi_2$.
The resulting shower predictions are then compared with the analytic
result at full colour~(FC).

In Fig.~\ref{fig:colour-2dcolour-qq} we show the results obtained with
PanGlobal($\betaps=0$) shower employing the segment, NODS and CFFE
colour schemes.
The emission density on the primary and secondary Lund plane are shown
in terms of the rapidity $\eta$ of the second emission, and its
azimuthal angle $\psi$.  These variables are defined according to our
generalised DIS Cambridge-Aachen algorithm~(Appendix~\ref{sec:decl}).
Specifically, for emissions in the primary Lund plane, we define $\psi
= \phi_2-\phi_1 +\pi$ (or $\psi = \phi_2-\phi_1 -\pi$, if
$\phi_2-\phi_1$ is larger than $\pi$).
For emissions in the secondary Lund plane we instead use
\begin{equation}
\psi = \arctan \frac{ \phi_2-\phi_1}{\eta_2-\eta_1} +\pi\,.
\end{equation}
Let us first focus on the ratio between the parton shower and the
analytic LC result, obtained using $C_A = 2C_F = 3$.
These are illustrated in the upper set of panels (coloured
green-blue).
For the majority of phase space, the value of this ratio is either $1$
or $8/9$, when the effective colour factor is either $C_A/2$ or $C_F$.
The lower set of panels (coloured red-white-blue) shows the ratio
between the parton shower and the analytic FC result.
As expected, the segment scheme assigns the correct colour factor
everywhere except for the region where the second gluon is close in
angle to the first emission.
Note however that these deviations integrate to $0$ after averaging over
the angular phase space,
therefore the all-order validations in
Sec.~\ref{sec:ao-test} would not be sensitive to this effect.
The NODS colour scheme corrects for this effect and produces the
analytic differential matrix element also in the region where a pair of
energy-ordered emissions occur at commensurate angles.
The last set of plots shows the deviation from the CFFE scheme, which
is currently implemented in standard dipole showers.  This
results in a wrong colour assignment over a region in phase
space that is logarithmically extended.  Hence this will result in a
wrong LL term at FC, as already pointed out in
Ref.~\cite{Dasgupta:2018nvj}.

Note that although here we have only shown results for the PanGlobal($\betaps = 0$)
shower, analogous results can be obtained for PanGlobal($\betaps = 0.5$) and PanLocal. 
Similar considerations also apply to the Dipole-$k_t$ shower.
However, the fraction of phase space that has the wrong subleading
colour corrections in the CFFE scheme is larger, as the emitter is
chosen by partitioning the dipole frame, instead of in
the Breit frame.

\section{All-order validation}
\label{sec:ao-test}

In this section, we investigate the all-order behaviour of the
PanScales showers for DIS over a broad range of observables,
targeting distinct classes of next-to-leading logarithmic effects, 
i.e.~those of the form $\as^n L^n$.
In Sec.~\ref{sec:dglap} we test the capability of the showers to
reproduce the DGLAP evolution of the parton distribution functions at
single-logarithmic (SL) accuracy, which probes nested emissions in the
hard-collinear region.
In Sec.~\ref{sec:multiplicity} we focus on the average particle multiplicity
at next-to-double-logarithmic accuracy (NDL), targeting nested
emissions in both the soft and collinear regions.
In Sec.~\ref{sec:global-obs} we compare the shower cumulative cross
section against the NLL predictions for several continuously-global
observables, which is the only test sensitive to both
double- and single-logarithmic terms in the 
Sudakov exponentiation (including running coupling
effects), 
created through soft and
collinear emissions.
Finally, in Sec.~\ref{sec:non-global-obs} we study a non-global
observable at SL accuracy, which probes the description of soft
large-angle emissions in the shower.
In Appendix~\ref{sec:asto0} we comment on the size of NNLL
corrections.

For each observable, we compare the shower's predictions to the known
resummation.
Subleading colour corrections are accounted for using the NODS scheme.
Although it is possible to include spin correlations (see
Appendix~\ref{sec:spin}), this is not done for the tests described
below as none of the observables is sensitive to them at our
targeted accuracy.

\subsection{DGLAP evolution}
\label{sec:dglap}

Here we test the capability of our showers to reproduce the
DGLAP~\cite{Gribov:1972ri,Dokshitzer:1977sg,Altarelli:1977zs}
evolution of parton distribution functions~(PDFs).
More specifically, we fix the starting point of the shower to be the
tree-level DIS process $q\, \gamma^* \to q $, where the incoming quark
$q$ is a down quark and has an energy fraction $x_{\dis}$, the photon
has a space-like momentum equal to $q_{\dis}$ with $Q^2 =
-q_{\dis}^2$, and we run our showers from the scale $Q$ until a given
transverse-momentum cutoff $p_{t,\rm{cut}} =Q e^L$.
This cutoff is applied to the scale used as the argument of the running
coupling, and it corresponds to the ordering-scale $v$ for
Dipole-$k_t$, and to $\kappa_\perp$ defined in
eq.~\eqref{eq:kappaperp} for the PanScales showers.
The flavour and energy fraction of the original quark extracted from
the proton can change throughout the shower evolution.
When the showering terminates the parton extracted from the proton
will have a flavour denoted by $i$, and an energy fraction $x\geq
x_{\dis}$.
The dominant term of the distribution over $i,x$ is SL ($(\alpha_s
L)^n$ with $L = \ln p_{t,\rm{cut}}/Q$).  The expected distribution
will take the form
\begin{equation}
\frac{1}{\sigma} \frac{d\sigma_i}{d x} =  \frac{1}{f_{q}(x_{\dis}, Q^2)} \int_{x_{\dis}}^1 \frac{{\rm d}z}{z} D_{q i} \left(z,\as L \right) f_i\left(\frac{x_{\dis}}{x}, p_{t,{\rm cut}}^2\right)\delta\left(\frac{x_{\dis}}{x} - z\right),
\end{equation}
where $f_j(y, \mu^2)$ is the density of partons of flavour $j$,
carrying momentum fraction $y$ at a factorisation scale $\mu$.
The (single-logarithmic) DGLAP evolution operator $D_{ij}(z,\as L)$ is
defined such that the parton density functions satisfy
\begin{align}
f_q(x,\mu^2) = \sum_j \int_{x}^1 \frac{{\rm d}z}{z}\, z\, D_{qj}(z,\alpha_s L)f_j\left(\frac{x}{z}, p_{t,{\rm cut}}^2\right)\,.
\end{align}

\begin{figure}[t]
  \begin{subfigure}{\textwidth}
    \includegraphics[width=\textwidth, page=1]{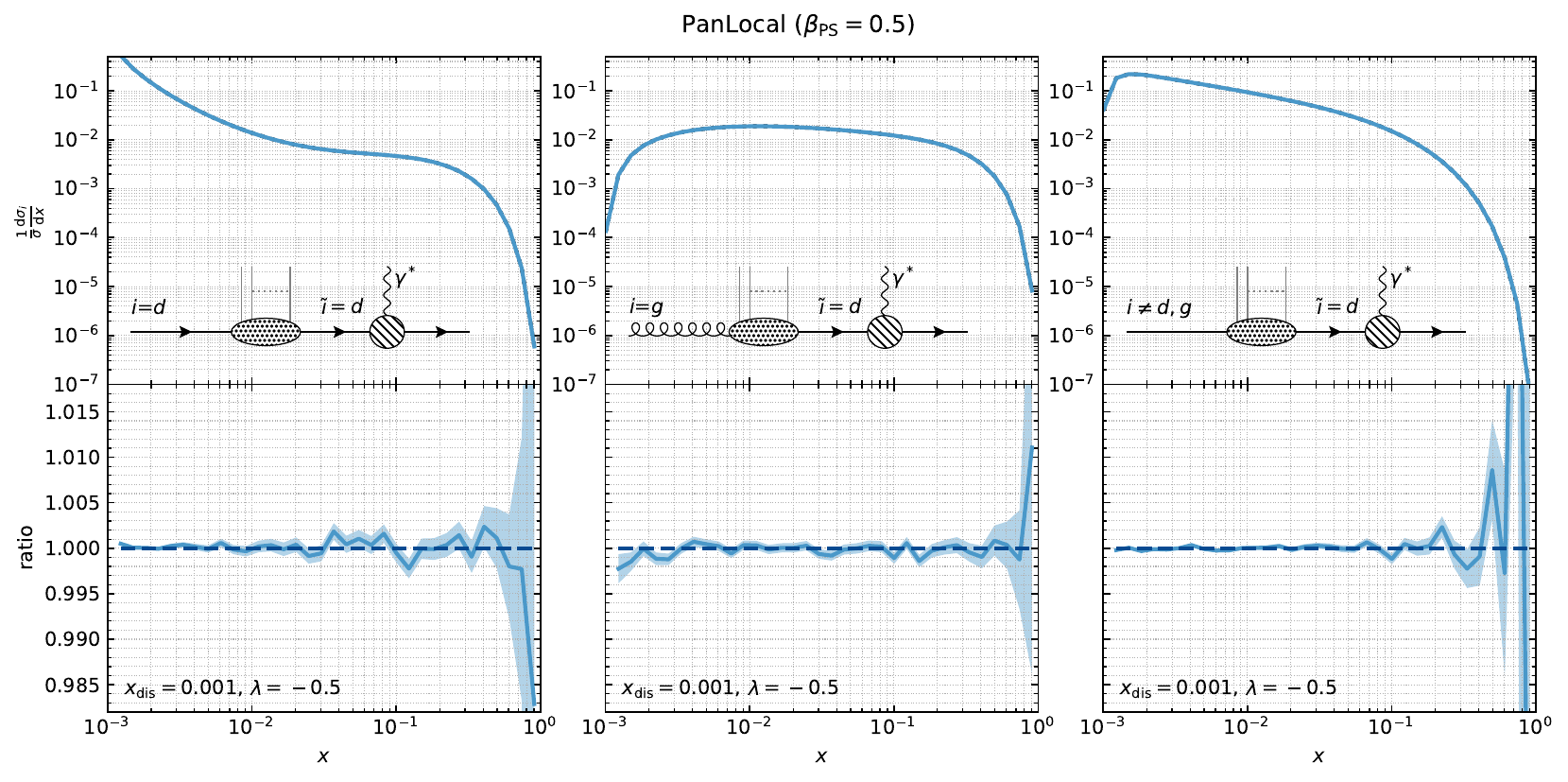}
  \end{subfigure}
  \begin{subfigure}{\textwidth}
    \includegraphics[width=\textwidth, page=2]{plots/plot-pdf-origin_dis_iflv-or_1_nloops_1_xdis_0.001-physical-colour.pdf}
  \end{subfigure}  
  \caption{The ratio of the DGLAP evolution produced by the
    PanLocal($\betaps=0.5$)~(upper panel) and
    PanGlobal($\betaps=0$)~(lower panel) showers versus the DGLAP
    evolution as calculated with HOPPET.  The results are shown as a
    function of the momentum fraction $x$ carried by the parton $i$
    extracted from the proton.  We fix the underlying Born DIS process to
    have a $d$ quark in the initial-state with $x_{\dis}=0.001$, and
    fix $\as= 5 \cdot 10^{-6}$ with $\lambda = \as L = -0.5$.
    The three columns show different extracted flavours $i$ that
    can lead to a $d$ quark with energy fraction $x_{\dis}$ entering
    the hard scattering process: we have a $d$ quark on the left
    panel, a gluon in the middle panel, and any other flavour in the
    right panel.}
  \label{fig:pdforigin}
  \end{figure}

In Fig.~\ref{fig:pdforigin} we illustrate the results for the
PanLocal($\betaps=0.5$) and PanGlobal($\betaps=0$) showers.
The other showers, PanGlobal($\betaps = 0.5$) and Dipole-$k_t$, lead to analogous
results.
The DGLAP reference prediction is obtained with HOPPET~\cite{Salam:2008qg}
at single logarithmic accuracy, probing only those initial conditions that
lead in the evolution to 
$\delta_{\itilde i} \delta(x-x_{\dis})$ at the hard scale $\muF=Q$, with
$x_{\dis}=0.001$ and $\itilde$ is a $d$ quark.
The shower runs are performed by fixing the underlying Born process to
contain a $d$ quark in the initial-state with $x_{\dis}=0.001$, and the
running of the strong coupling is performed at one loop. 
We show results obtained by setting $\as(Q)= 5 \cdot 10^{-6}$ and a
large value of $L= -10^5$ ($\lambda = -0.5$), which sets any terms
beyond SL accuracy to zero.
To further speed up the calculation we discard radiation with a
momentum fraction below some finite but small threshold $e^{-11}$, and
only keep radiation with an absolute rapidity larger than $18$
(i.e.~very collinear to the initial- or final-state quark).
We have verified that these cuts do not impact the results.
Further details on the treatment of the PDFs may be found in
Appendix~A of Ref.~\cite{vanBeekveld:2022ukn}.
We see that agreement with the HOPPET predictions is obtained for all
showers to within the statistical accuracy (below $0.1\%$ for the
majority of the $x$ range).

\subsection{Particle multiplicity}
\label{sec:multiplicity}

In this section, we examine the shower's ability to reproduce the
analytic prediction of the average particle multiplicity given the
shower's transverse-momentum cutoff.
Although particle multiplicity is not an infrared-safe quantity, the
resummation structure of particle multiplicity, defined with an
infrared cutoff, can be related to that of subject multiplicity at NDL,
which is theoretically well-defined, and does not depend on the 
jet algorithm (as long as it is infrared safe). 
The logarithmic accuracy of particle multiplicity needs to be
determined at the level of the distribution rather than the logarithm
of the distribution.
For such non-exponentiating observables, one writes
\begin{equation}
	\label{eq:mult-at-NDL}
	\langle N_{\rm NDL}(\as, \xi)\rangle = h_1(\xi) + \sqrt{\as} h_2(\xi)\,,
\end{equation}
with $\xi = \alpha_s L^2$ and where $h_1$ collects double-logarithmic
(DL) $\mathcal{O}(\alpha_s^n L^{2n})$ terms, and $\sqrt{\alpha_s} h_2$
the next-to-double logarithmic~(NDL) $\mathcal{O}(\alpha_s^n L^{n})$
terms.
At the desired level of accuracy (NDL), 
the average multiplicity obtained by 
counting the number of emissions generated by the
shower with a transverse-momentum cutoff $p_{t, \rm cut} = Q e^L$ (or
equivalently, a shower that uses a strong coupling equal to zero below
a given value of $p_{t, \rm cut}$) is equivalent to that obtained
after using a well-defined jet algorithm. 
The logarithm that is resummed then takes the form $L = \ln p_{t, \rm
  cut}/Q$.

For the process at hand, if we fix the Born flavour to be
$\itilde=q$, we have~\cite{Catani:1991pm, Catani:1993yx, Medves:2022ccw}
\begin{subequations}
	\begin{align}
		h_1 (\xi) &=   \frac{2 C_F}{C_A} \left[\cosh \nu-1\right] + N_b \,,\\
		h_2(\xi) &= \sqrt{\frac{1}{2\pi C_A}}\sinh \nu \,\frac{Q^2 {\partial} \ln f_{q}(x_{\dis}, Q^2)}{\partial Q^2}+
			 \frac{2C_F}{C_A} \Bigg[ \frac{\beta_0}{2} \sqrt{\frac{\pi}{2 C_A}} \left(\nu \cosh \nu + (\nu^2 -1)\sinh\nu\right) \nonumber \\
			& \hspace{1cm}  +\sqrt{\frac{C_A}{2 \pi}}B_{gq}\left(\frac{2C_F - C_A}{C_A}\nu \cosh\nu + \frac{5C_A - 6C_F}{C_A}\sinh \nu + 4\frac{C_F - C_A}{C_A}\nu\right)  \nonumber \\
			&  \hspace{1cm} +  \sqrt{\frac{C_A}{2 \pi}}\left(2 B_{qq}\sinh\nu + B_{gg}\left(\nu \cosh \nu - \sinh \nu\right)\right) \Bigg], \label{eq:NDLmult} 
	\end{align}
\end{subequations}
with $N_b = 1$ the number of final-state partons at LO, and
\begin{align}
\nu = \sqrt{\frac{2 C_A \xi}{\pi}}\,, \quad B_{gg} = -\frac{11}{12}\,,\quad B_{qq} = -\frac{3}{4}\,, \quad B_{gq} = \frac{2n_F}{3C_A}\,, \quad \beta_0 = \frac{11C_A - 2n_F}{12\pi}\,.
\end{align}
As before, $f_{q}$ denotes the PDF of the incoming parton at LO, which 
carries an energy fraction $x_{\dis}$.

At DL accuracy, this test probes the soft-collinear nested structure
of the shower.
To reproduce the average multiplicity at NDL, the parton shower must
also correctly incorporate hard-collinear corrections to the splitting
functions (corresponding to the second and third line of
Eq.~\eqref{eq:NDLmult}), the running of the coupling constant (first
line of Eq.~\eqref{eq:NDLmult}), the DGLAP evolution of the PDF (last
line of Eq.~\eqref{eq:NDLmult}), and the colour terms.
The treatment of these contributions is the same across all showers,
hence we expect to see agreement.
The parton shower correctly reproduces the analytic expectation
at NDL if the ratio
\begin{equation}
	\frac{ \langle N_{\rm shower}(\as, \xi)\rangle - \langle N_{\rm NDL}(\as, \xi)\rangle }{\langle N_{\rm NDL}(\as, \xi)\rangle - \langle N_{\rm DL}(\as, \xi)\rangle}\,,
	\label{eq:mult_test}
\end{equation}
vanishes in the $\alpha_s \to 0$ limit,
where $\xi$ is kept fixed and $\langle N_{\rm DL}(\as, \xi)\rangle$ corresponds
to Eq.~\eqref{eq:mult-at-NDL} with $h_2$ set to $0$. 
\begin{figure}[t]
      \centering
      \includegraphics[width=0.75\textwidth]{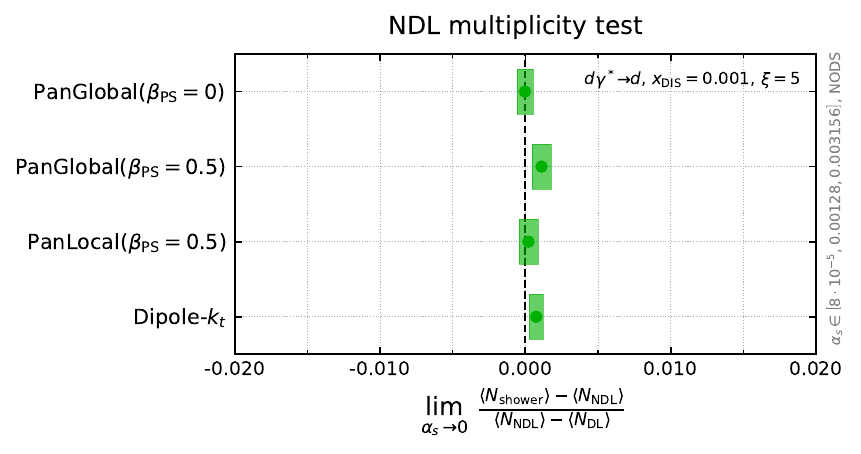}
      \caption{Extrapolation of $
      	\frac{ \langle N_{\rm shower}(\as, \xi)\rangle - \langle N_{\rm NDL}(\as, \xi)\rangle }{\langle N_{\rm NDL}(\as, \xi)\rangle - \langle N_{\rm DL}(\as, \xi)\rangle}$ for $\as \to 0$ at fixed value of $\xi = \as L^2=5$
        for several showers considering the DIS process $d \gamma^* \to d$
        with $x_{\rm DIS}=0.001$.
      }
      \label{fig:mult}
  \end{figure}
To extract the $\as \to 0$ limit, we run the PanScales showers for the
DIS process $q \gamma^* \to q$, where we set $q=d$, $x_{\dis}=0.001$, $\xi = 5$ and sample $\as = \left\{8 \cdot
10^{-5}, \,  0.00128, 0.003156\right\}$, performing a
quadratic polynomial extrapolation to get the $\alpha_s \to 0$ result. Systematic uncertainties are
estimated performing an alternative extrapolation with $\as = \, 0.00512$ instead of $\as = 0.003156$,
and added in quadrature to the statistical uncertainty. 
Like in the previous section, the running of the coupling constant is
performed at one loop, as the $2$-loop running  only enters at NNDL accuracy.
Subleading colour corrections are included using the NODS scheme.
The obtained result is shown in Fig.~\ref{fig:mult}. We notice that all the showers are consistent with the NDL
expectation, with an uncertainty well below 0.1\%.

\subsection{Continuously-global observables}
\label{sec:global-obs}

This section details tests on a range of global~\cite{Banfi:2004yd}
observables.  The cumulative distribution of such observables,
i.e.~the probability that the observable $O$ takes a value smaller
than $e^{L}$, can be written as a function of $\lambda = \alpha_s L$,
and takes the form
\begin{align}
	\label{eq:global-obs-res}
\Sigma(O < e^L) = H(\alpha_s)\exp\left[-L\, g_1(\lambda) + g_2(\lambda) + \dots \right] + \dots\,,
\end{align}
where $g_1$ contains the LL, and $g_2$ the NLL contribution.
The function $H(\alpha_s)$ is the hard function, which can be set equal to $1$
at our targeted NLL accuracy.
To test the accuracy of the shower we examine
\begin{align}
\lim_{\alpha_s \to 0}\frac{\Sigma_{\rm PS}(\lambda) - \Sigma_{\rm NLL}(\lambda)}{\Sigma_{\rm NLL}(\lambda)}\,,
\end{align}
at a fixed value of $\lambda$, which should tend to $0$ if the shower is NLL accurate.
To extract the $\as \to 0$ limit, we need to run the shower
at very small $\as$ values.
We are able to do so thanks to the numerical techniques developed in
Refs.~\cite{Dasgupta:2020fwr,Hamilton:2020rcu,vanBeekveld:2022ukn}.
These techniques assume that in the soft-collinear limit, the observable
scales as
\begin{align}
O \sim \frac{p_T}{Q}e^{-\betaobs |\eta|}\,,
\end{align}
where $ \betaobs \geq 0$ is a constant number.
This implies that we
can only consider continuously-global observables~\cite{Banfi:2004yd} in our tests.

Keeping this technical limitation in mind, to test our showers, 
we introduce three sets of
observables, parameterised by $\betaobs \in [0, 0.5, 1]$:
\begin{subequations}
	\label{eq:generic-global}
\begin{align}
	S_{p,\, \betaobs } & =  \sum_{j\in {\rm partons}}  \frac{k_{t,j} e ^{-\betaobs  |y_j| }}{Q}\,, \\
	S_{j,\, \betaobs } & =  \sum_{j\in {\rm jets}}  \frac{k_{t,j} e ^{-\betaobs  |y_j| }}{Q}\,, \\ M_{j, \,\betaobs } & =  \max_{j\in {\rm jets}}  \frac{k_{t,j} e ^{-\betaobs  |y_j| }}{Q}\,.
\end{align}
\end{subequations}
For the particle observable $S_{p,\, \betaobs }$, the sum over $j$
runs over all partons, and $k_{t,j}$, $y_j$ correspond to the
transverse momentum and the rapidity of the parton in the Breit frame.
Jets are defined with the algorithm detailed in
Appendix~\ref{sec:decl}.  For the two jet observables, the sum over
$j$ runs over all jets found inside the collection of beam jets and
the final-state macro-jet. The $k_{t,j}$ and $y_j$ are their primary
Lund-plane coordinates. The analytic expectations for the observables
quoted in Eq.~\eqref{eq:generic-global} are collected in
Appendix~\ref{sec:resummation}.
Note that $S_{p,\, \betaobs }$ will only be computed for $\betaobs >
0$, i.e.~when its NLL prediction corresponds to the one of a standard
additive observable and is equivalent to $S_{j,\,\betaobs}$. For
$\betaobs>0$ the contribution of the original hard final-state leg to
$ S_{p,\, \betaobs }$ is subleading and can be neglected, thus
yielding the same result as $S_{j,\, \betaobs}$. This is no longer the
case for $\betaobs=0$, as recoil effects contribute at NLL.
We stress that these observables are not directly measurable,
because they are built from emissions in both the remnant and the
current hemisphere, but they have the property of having a remarkably
simple resummation structure at NLL.
For this reason, they can be easily used to test the logarithmic
accuracy of showers, but they can also be employed as resolution
variables to build slicing methods~\cite{Catani:2007vq,Stewart:2010tn}
or NNLO+PS matching
prescriptions~\cite{Hamilton:2012rf,Alioli:2013hqa,Monni:2019whf}.

We further test our showers with observables that are
phenomenologically accessible.  To remove contamination from the
fragmenting beams in the remnant hemisphere, allowing for a better
experimental measurement, event shapes in DIS are often defined in the
current hemisphere $\mathcal{H}_c$.
Examples of such observables, which are
continuously-global and satisfy the recursive infrared-safety
requirement of Ref.~\cite{Banfi:2004yd}, are
\begin{subequations}
	\begin{align}
		B_{zE} =& \frac{\sum_{i \in \mathcal{H}_c} |\vec{p}_{\perp,i}|}{2\sum_{i \in \mathcal{H}_c} |\vec{p}_{i}|}\,, \label{eq:BzE}\\
		B_{zQ} =& \frac{\sum_{i \in \mathcal{H}_c} |\vec{p}_{\perp,i}|}{Q}\,,\label{eq:BzQ}\\
		\tau_{zQ} =& 1-\frac{2\sum_{i \in \mathcal{H}_c} |p_{z,i}|}{Q}, \label{eq:tauQ}
	\end{align}
	\label{eq:broadenings}
\end{subequations}%
\hspace{-5pt}where all the quantities are defined in the Breit frame, and the
suffix $z$ implies they are measured with respect to the photon axis.
The analytic predictions for the two definitions
of broadening, $B_{zE}$ and $B_{zQ}$, can be found in Ref.~\cite{Dasgupta:2001eq}.
They are examples of $\betaobs = 0$ observables, and their NLL 
prediction is identical.
For $\tau_{zQ}$, which is a
$\betaobs=1$ observable, these predictions were computed in
Ref.~\cite{Antonelli:1999kx}.  These results are also summarised in
Appendix~\ref{sec:resummation}.

\begin{figure}[t]
\begin{subfigure}{0.5\textwidth}
  \centering
  \includegraphics[height=0.3\textheight, page=3]{plots/plot-broadening.pdf}%
  \caption{}
  \label{fig:bzE_PG0}
\end{subfigure}%
\begin{subfigure}{0.5\textwidth}
  \centering
  \includegraphics[height=0.3025\textheight, page=1]{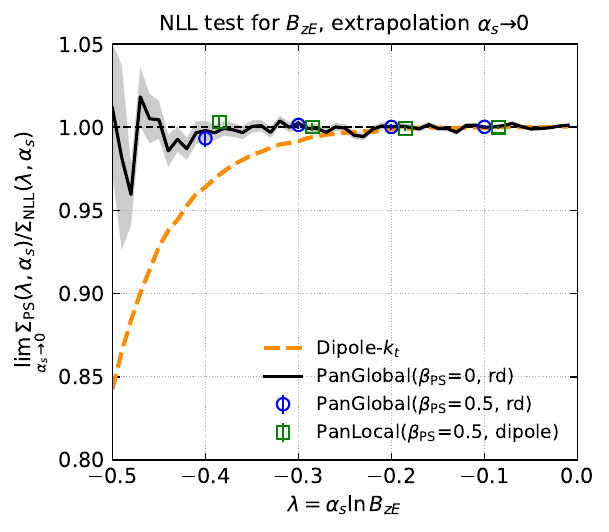}%
  \caption{}
  \label{fig:bzE_all}
\end{subfigure}
\caption{Cumulative distribution for the broadening normalised with
  respect to the energy in the current hemisphere, $B_{zE}$, for the
  process $q\gamma^* \to q$ with $x_{\dis}=0.2$.  In the left panel we
  show PanGlobal($\betaps=0$) results, obtained with finite values of
  $\alpha_s$, as well as the $\alpha_s \to 0$ extraction.  In the
  right panel, we illustrate the $\alpha_s \to 0$ extraction for all
  the showers: the ratio with the analytic prediction must be 1 if the
  shower is NLL.  Note that the values for the PanLocal shower have
  been shifted to the left to improve readability, i.e.~their probed
  $\lambda$ values coincide with those for PanGlobal($\betaps=0.5$).
}
	\label{fig:bzE}
\end{figure}

In Fig.~\ref{fig:bzE_PG0} we show the ratio of the
PanGlobal($\betaps=0$) prediction to the NLL result for the cumulative
distribution of the broadening normalised with respect to the energy in
the current hemisphere, $B_{zE}$, for increasingly smaller values of $\as$,
\begin{align}
	\label{eq:asglobal}
\alpha_s = \{ 0.0015625, 0.003125, 0.00625 \}
\end{align}
and $\lambda
\geq -0.5$. The coloured band represents only the statistical
uncertainty.
More information on how these results are obtained
is given in Appendix~\ref{sec:asto0}.
In black we illustrate the result of the $\alpha_s \to 0$
extrapolation, which has been performed with a quadratic
interpolation. 
In Fig.~\ref{fig:bzE_all} we summarise the $\alpha_s\to 0$
extractions for all the showers. From this figure it is clear
that all the new PanScales showers for DIS processes
reproduce the analytic expectation, 
while we observe deviations for Dipole-$k_t$ reaching 
up to 15\% for $\lambda=-0.5$.
We obtain identical results for $B_{zQ}$, not shown here.

\begin{figure}[t]
	\centering
	\includegraphics[width=0.9\textwidth, page=2]{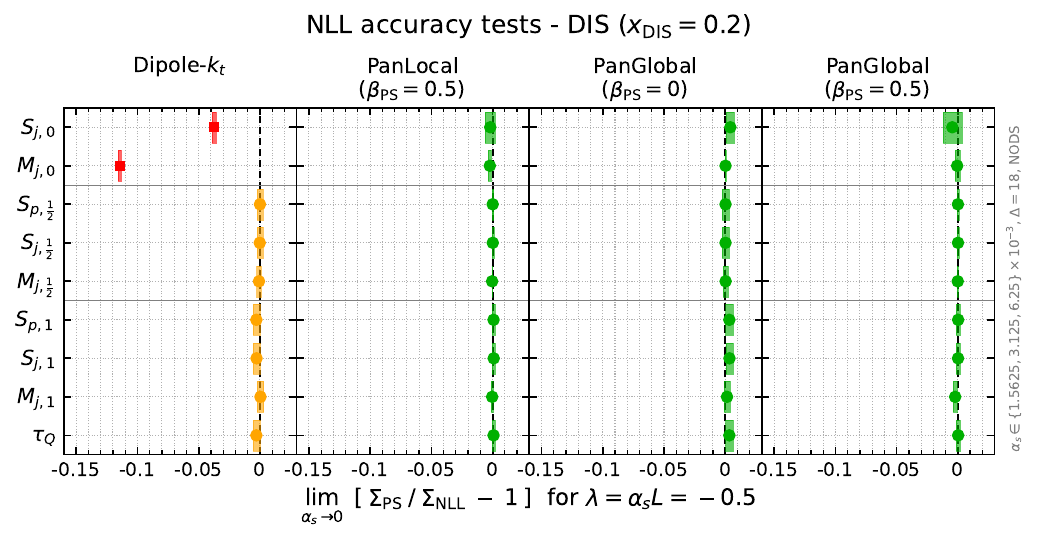}
	\caption{The $\alpha_s \to 0$ extrapolation of the deviation
          of the shower result from the NLL expectation for the
          process $q\gamma^* \to q$ with $x_{\dis}=0.2$ and $\lambda =
          -0.5$ for several continuously-global observables.
          The red colour signals a deviation from the analytic
          expectation larger than two standard deviations, while
          the green colour signals an agreement.
          When numerical agreement between the shower and the analytic
          resummation is found, but fixed-order issues are known, we
          use the amber colour.
        }
	\label{fig:global-obs}
\end{figure}

The shower's expectation for the other observables are summarised in
Fig.~\ref{fig:global-obs}, where we show the ratio with the NLL
result for the cumulative distribution $\Sigma(O < e^L)$ in the
$\alpha_s \to 0$ limit with $\lambda = -0.5$.
Here, the central value is again obtained using the
$\as$ values in eq.~\eqref{eq:asglobal}, 
but the uncertainty is given by
summing in quadrature the statistical uncertainty and the difference between
the central value and the one obtained 
by performing the extrapolation on the set of $\as$ 
values where $\as = 0.003125$ is replaced
with $\as = 0.0125$. 

The PanScales showers agree with the analytic expectations for all the
observables.
 Dipole-$k_t$ leads to manifestly wrong all-order results at NLL for
 $\betaobs=0$ observables, with a 4\% deviation for
 $S_{j,0}$ and a 12\% deviation for $M_{j,0}$.
Like in Refs.~\cite{Dasgupta:2020fwr,vanBeekveld:2022ukn}, we observe
that despite the fixed-order issue that we highlighted in
Sec.~\ref{sec:fo-test-contours}, Dipole-$k_t$ seems to reproduce the
correct analytic expectations for $\betaobs>0$.
The
NLL-violating terms manifest themselves as super-leading logarithms
that violate the exponentiation, as observed in
Ref.~\cite{Dasgupta:2020fwr}, which however resum to 0 in the
all-orders limit for $\betaobs>0$.

\subsection{Non-global logarithms}
\label{sec:non-global-obs}

Many observables are sensitive to radiation in a restricted portion of
the Lund plane.  In the context of DIS, these observables are, for
example, the current jet mass, the $C$-parameter, the thrust with
respect to the current-hemisphere thrust-axis~\cite{Dasgupta:2002dc}
and $Q_t$, i.e. the transverse momentum of the system comprising the
partons in the current hemisphere~\cite{Dasgupta:2006ru}.
For what concerns Higgs production in VBF, isolation criteria can be used
to reduce this production mode from gluon fusion~\cite{Buckley:2021gfw}.
The resummation for these observables naturally involves non-global
logarithms~(NGLs) \cite{Dasgupta:2001sh,Dasgupta:2002bw}, which can be
correctly reproduced at leading-colour single-logarithmic (SL) accuracy only 
by dipole showers~\cite{Banfi:2006gy}.

To assess the ability of our showers to reproduce such NGLs, we
consider the scalar sum of the transverse
momenta of the partons in a rapidity slice
\begin{equation}
S_{\Delta}^{\rm \sss slice} = \frac{\sum_{i} k_{t,i} \Theta(|\eta_i|< \Delta)}{Q} \equiv \frac{k_{t,\text{slice}}}{Q}\,,
  \end{equation}
where the transverse momentum $k_{t,i}$ and the rapidity $\eta_{i}$
of the partons are defined in the Breit frame.
NGLs for this observable are single-logarithmic
terms of the form $\lambda^n = \as^n L^n$, created by soft large-angle
emissions near the edge of the slice, i.e.~near $y=\pm \Delta$. 

\begin{figure}[t]
  \centering
   \begin{subfigure}{0.5\textwidth}
     \includegraphics[height=0.4\textheight, page=6]{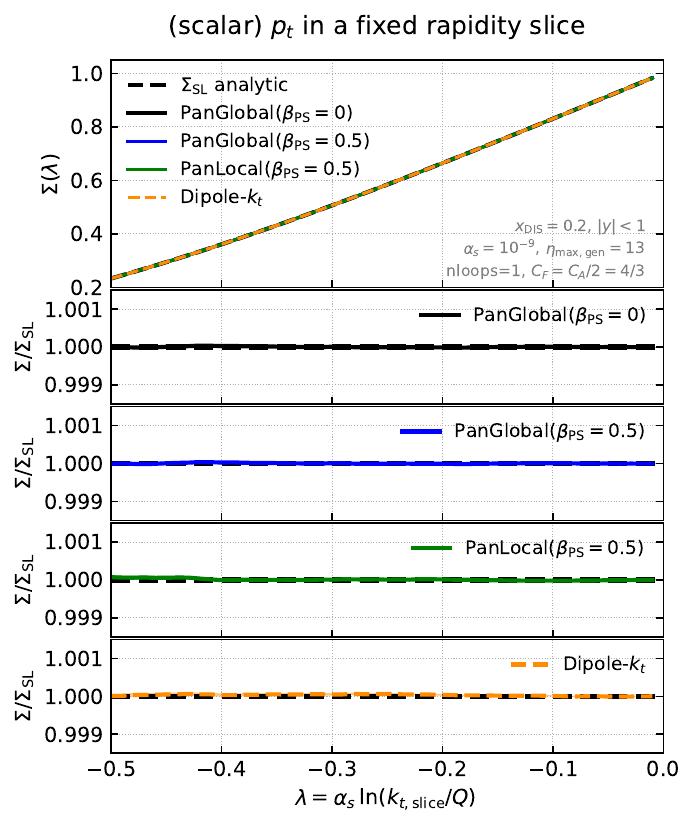}\hfill
     \caption{}
     \label{fig:slice_alllambdas}
   \end{subfigure}
      \begin{subfigure}{0.5\textwidth}
     \includegraphics[height=0.4\textheight, page=5]{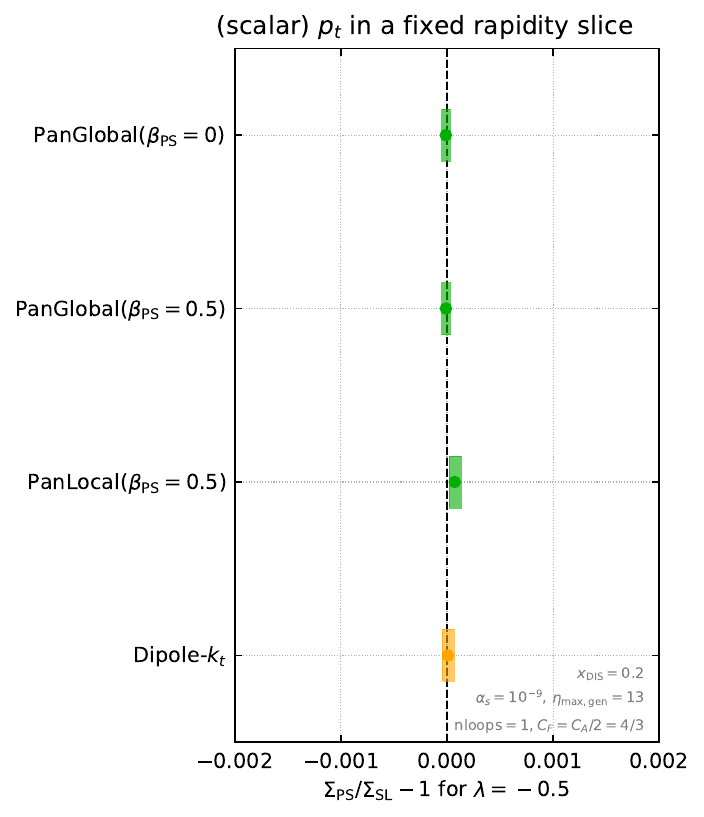}\hfill
     \caption{}
     \label{fig:slice_lambda5}
   \end{subfigure}
      \caption{
        (a) Cumulative distribution for the transverse momentum in a
        rapidity slice of $|y|<1$ as a function of $\lambda= \as \ln
        \frac{k_{t,\text{slice}}}{Q}$ for the PanScales showers and
        Dipole-$k_t$. The top panel shows the expected (black dashed)
        and the showers (solid) results, while the bottom panels show
        the ratio between the shower and the analytic prediction for
        each of the showers.
        (b) Relative difference between the shower $\Sigma_{\rm PS}$
        and the expected single-logarithmic result ($\Sigma_{\rm SL}$)
        for a fixed value of $\lambda = -0.5$ for all the PanScales
        showers. Colour coding is like in Fig.~\ref{fig:global-obs}.
      }
      \label{fig:slice}
\end{figure}

Fig.~\ref{fig:slice_alllambdas} shows the comparison between the
PanScales shower predictions with the expected results for $-0.5 \leq
\lambda < 0$ and $|y|<1$, while in Fig.~\ref{fig:slice_lambda5} we
show the results for $\lambda = -0.5$.
We generate our reference calculation in the large-$\nc$ limit (with
$C_F=C_A/2=3/2$) from the code developed for
Ref.~\cite{Caletti:2021oor}, which uses the strategy of
Ref.~\cite{Dasgupta:2001sh}.
The shower predictions are obtained running with $\as = 10^{-9}$, so
that NSLs are numerically negligible.
To reduce the parton multiplicity, without affecting the observable
under consideration, we impose a rapidity and a soft-emission cut, vetoing
radiation with $|\eta|>13$ and $\ln k_t < \ln k_{t,\text{slice}} -18$.
Furthermore, like in Secs.~\ref{sec:dglap} and~\ref{sec:multiplicity},
the running of the coupling constant is performed at one loop, with
$K=0$, as these effects only enter at NSL/NNDL.

In all cases, we notice an excellent agreement between the PanScales
showers and the correct SL distribution.
This is true also for Dipole-$k_t$, despite fixed-order issues,
exactly like in the case of continuously-global event shapes with
$\betaobs>0$.


\section{Phenomenological results Higgs production in VBF}
\label{sec:pheno}

We now move to the PanScales showers for VBF, presenting some
exploratory phenomenological results, and explaining more details on
the implementation of such process in our framework.  
This channel provides a clean
experimental signature and is therefore an ideal
environment to study the Higgs boson.\footnote{The VBF
channel is also used to search for di-Higgs production at the LHC~\cite{ATLAS:2020jgy}. }
As already explained in Sec.~\ref{sec:VBF-shower}, 
at NLO the VBF channel can be seen
as two independent DIS processes for each hadronic sector.
The two jets that are formed after the initial-state
quark emits a vector boson are typically produced with a large absolute
rapidity.
Colour coherence then results in little jet activity in 
the central rapidity region; radiation will be primarily concentrated around
the two hard jets at (opposite) large rapidities. 

For the phenomenological studies we produce the Higgs boson via the $ZZ$ channel 
in a VBF topology with a centre-of-mass energy of 13.6~TeV. 
We run all showers using the NODS colour scheme, 
even though the CFFE scheme is the one adopted by standard dipole showers,
to more faithfully gauge the kinematic differences between the LL and NLL showers.
The hard process is obtained from 
\textsc{Pythia8.3}~\cite{Bierlich:2022pfr} at
LO accuracy,
using the default values for the electroweak parameters and the Higgs mass.
We use the NNPDF~4.0 LO PDF set with perturbative charm content~\cite{NNPDF:2021njg} 
(LHAPDF label 332500~\cite{Buckley:2014ana}),
corresponding to $\as(m_Z) = 0.118$. 
The default factorisation and renormalisation scale used in \textsc{Pythia8.3} to generate the hard process is 
\begin{equation}
\mu_{F,h} = \mu_{R,h} \equiv  \mu_h = \sqrt[1/3]{m_{T,H} m_{T, V_1} m_{T, V_2}}, \quad m_{T,i}=\sqrt{p_{\perp,i}^2 + m_i^2}\,,
 \end{equation}
where $V_{1,2}$ denote the vector-bosons exchanged in the $t$-channel
propagators for each of the two hadronic sectors, and $m_{H, V_1, V_2}$ the Higgs/vector-boson masses.
However, as discussed in Sec.~\ref{sec:VBF-shower}, another choice
would be to use the virtualities of the exchanged boson as two
independent scales for the two hadronic sectors.
To this end, we apply the reweighting factor
\begin{align}
	w = \frac{f_{i_1}\left(x_1, \mu_1\right)\, f_{i_2}\left(x_2, \mu_2\right)}{f_{i_1}\left(x_1, \mu_h\right)\, f_{i_2}\left(x_2, \mu_h\right)}\,,
\end{align}
where $f_{i_{1,2}}$ is the PDF of the incoming quark $i_{1,2}$, which
carries an energy fraction $x_{1,2}$, and $\mu_{1,2} =
\sqrt{Q_{1,2}^2}$.
We start the parton shower at a distinct scale for 
each of the hadronic sectors, i.e.
\begin{equation}
	v^2_{\max, i} = Q_i^2 \frac{1-x_i}{x_i}\,.
\end{equation} 
In practice, we use the maximum of the two values as common
starting scale, and we perform a veto to ensure that $v < v_{\max, i}$ in each hadronic
section.

We estimate the uncertainty stemming from renormalisation scale variations 
using a modified scheme for $\as$~\cite{Mrenna:2016sih, vanBeekveld:2022ukn}, that is
\begin{align}
	\as\left(\mu_R^2\right) \left(1 + \frac{\as\left(\mu_R^2\right) }{2\pi}K + 2\as\left(\mu_R^2\right)b_0(1-z)\ln x_R\right)\,, \quad \mu_R = x_R \mu_{R,0}\,,
	\label{eq:xRmuR}
\end{align}
with $\mu_{R,0} = \rho v {\rm e}^{\betaps|\bar{\eta}_{Q_i}|}$ the central scale. 
The factor $z$ is the fraction of the emitter-momentum carried away by the radiation.
With this, $1-z$ ensures that scale compensation at NLL is present for soft emissions,
but not for hard emissions. 
We should omit this term for
the LL-accurate shower Dipole-$k_t$, but we do include the CMW factor $K$ by default.
Renormalisation-scale variations are probed taking $x_R \in \{0.5,1,2\}$, and
the infrared cutoff of the shower is implemented such that $\as(\mu_R^2) = 0$ 
for $\mu_R < x_R \times 0.5$~GeV. 
Factorisation-scale uncertainties are probed independently from the renormalisation scale, 
and are assessed using
\begin{align}
	\ln \mu_F \equiv \ln\left(x_F \mu_{F,0}\right) = \ln Q_i + \frac{1}{1+\betaps} \ln\frac{v_i}{Q_i} + \ln x_F\,,
	\label{eq:lnmuF}
\end{align}
with $x_F \in \{0.5,1,2\}$. 
In our results we show the $7$-point scale variation, obtained
by taking $(x_R,x_F) \in \{(1,1),(0.5,1),(1,0.5),(0.5,0.5),(2,1),(1,2),(2,2)\}$.
It is important to note that stress that such variations typically do not capture all sources of uncertainty.
Indeed, the showers feature different recoil schemes and 
evolution variables, which lead to subleading (uncontrolled)
corrections, whose uncertainty is not captured in the above approach.

\begin{figure}[t]
	\begin{subfigure}{\textwidth}
		\centering
		\mbox{
			\includegraphics[height=6.5cm, page=5]{plots/vbf-pheno-pThiggsmin0.pdf}
			\includegraphics[height=6.5cm, page=6]{plots/vbf-pheno-pThiggsmin0.pdf}}
	\end{subfigure}
\begin{subfigure}{\textwidth}
	\centering
	\mbox{
		\includegraphics[height=6.5cm, page=1]{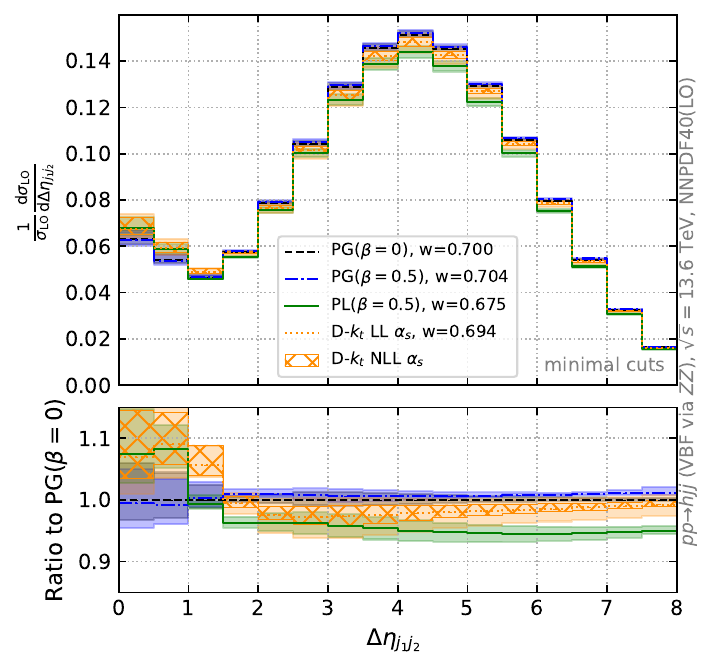}
		\includegraphics[height=6.5cm, page=2]{plots/vbf-pheno-pThiggsmin0.pdf}}
\end{subfigure}
	\caption{
		Distributions normalised to the total (before cuts)
		LO cross section $\sigma_{\rm LO}$ for $m_{j_1j_2}$ (top)
		and $\Delta \eta_{j_1 j_2}$ (bottom)
		after applying
		the minimal selection cuts (left) and the VBF cuts (right).
		We show PanGlobal($\betaps = 0$), black dashed,
		PanGlobal($\betaps=0.5$), blue dash-dotted,
		PanLocal($\betaps = 0.5$), green solid and Dipole-$k_t$,
		orange, dotted.
		The label $w$ indicates the total area under the curve.  The
		vertical lines on the central value of each bin indicate the
		statistical uncertainty, which is mostly negligible and
		therefore not visible.  The band indicates the
		renormalisation/factorisation scale uncertainty, obtained as
		explained after eq.~\eqref{eq:lnmuF}.
		For Dipole-$k_t$ the hashed band indicates the uncertainty obtained
		using the strong coupling for the shower as in eq.~\eqref{eq:lnmuF},
		whereas the solid band indicates the uncertainty obtained without the 
		NLL scale-compensating term of eq.~\eqref{eq:lnmuF}. 
		The bottom panel 
		shows the ratio with respect to the PanGlobal($\betaps = 0$)
		result.  }
	\label{fig:inclusive-dist-VBF}
\end{figure}

We use the anti-$k_T$ algorithm~\cite{Cacciari:2008gp} with $R=0.4$, implemented in 
\textsc{FastJet}~\cite{Cacciari:2011ma} to cluster jets with the definition
$p_{T,j} > 25$~GeV, $|\eta_{j}| < 4.5$, and consider two setups:
\begin{itemize}
	\item ``minimal cuts'', where we require the presence of two resolved jets;
	\item ``VBF cuts'', where we require that the two leading jets (i.e.~those two with the largest transverse momenta)  are  separated by a rapidity $\Delta \eta_{j_1 j_2} > 4.5$, have a dijet invariant mass of at least $m_{j_1 j_2} > 600$~GeV and lie in opposite hemispheres ($\eta_{j_1} \cdot \eta_{j_2} < 0$).
\end{itemize}
The results shown below do not include the simulation of beam remnants, hadronisation
or multi-parton interaction. 
We have the option to incorporate them into our framework trivially
using \textsc{Pythia8.3}, but we have made the decision not to do so.
This choice enables us to provide a clearer comparison between the NLL PanScales showers and 
the LL-accurate Dipole-$k_t$ shower. 
Note that comparisons with other publicly-available LL showers are not
being performed here, but we should stress that since we consider also
regions where the prediction is not necessary dominated by logarithmic
enhancements, differences between publicly-available showers and
Dipole-$k_t$ can be sizeable.

We first consider inclusive observables, i.e.~those observables that are
non-vanishing at LO. 
The value of these observables is primarily set by the hard scattering process,
and the shower should impact these observables only marginally. 
In Fig.~\ref{fig:inclusive-dist-VBF}  we show two such observables: 
the invariant mass of the two leading jets $m_{j_1 j_2}$ and
the rapidity seperation between them $\Delta \eta_{j_1 j_2}$.
Uncertainties from renormalisation/factorisation scale variations 
are noticably smaller in the NLL showers than in Dipole-$k_t$. This
is a direct consequence of including the scale-compensating term of
eq.~\eqref{eq:xRmuR}. 
In general, the scale uncertainties are small, except in the regions
where $\Delta \eta_{j_1 j_2}$ and $m_{j_1 j_2}$ are small 
(see~Fig.~\ref{fig:inclusive-dist-VBF}, left).
These regions are contaminated by events where one of the two 
tagged jets actually originates from the shower hardest emission. 
Indeed, because of this effect, the scale uncertainties here are larger than
 in the bulk of the distribution.  
With the exception of these regions, which are excluded after applying
the VBF cuts, it can be observed that differences between the NLL
showers are of the order of $\mathcal{O}(5-10\%)$, i.e. commensurate
with known NLO corrections. 
However,  the spread of the
NLL showers is much larger than their scale variation, and the latter 
indeed can then not
be used to accurately and reliably capture the differences between the showers.
The effect of adding the scale-compensating term in eq.~\eqref{eq:lnmuF} is sizeable:
uncertainty stemming from scale variations increases by roughly a factor of two
when turning off those scale-compensating terms.
We do want to stress that turning off those terms underestimates the scale uncertainty
for the Dipole-$k_t$ shower. 
As this is not an NLL shower, and the scale-compensating term originates from an NLL prediction,
we are not strictly allowed to turn off the compensating effect.
We also notice that the LL shower lies in between the prediction of the PanScales NLL
showers.

 \begin{figure}[t]
 	\begin{subfigure}{\textwidth}
           	\centering
 		\mbox{\includegraphics[height=6.5cm, page=3]{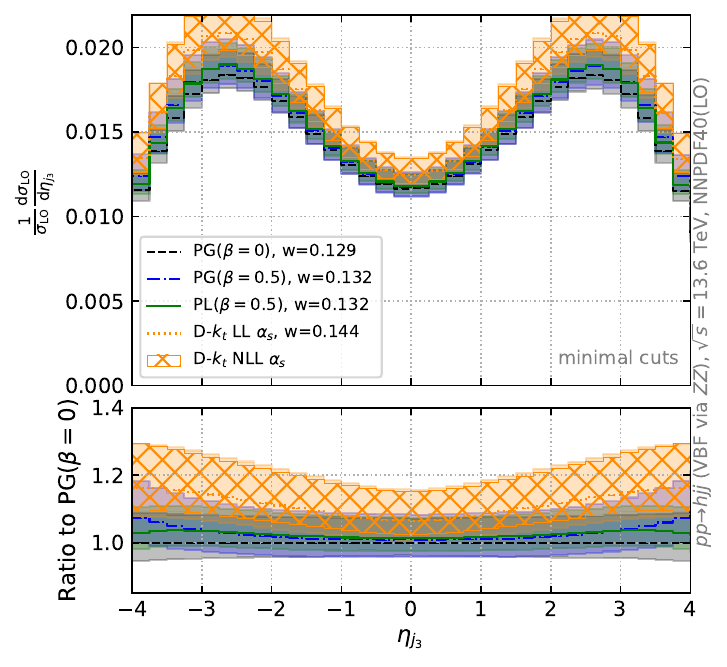}
 		\includegraphics[height=6.5cm, page=4]{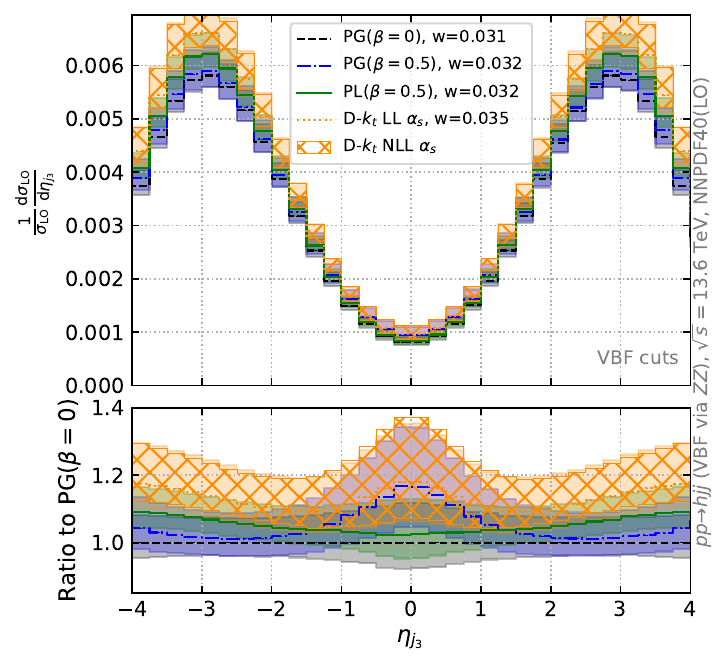}}
 \end{subfigure}
 	\begin{subfigure}{\textwidth}
 		\centering
 		\mbox{
 			\includegraphics[height=6.5cm,page=8]{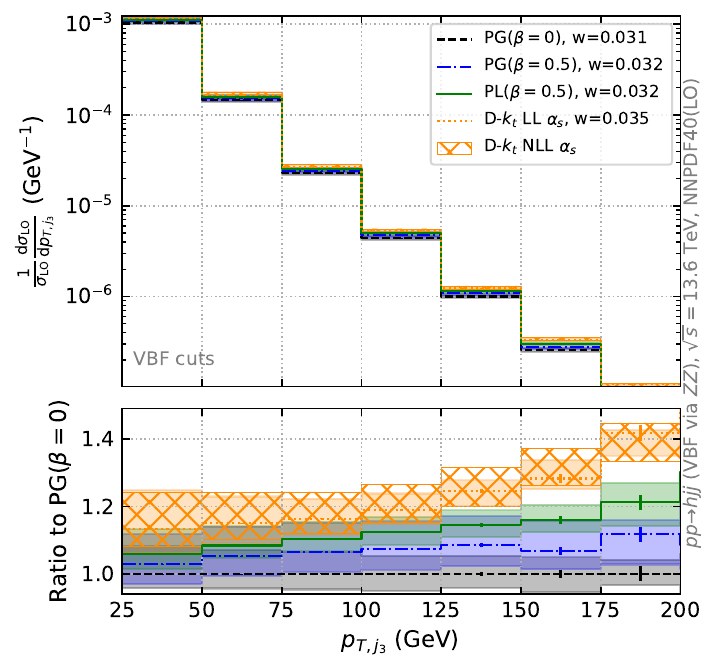}
 			\includegraphics[height=6.5cm, page=10]{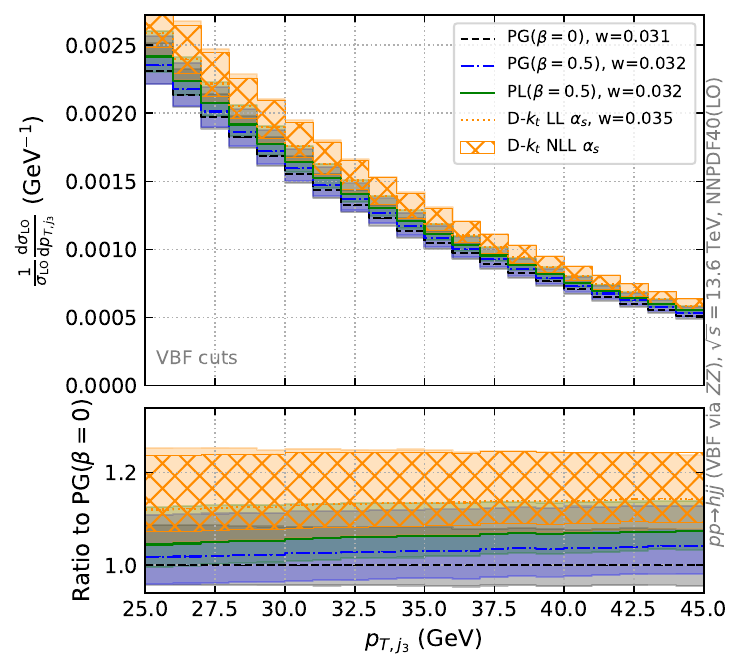}}
 	\end{subfigure}
 	\caption{As Fig.~\ref{fig:inclusive-dist-VBF} but for $\eta_{j_3}$ (top), and $p_{T,j_3}$ (bottom).
 	For the latter we only show the result after applying the VBF cuts, and the
 	right bottom panel shows a magnification of the (relatively) small transverse-momentum region.  }
 	\label{fig:exclusive-dist-VBF-eta3-ptj3}
 \end{figure}
 
We now turn to observables that can only be defined in the presence of a third jet,
and therefore are a clean probe of the shower's behaviour. 
In particular, we consider the pseudo-rapidity of the third jet,
$\eta_{j_3}$, and
the transverse momentum of the third jet $p_{T,j_3}$, as shown in 
Fig.~\ref{fig:exclusive-dist-VBF-eta3-ptj3}.
Colour coherence predicts a suppression of radiation in the central rapidity region.
All the showers considered here indeed show this behaviour, as can
be observed in the top panel of Fig.~\ref{fig:exclusive-dist-VBF-eta3-ptj3}.
In contrast to the previously-considered inclusive observables, we now
see that the difference between using an LL effective running coupling in 
the shower or an NLL one is minimal. 
Indeed, the scale uncertainty bands are of roughly equal size for the two Dipole-$k_t$
results. 
This is related to the fact that these observables are all dominated by the shower's
hardest emission. 
For this emission, the uncertainty stemming from $\mu_F$ variations exceed that
of that for $\mu_R$ variations.
In addition, we note that Dipole-$k_t$ typically lies above the predictions of the NLL showers. 
This can be mostly traced back to a diference in normalisation, as seen
by comparing the weights $w$ of the histograms of the NLL showers versus
that of Dipole-$k_t$.
These normalisation differences are due to the fact that
Dipole-$k_t$ tends to show a higher rate of $3$-jet events than the other showers.
This rate is controlled by the hard-emission phase-space region.
However, since we do not include matching, this region is not controlled 
in our showers, and we leave a further detailed investigation of the differences between the 
LL and NLL showers for future work. 

At this stage, we also refrain from performing a comparison with
experimental data, for  similar reasons: (I) NLO
matching is not yet implemented; (II) we have not yet studied the
interplay between our showers and non-perturbative effects; (III) we
have not yet tuned our showers.  We thus leave a theory-data
comparison for future work.

\section{Conclusions}
\label{sec:conc} 
In this work, we have introduced new NLL-accurate
dipole showers for processes involving the exchange of a
colour-singlet in the $t$-channel, such as DIS, VBF and VBS.
The latter two processes are handled following a factorised
approach, i.e.~neglecting non-factorisable corrections between the two
hadronic sectors.
The main novelty of these showers, with respect to the PanScales
showers for hadron collisions introduced in
Refs.~\cite{vanBeekveld:2022zhl,vanBeekveld:2022ukn}, is that the
transverse-momentum recoil due to initial-state radiation is
smoothly redistributed primarily to partons in the current hemisphere
(i.e.~anti-parallel to the direction of the incoming proton in the
Breit frame).
This feature ensures that partons in the remnant hemisphere remain mostly
unaffected, which is required from colour coherence.
Furthermore, compared to standard showers for DIS and
VBF/VBS, our showers differ in the choice of the dipole
partitioning: this is not done in the dipole frame, but
in the Breit frame.
This, combined with a global recoil scheme or with a careful choice
of the ordering scale, prevents soft gluons from taking unphysical
recoil. 

We have carried out a number of fixed-order tests, focusing on DIS, such as 
analysing the phase-space contours for two emissions with commensurate softness
(Section~\ref{sec:fo-test-contours}), 
and colour/spin
matrix-element comparisons (Section~\ref{sec:colour} and Appendix~\ref{sec:spin}), 
related to the PanScales conditions needed to achieve NLL accuracy.
All-order validations of our new showers for DIS have also been carried out for a variety of
observables.
These include tests of the DGLAP evolution (Section~\ref{sec:dglap}),
jet-multiplicity (Section~\ref{sec:multiplicity}),
DIS continuously-global event-shapes (Section~\ref{sec:global-obs}), 
and 
the scalar sum of transverse momenta in a fixed rapidity slice (Section~\ref{sec:non-global-obs}).
In our comparisons, we introduced new
continously-global event shapes, which rely on the use of
Lund Plane coordinates and a jet algorithm specific for DIS.
All these tests were carried out including subleading colour corrections, 
except for those of the non-global observable. 
The PanLocal shower with $\betaps = 0.5$ and PanGlobal showers with $\betaps = 0,0.5$
sucessfully pass the fixed- and all-order NLL accuracy tests.
We have compared these showers to a `standard' transverse-momentum ordered shower, 
Dipole-$k_t$, which has fully local transverse-momentum recoil. 
Differences between the LL-accurate Dipole-$k_t$ and the new
NLL-accurate showers can grow up to $15\%$ for phenomenologically
relevant continuously-global event-shape observables such as the current-hemisphere
broadening.\footnote{The NLL test for the Dipole-$k_t$ shower has been performed implementing subleading colour corrections 
with the NODS colour scheme, despite the fact that the colour-factor-from-emitter scheme is
the standard choice for such showers. } 
The SL accuracy of our showers established for DIS proceses,
automatically propagates to VBF/VBS processes
in the factorised approach. 
This is the first time such accuracy is
achieved for VBF/VBS global and non-global observables.
The neglected non-factorisable contributions are SL
subleading colour corrections,
and are typically further surpressed after applying VBF cuts. 

In Section~\ref{sec:pheno} we present an exploritory
phenomenological application of our newly developed showers for Higgs production
in VBF at $\sqrt{s}=13.6$~TeV.
For each of the two hadronic sectors individually, 
we choose separate values for the shower starting, renormalisation, and factorisation scales. 
While we examined the impact of variations in the renormalisation and factorisation scales, 
it is important to note that such variations typically do not capture all sources of uncertainty.
Indeed, the fact that we have developed not one, but
several showers, is important for a realistic estimate of shower uncertainties,
as the spread of predicitions obtained with our PanScales showers
is not captured by the scale uncertainty. 
For inclusive observables, the LL shower is contained in the spread of our newly-developed
NLL ones. 
However, for exclusive observables, like the rapidity or transverse momentum 
of the third jet, we find that the Dipole-$k_t$ predictions typically overshoot
the NLL showers.
This discrepancy can be attributed to the fact that the Dipole-$k_t$ 
shower typically produces a higher rate of $3$-jet events compared to our NLL showers.
Next steps involve matching the NLL-accurate showers to the NLO
fixed-order results, where care needs to be taken not to compromise the
NLL accuracy~\cite{Hamilton:2023dwb}, as well as including heavy-quark
mass effects to handle $t$-channel single-top production.

\appendix


\section*{Acknowledgements}

We are grateful to our PanScales collaborators (%
Mrinal Dasgupta,
Fr\'ed\'eric Dreyer,
Basem El-Menoufi,
Keith Hamilton,
Jack Helliwell,
Alexander Karlberg,
Rok Medves,
Pier Monni,
Gavin Salam,
Ludovic Scyboz,
Alba~Soto-Ontoso,
Gregory Soyez,
Rob Verheyen, and
Scarlett Woolnough%
),
for their work on the code, comments on the manuscript, the underlying
philosophy of the approach, and the adaptations of the PanGlobal shower. 
In particular, we want to thank Pier Monni and Mrinal Dasgupta for
having shared with us their knowledge on
resummation for DIS event shapes, Gavin Salam for help with the
implementation of jet algorithm for DIS, Gregory Soyez for
frequent discussions on the technical details of the NLL tests, and
Alexander Karlberg for pointing out relevant VBF literature. 
This work was supported
by a Royal Society Research Professorship
(RP$\backslash$R1$\backslash$180112) (MvB),
by the European Research Council under the European Union’s
Horizon 2020 research and innovation programme (grant agreement No.\
788223, PanScales) (MvB and SFR), 
and by the Science and Technology Facilities Council under ST/T000864/1 (MvB).

\section{Description of the momentum-conservation restoring boost}
\label{sec:disboost}
This appendix details the derivation of the boost $\Lambda^{\mu \nu}$,
introduced in Sec.~\ref{sec:panshowers}.
For PanGlobal this boost acts on the collection of final-state
partons $\bar{p}_{\sss X}^{\mu}$ to restore momentum conservation. 
For the PanLocal shower it in addition also acts on the incoming
parton $\bar{p}_1^\mu$ to restore momentum conservation
after the collection of partons
are rotated so that $\bar{p}_1^\mu$ is aligned with the direction of $n_1^\mu$. 
We demand that the boost preserves the invariant mass of
$\bar{p}_{\sss X}^{\mu}$, which implies
\begin{align}
\Lambda^{\mu \nu}\left(\bar{p}_{\sss X, \nu}\right) \equiv  p_{\sss X}^{\mu} = n_2^{\mu} +\frac{\bar{p}_{\sss X}^2}{Q^2} n_2^{\mu}\,.
\label{eq:boost1}
\end{align}
Using a Sudakov decomposition of $\bar{p}_{\sss X}$ along
the directions of $n_1^\mu$ and $n_2^\mu$, eq.~\eqref{eq:boost1} can be written as
\begin{equation}
\Lambda^{\mu \nu}\left(\beta n_{2,\nu} + \frac{\bar{p}_{\sss X}^2 +p_t^2}{\beta Q^2} n_{1,\nu} + p_{\perp,\nu}\right) = n_{2}^\mu + \frac{\bar{p}_{\sss X}^2}{Q^2} n_{1}^\mu\,,
\label{eq:direct}
\end{equation}
with $p_\perp^2=-p_t^2<0$, $n_1^2=n_2^2=0$, $2n_1\cdot n_2=Q^2$ and $p_\perp \cdot n_{1,2}=0$.
We aim to design a boost that acts as a rescaling for 
momentum components along $n_1^{\mu}$,
and assigns all the transverse momentum recoil to the $n_2^{\mu}$ direction.%
\footnote{The construction of this boost shares similarities with the
boost used in Deductor for $pp\to Z/h$ collisions~\cite{Nagy:2009vg}.}
Inverting eq.~\eqref{eq:direct}, we have
\begin{equation}
	\label{eq:invert}
\left[\Lambda^{-1}\right]^{\mu \nu}\left(n_{2,\nu} + \frac{\bar{p}_{\sss X}^2}{Q^2} n_{1,\nu}\right) = \beta n_{2}^{\mu} + \frac{\bar{p}_{\sss X}^2 +p_t^2}{\beta Q^2} n_{1}^{\mu} + p_{\perp}^{\mu}\,.
\end{equation}
We now may examine the action of the inverse boost on the $n_1^{\mu}$ and $n_2^{\mu}$ 
components individually. 
The direction $n_2^\mu$ absorbs all the transverse-momentum recoil, but also must stay massless, meaning it also must absorb a component in the $n_1^{\mu}$ direction. 
To satisfy these constraints we infer that 
\begin{equation}
\left[\Lambda^{-1}\right]^{\mu \nu}n_{2,\nu} \equiv \beta n_2^\mu + \frac{p_t^2}{\beta Q^2} n_1^\mu + p_\perp^\mu\,.
\end{equation}
To satisfy eq.~\eqref{eq:invert} we then have to require 
\begin{equation}
\left[\Lambda^{-1}\right]^{\mu \nu}n_{1,\nu} \equiv \frac{Q^2}{\bar{p}_{\sss X}^2}\left(\beta n_2^\mu + \frac{\bar{p}_{\sss X}^2 +p_t^2}{\beta Q^2} n_1^\mu + p_\perp^\mu - \left[\Lambda^{-1}\right]^{\mu \nu}n_{2,\nu} \right)=\frac{1}{\beta} n_{1}^\mu\,.
\end{equation}
We now have established the action of the inverse boost on the
$n_1^{\mu}$ and $n_2^{\mu}$ directions.  What remains is to determine
what happens when the inverse boost acts on a generic
transverse-momentum component $q_{\perp}^{\mu}$ (with $q_\perp \cdot
n_{1,2}=0$).
We require that the action of the inverse boost on such a
perpendicular component is only allowed to bring in a component in the
direction of $n_1^{\mu}$.
Furthermore, the perpendicular component
does not get rescaled, so that its norm is preserved:
\begin{equation}
\left[\Lambda^{-1}\right]^{\mu \nu} q_{\perp,\nu} = q_\perp^\mu + A n_1^\mu\,.
\end{equation}
To find the value of $A$, we require that the invariant mass of generic vector $q^{\mu} = a_q n_1^{\mu} + b_q n_2^{\mu} + q_\perp^{\mu}$ is preserved by the boost. This leads to 
\begin{equation}
A=-\frac{ 2 q_\perp \cdot p_\perp}{\beta Q^2}\,.
\end{equation}
Note that we cannot introduce a component along $n_2^\mu$ in the action of the inverse boost
on a transverse component.
To see this, consider the action of the inverse
boost on a generic transverse component $q_\perp^\mu$ where
we now assign a $n_2^\mu$ component, 
\begin{align}
\left[\Lambda^{-1}\right]^{\mu \nu} q_{\perp,\nu} = B n_2^\mu + q_\perp^\mu\,.
\end{align}
With this definition, the action of the inverse
boost on a generic four-momentum $q^\mu$ would
become
\begin{align}
\left[\Lambda^{-1}\right]^{\mu \nu} q_{\nu} = \left(\frac{a_q}{\beta} + \frac{b_q\, p_t^2}{\beta Q^2}\right)n_1^\mu + \left(b_q\beta + B \right)n_2^\mu + b_q p_\perp^\mu + q_\perp^\mu\,.
\end{align}
Requiring that the boost leaves the invariant
mass of $q^\mu$ unchanged gives us
\begin{align}
B = \frac{2 p_\perp \cdot q_\perp}{a_q Q^2 + b_q p_t^2}b_q \beta\,.
\end{align}
This is not linear in $q^\mu$, which means that 
if we would consider $l^\mu = q^\mu + q^{\prime \mu}$,
we would have $\left[\Lambda^{-1}\right]^{\mu \nu}l_\nu \neq \left[\Lambda^{-1}\right]^{\mu \nu}q_\nu + \left[\Lambda^{-1}\right]^{\mu \nu}q'_\nu$. 
Therefore, $B$ needs to be set to zero.

The action of $\left[\Lambda^{-1}\right]^{\mu \nu}$ is now fully
specified, and we can invert it to find $\Lambda^{\mu \nu}$.  We
then obtain
\begin{subequations}
\begin{align}
 \Lambda^{\mu \nu}n_{1,\nu} &= {\beta} \, n_{1}^\mu\,, \\
 \Lambda^{\mu \nu}q_{\perp,\nu} &= q_{\perp}^\mu +\frac{ 2 q_\perp \cdot p_\perp}{Q^2} n_1^\mu\,, \\
   \Lambda^{\mu \nu}n_{2,\nu}& = \frac{1}{\beta} \left(n_2^\mu +\frac{p_t^2}{Q^2} n_1^\mu -p_{\perp}^\mu\right),
\end{align}
\end{subequations}
so that the final form of the boost reads
\begin{align}
 \Lambda^{\mu \nu} =& g^{\mu \nu} + \frac{2 n_1^\mu}{Q^2} \left[ (\beta-1) n_2^\nu + \frac{p_t^2}{\beta Q^2} n_1^\nu + p_\perp^\nu\right] + \frac{2 n_2^\mu n_1^\nu}{Q^2} \frac{1-\beta}{\beta}-\frac{2 p_{\perp}^\mu n_1^\nu}{\beta Q^2}\,.
\end{align}

\section{Local rescaling factors for PanGlobal}
\label{sec:resc}
In this section we derive the rescaling factors $r_{\rm \sss L}$
appearing in the PanGlobal map of eq.~\eqref{eq:mapping-PG},
distiguishing between final-final and initial-final dipoles.
We also comment on the implementation for hadron-hadron colliders, which also contains initial-initial dipoles.

\subsection{Final-final dipoles}
For an emission off a final-final dipole we have,
\begin{subequations}
	\label{eq:mapping-PG-FF}
\begin{align}
	 \bar p_i^{\mu} & = r_{\rm \sss L} (1- a_k) \tilde{p}_i^{\mu}\,, \\ 
	\bar p_j^{\mu} &=   r_{\rm \sss L} (1- b_k) \tilde{p}_j^{\mu}\,, \\
  \bar p_k^{\mu} & =  r_{\rm \sss L}(a_k \tilde{p}_i^{\mu}  + b_k \tilde{p}_k^{\mu}  + k_{\perp}^{\mu}) \,,
\end{align}
\end{subequations}
and hence the final-state partonic momentum now reads
\begin{equation}
\bar{p}_{\sss X}^{\mu} = \tilde{p}_{\sss X}^{\mu} + p_i^{\mu}+p_j^{\mu}+p_k^{\mu} -\tilde{p}_i^{\mu}-\tilde{p}_j^{\mu} =  \tilde{p}_{\sss X}^{\mu}+ (r_{\rm \sss L}-1)(\tilde{p}_i^{\mu}+\tilde{p}_j^{\mu}) + r_{\rm \sss L} k_\perp^{\mu}.
\end{equation}
As discussed in Ref.~\cite{DOUBLESOFT}, when $\tilde{p}_i^{\mu}$ and
$\tilde{p}_j^{\mu}$ are close in angle, $k_\perp \cdot Q$ can be
large, despite $k_\perp^2$ being small.
Practically for our DIS map, this means that when $k_\perp^\mu$ carries a
large component aligned along the $x$ or $y$ axis, or along $n_2^\mu$,
the boost applied to the collection of the final-state partons is
substantial, while when $k_\perp^\mu$ carries a large component along
$n_1^\mu$, we need to introduce a large rescaling for the intial-state
parton.
In all of these cases, we would produce undesired correlations with
other partons in the event, that we prevent by adopting the
energy-preserving solution proposed by Ref.~\cite{DOUBLESOFT}.
In particular, in this case we impose that $\bar{p}_{\sss X} \cdot Q =
\tilde{p}_{\sss X} \cdot Q$, leading to
\begin{equation}
  r_{\rm \sss L} = \frac{(\tilde{p}_i+\tilde{p}_j)\cdot
    Q}{(\tilde{p}_i+\tilde{p}_j+ k_\perp )\cdot Q}
  =\frac{\tilde{s}_i+\tilde{s}_j}{\tilde{s}_i+\tilde{s}_j+2 k_\perp
    \cdot Q}.
  \label{eq:rLFF}
 \end{equation}
This guarantees that all the components of the momentum
imbalance due to the map, i.e.
\begin{equation}
 \bar{p}_{\sss X}^{\mu} -q_{\dis}^{\mu} -\tilde{p}^{\mu} _1=
  \bar{p}_{\sss X}^{\mu} - \tilde{p}_{\sss X}^{\mu}  = (r_{\rm \sss L}-1)(\tilde{p}_i^{\mu}+\tilde{p}_j^{\mu}) + r_{\rm \sss L} k_\perp^{\mu},
  \label{eq:imbalance}
\end{equation}
are proportional to $|k_\perp|$ (or   $|k_\perp|^2$), and hence are small not only when
$a_k$ are $b_k$ small, but also when $\tilde{s}_{ij}$ is small,
i.e.~in the triple-collinear limit. 

\subsection{Initial-final dipoles}
For an initial-final dipole, we have
\begin{align}
	 \bar p_i^{\mu} & = r_{\rm \sss L} (1+a_k) \tilde{p}_i^{\mu}\,, \\ 
	\bar p_j^{\mu} &=   r_{\rm \sss L} (1- b_k) \tilde{p}_j^{\mu}\,, \\
  \bar p_k^{\mu} & =  r_{\rm \sss L}(a_k \tilde{p}_i^{\mu}  + b_k \tilde{p}_k^{\mu}  + k_{\perp}^{\mu}) \,.
\end{align}
We now will proof that we may safely set $r_{\rm \sss L} = 1$, and do not encounter the issue in the triple-collinear configuration that arises for FF dipoles.  
Using $\tilde{s}_i = 2 \tilde{p}_i \cdot Q = M_{\sss X}^2 + Q^2$, the partonic final-state momentum after the mapping becomes
\begin{align}
  \bar{p}_{\sss X}^{\mu}  =& \tilde{p}_{\sss X}^{\mu}  - \tilde{p}_j^{\mu}  + p_j^{\mu}  + p_k^{\mu} 
  = \frac{\tilde s_i(1+a_k r_{\rm \sss L})-Q^2 }{Q^2} n_1^\mu + n_2^\mu  + r_{\rm \sss L} k_\perp^{\mu} + (r_{\rm \sss L}-1) \tilde{p}_j^{\mu}.
  \label{eq:b8}
\end{align}
We now aim to write Eq.~\eqref{eq:b8} in terms of $n_1^{\mu}$, $n_2^{\mu}$ and a perpendicular component.
To this end, parameterising the momenta in the DIS frame, we  write
\begin{align}
  \tilde{p}_i ^\mu & =   \frac{\tilde s_i}{2Q} \left(0,0,-1;1\right) =  \frac{\tilde s_i}{Q^2} n_1^\mu\,, \\
  \tilde{p}_j ^\mu & =   \frac{\tilde s_j}{2Q} \left(\sin\theta_{ij}\sin \phi_j,\sin\theta_{ij}\cos \phi_j  ,-\cos \theta_{ij}; 1\right) \\
  & = \frac{\tilde{s}_{j}(1+\cos\theta_{ij})}{2Q^2}n_1^\mu + \frac{\tilde{s}_{j}(1-\cos\theta_{ij})}{2Q^2}n_2^\mu + p_{j,\perp}^\mu\,, \nonumber 
  \end{align}
with
\begin{equation}
  p_{j,\perp}^\mu = \frac{\tilde s_j \sin \theta_{ij}}{2Q}  \left(\sin \phi_j,\cos \phi_j,0;0\right).
\end{equation}
The vectors
\begin{equation}
\hat{k}_{\perp,1}^\mu = \left(\cos \phi_j, -\sin \phi_j, 0,0 \right), \quad \hat{k}_{\perp,2}^\mu =  \left(\sin \phi_j,\cos \phi_j,-\frac{\sin\theta_{ij}}{1-\cos\theta_{ij}};\frac{\sin\theta_{ij}}{1-\cos\theta_{ij}}\right),
\end{equation}
form a basis of space-like vectors with norm $-1$, orthogonal to $\tilde{p}_i^\mu$ and $\tilde{p}_j^\mu$, and we define
\begin{equation}
  k_\perp^\mu =  k_t (\hat{k}_{\perp,1}^\mu \sin \phi_k + \hat{k}_{\perp,2}^\mu \cos \phi_k),
\end{equation}
with $k_t = \sqrt{a_k b_k \tilde{s}_{ij}}$.
Note that $\hat{k}_{\perp,1}^\mu$ is orthogonal to both $n_1^\mu$ and $n_2^\mu$, while $\hat{k}_{\perp,2}^\mu$ carries a longitudinal component. 
To extract this longitudinal piece, we further decompose
\begin{equation}
\hat{k}_{\perp,2}^\mu =   \hat{k}_{\perp,2}^{\prime \mu} + \frac{2 \sin\theta_{ij}}{Q(1-\cos\theta_{ij})} n_1^\mu, 
\text{ with } \hat{k}_{\perp,2}^{\prime \mu}= \left(\sin \phi_j,\cos \phi_j,0;0\right)=   \frac{p_{j,\perp}^\mu}{\sqrt{- p_{j,\perp}^2}}\,,
\end{equation}
where $ \hat{k}_{\perp,2}^{\prime \mu}$ is now orthogonal to both $n_1^{\mu}$ and $n_2^{\mu}$. 
We then write
\begin{equation}
	r_{\rm \sss L} k_\perp^{\mu} + (r_{\rm \sss L}-1) \tilde{p}_j^{\mu} =\Delta \alpha \, n_1^\mu + \Delta \beta \, n_2^\mu+p_\perp^\mu,
\end{equation}
with
\begin{align}
	\Delta \alpha   &=  \frac{ (r_{\sss \rm L} -1 )\tilde{s}_{ij}(1+\cos\theta_{ij})}{2Q^2} + \frac{2 r_{\sss \rm L}  k_t \sin\theta_{ij}\cos\phi_k}{Q(1-\cos\theta_{ij})}\,,\\
	\Delta \beta &=  \frac{ (r_{\sss \rm L} -1 ) \tilde{s}_{ij}(1-\cos\theta_{ij})}{2Q^2}\,, \\
	p_{\perp}^\mu &=   r_{\sss \rm L} k_t \sin \phi_k\hat{k}_{\perp,1}^{\mu} + \left[ (r_{\sss \rm L} -1 ) \frac{\tilde s_j \sin \theta_{ij}}{2Q} + r_{\sss \rm L} k_t \cos \phi_k\right] \hat{k}_{\perp,2}^{\prime, \mu}\,,
\end{align}
so that
\begin{equation}
	\bar{p}_{\sss X}^\mu = \frac{\bar{p}_{\sss X}^2 + p_t^2}{(1+\Delta \beta) Q^2} n_1^\mu + (1+\Delta \beta) n_2^\mu + p_\perp^\mu,
\end{equation}
with $p_t^2=-p_\perp^2$.  For $\theta_{ij} \to 0$, i.e.~in the
triple-collinear limit, we note that $\Delta \beta \to 0$ and $p_t \to
0$, thus $\bar{p}_{\sss X}^\mu \approx \frac{\bar{p}_{\sss X}^2}{Q^2}
n_1^\mu +n_2^\mu$. The action of the momentum-conserving boost is
therefore minimal, as it needs to act on $\bar{p}_{\sss X}^\mu$ such that
$\Lambda^{\mu}_{\phantom{\mu}\nu} \bar{p}_{\sss X}^\nu =
\frac{\bar{p}_{\sss X}^2}{Q^2} n_1^\mu +n_2^\mu \simeq \bar{p}_{\sss X}^\mu$. 
In other words, we can safely use $r_{\rm \sss L}=1$ without introducing any
long-distance correlations. Indeed, for $r_{\rm \sss L}=1$ we simply obtain for the momentum imbalance after the mapping
\begin{align}
	\bar{p}_{\sss X}^\mu- q_{\dis}^\mu- \bar{p}_i^\mu & =   \frac{2 k_t \sin\theta_{ij}}{Q(1-\cos\theta_{ij})} n_1^\mu + k_t (\hat{k}_{\perp,1}^\mu \sin \phi_k + \hat{k}_{\perp,2}^{\prime, \mu} \cos \phi_k)\,.
\end{align}
In the triple-collinear limit this becomes
\begin{align}
\bar{p}_{\sss X}^\mu- q_{\dis}^\mu- \bar{p}_i^\mu & \approx \frac{4}{Q^2} \sqrt{a_k b_k \tilde{s}_i \tilde{s}_j} \, {\rm Sign}(\theta_{ij}) n_1^\mu +\mathcal{O}(\theta_{i j})\,.
\end{align}
This four-vector is aligned along $n_1^\mu$, and the imbalance is
reabsorbed locally within the dipole by rescaling $\bar p_i$, without
affecting any other partons.

\subsection{Initial-initial dipoles and extention to hadron-hadron colliders}
The PanGlobal shower for hadron collisions of
Refs.~\cite{vanBeekveld:2022zhl,vanBeekveld:2022ukn} features the same issues as the previously discussed DIS variant, and the final-state variant presented in Ref.~\cite{Dasgupta:2020fwr}. 
For the $pp$ variant, the transverse-momentum imbalance is fully absorbed
by the hard system (which coincides with the colour singlet in
Ref.~\cite{vanBeekveld:2022zhl}), while the two initial-state partons
are rescaled to ensure that the invariant mass of this system is unchanged. 
It is straightforward to show that for initial-initial dipoles, we do not need to
modify the original proposal, as no triple-collinear configuration can
arise, since the two initial-state particles are never close in angle. 
Like for DIS, for emissions off initial-final
dipoles, the longitudinal rescaling ensures that in the
triple-collinear limit no parton, besides the initial-state one
contained in the emitting dipole, is subject to modifications, thereby
enabling us to use once again the original map.
However, for final-final dipoles, we need to introduce a local rescaling $r_{\rm \sss L}$,
as the triple-collinear configuration can result in a substantial modification to the
hard system. 
For these types of emissions, we employ the definition of the rescaling
given in eq.~\eqref{eq:rLFF}, where now $Q^\mu$ represents the
momentum of the hard system.  The rescaling and boost procedure
then follows that detailed in
Ref.~\cite{vanBeekveld:2022zhl}.
The modifications presented here do not alter the
logarithmic accuracy of the shower.

\section{Cambridge/Aachen algorithm for DIS and Lund variables}
\label{sec:decl}
In this section, we describe the exclusive Cambridge/Aachen algorithm for DIS.  The
$e^+e^-$ variant was introduced in Ref.~\cite{Dokshitzer:1997in},
while a DIS variant was proposed in Ref.~\cite{Wobisch:1998wt}.  This
algorithm is similar to the $k_\perp$ algorithm for DIS introduced of
Ref.~\cite{Dokshitzer:1997in}, but using an angular distance to
determine the cluster sequence.

Conversely to the original proposal, here we do not employ a
resolution variable (or equivalently, we set $y_{\rm cut}=0$).
Instead, the clustering procedure we use is
\begin{itemize}
\item For every final-state parton $i$, define the angular
  distance with respect to the beam $B$
\begin{equation}
  d_{\sss iB} = 1-\cos \theta_i\,,
  \end{equation}
where $\theta_i$ is the angle between $i$ and $B$ in the Breit frame.
In addition, define the angular distance between each pair of partons $i$
and $j$, again in the Breit frame,
\begin{equation}
  d_{\sss ij} = 1 -\cos \theta_{ij}\,.
\end{equation}
\item 
Find the smallest distance among $\left\{d_{\sss iB}, d_{\sss
  ij}\right\}$.  If it is of the type $d_{\sss iB}$, we remove $i$ from the
  list of partons, and make it a candidate jet.  If instead it is of the type $d_{\sss ij}$,
  we remove $i$ and $j$ from the list, and insert a new pseudo-parton with total momentum $p_i^\mu+p_j^\mu$.
\item Repeat the procedure until only one pseudo-parton remains in
  the list: this pseudo-parton forms its own candidate jet.
\end{itemize}

We have now a list of candidate jets, among which we need to find
the final-state macro jet, which contains all the radiation from
the original final-state quark that had a momentum $n_2^\mu$.
Each jet momentum $p_j^\mu$ can be written as
\begin{equation}
p_j^\mu = \alpha_j P^\mu + \beta_j n_2^\mu + p_{\perp, j}^\mu\,,
\end{equation}
where $P^\mu$ is the incoming proton momentum.  We label as
final-state macro jet the one with the largest $\beta_j$ value,
i.e.~the one that retains the largest fraction of the light-cone
component of the original final-state quark in the Breit frame.
This jet can easily be found by searching for the jet whose momentum
$p_j^\mu$ yields the largest value of $p_j\cdot P$.

After this clustering algorithm, the event contains a collection of
initial-state/beam jets and one final-state macro jet.
To calculate our event shapes we need to define the primary Lund-plane
variables.  For each initial-state/beam jet, we define the primary
Lund-plane coordinates as
\begin{equation}
\eta_{i} = -\frac{1}{2}\ln\frac{1- \cos\theta_{i}}{1+\cos\theta_{i}}\,,  \qquad k_{t, i}
= E_i \sin\theta_{i}\,.
\end{equation}
Note that these exactly correspond to the pseudorapidity and
transverse-momentum of $i$ in the Breit frame.
The negative sign of the pseudorapidity is due to the convention to orientate
the incoming beam along the negative $z$ axis in the Breit frame.

For the final-state macro jet a different procedure
is adopted. 
We consider the clustering sequence starting from the
last recombination,
and consider the energies of the two pseudo-jets
that were combined, $E_i$ and $E_j$,
where $E_i > E_j$.
The softer pseudojet, which we have labelled with $j$, is then
promoted to be a jet, with Lund coordinates
\begin{align}
\eta_j = \frac{1}{2}\ln\frac{1-\cos\theta_{ij}}{1+\cos\theta_{ij}}\,, \quad k_{t,j} = E_j \sin\theta_{ij}\,.
\end{align}
We iterate this procedure for the most energetic pseudojet.

The procedure is terminated when the most-energetic pseudojet has
no further children, i.e.~it is a single parton
(notice that this parton is not associated with any jet).
The secondary Lund plane is defined by the jet (found inside the
final-state macrojet, or among the collection of initial-state jets)
that has the largest value for the Lund variable $k_{t}$.

\section{Resummation formulae for continuously-global observables}
\label{sec:resummation}
We consider a continuously-global observable~\cite{Banfi:2004yd} that
in the soft-collinear limit behaves as
\begin{equation}
O \sim \frac{p_T}{Q} e^{-\betaobs |\eta|}.
\end{equation}
The cumulative cross section at NLL accuracy can be written as
\begin{equation}
\label{eq:master-nkll}
 \Sigma_{\rm NLL}(O<e^{L}) = \exp\left[-L g_1(\bar\lambda) + g_2(\bar \lambda)\right], 
\end{equation}
with $\bar\lambda = -\as b_0 L$ (and $L<0$) and
\begin{align}
  \label{eq:b0}
b_0 = \frac{11 C_A - 4 n_f T_R}{12 \pi}.
\end{align}
The $g_1$ function contains the LL terms.
For processes with two hard legs it reads
\begin{subequations}
\begin{align}
    g^{\beta_\text{obs} = 0}_1&= \frac{C_i}{\pi b_0\bar\lambda}\left(2 \bar\lambda + \ln(1-2\bar\lambda)\right),
    \label{eq:g1-b0} \\
    g^{\betaobs \neq 0}_1 &=  \frac{C_i}{\pi b_0\bar\lambda\beta_\text{obs}}\left((1 + \betaobs - 2 \bar\lambda) \ln\left(1-\frac{2\bar\lambda}{1+\beta_\text{obs}}\right) - (1-2 \bar\lambda)\ln(1-2\bar\lambda)\right)
    \label{eq:g1-bn0},
\end{align}
\end{subequations}
where $C_i$ is the Casimir factor ($C_i = C_F$ for quark radiators,
$C_A$ for gluon radiators).
The NLL term $g_2$ can be written as 
\begin{equation}
g^{\beta_\text{obs}}_2 = \ln \frac{f_i\left(x_{\dis},  Q^{2}
  e^{2L/(1+\beta_{\text{obs}})}\right) }{f_i\left(x_{\dis},  Q^{2}
\right)} + \bar{g}^{\beta_\text{obs}}_2 + \ln \mathcal{F}_\text{obs}\,,
\end{equation}
where $f_i$ is the PDF of the incoming parton.
We use $\bar{g}^{\beta_\text{obs}}_2$ to denote the universal
$\beta_\text{obs}$-dependent term originating from soft and
hard-collinear emissions, and $\mathcal{F}_\text{obs}$ is an
observable-dependent correction.  We have
\begin{subequations}
\begin{align}
  \bar{g}^{\betaobs = 0}_2=&\frac{C_i}{\pi b_0^2}\Bigg[b_0 B^{(1)}_i \ln(1-2\bar{\lambda}) - \frac{K}{2\pi}\left(\frac{2\bar{\lambda}}{1-2\bar{\lambda}}+\ln(1-2\bar{\lambda})\right)   \label{eq:g2-b0}   \\
 &\quad\quad\quad\quad\quad\quad + \frac{b_1}{b_0}\left( \frac{2\bar{\lambda}+\ln(1-2\bar{\lambda})}{1-2\bar{\lambda}}+\frac{1}{2}\ln^2(1-2\bar{\lambda})\right)\Bigg],
\nonumber \\
  \bar{g}^{\betaobs \neq 0}_2=& \frac{C_i}{\pi b_0^2 \beta_\text{obs}}\Bigg[   \frac{K}{2\pi}\left(\ln(1 - 2\bar\lambda)-(1+\beta_\text{obs})\ln\left(1 - \frac{2\bar\lambda}{1+\beta_\text{obs}}\right)\right)       \label{eq:g2-b1} \\
& \quad\quad\quad\quad +b_0\betaobs B^{(1)}_i\ln\left(1-\frac{2\bar{\lambda}}{1+\beta_{\text{obs}}}\right) - \frac{b_1}{b_0}\Bigg(\frac{1}{2}\ln^2(1-2\bar\lambda)  + \ln(1-2\bar\lambda)  \nonumber \\
    & \quad\quad\quad\quad- \frac{1}{2}(1+\beta_\text{obs})\ln^2\left(1 - \frac{2\bar\lambda}{1+\beta_\text{obs}}\right) - (1 + \betaobs )\ln\left(1 - \frac{2\bar\lambda}{1+\beta_\text{obs}}\right)  \Bigg)\Bigg], \nonumber
\end{align}
\end{subequations}
with
\begin{align}
b_1 =    \frac{17 C_A^2 - 10 C_A n_f T_R - 6 C_F n_f T_R}{24\pi^2}\,, &\qquad 
K =  \left(\frac{67}{18}-\frac{\pi^2}{6}\right) C_A-\frac{10}{9} n_f T_R\,, \\
B^{(1)}_{q} = -\frac{3}{4} \,,& \quad B^{(1)}_{g} = \frac{-11 C_A + 4n_F T_R}{12 C_A} \,. \nonumber
\end{align}
Finally, we come to the observable-dependent correction.  For the
max-type observable $M_{j,\,\betaobs }$ (defined in
eq.~\eqref{eq:generic-global}), we have $\ln
\mathcal{F}_\text{obs}=0$, while for $S_{p/j,\,\betaobs }$, (eq.~\eqref{eq:generic-global}), this correction reads
\begin{align}
\label{eq:f-sum-obs}
\ln \mathcal{F}_{S_{j,\,\beta_\text{obs}}} = -\gamma_E R'(\bar\lambda) - \ln\Gamma\left(1 + R'(\bar\lambda)\right),
\end{align}
with $R'(\bar\lambda)$ defined as $\partial_L\left(L g_1(\bar\lambda)\right)$,
\begin{subequations}
\begin{align}
    R'_{\betaobs = 0}(\bar\lambda) &= \frac{4C_i}{\pi b_0}
                                    \frac{\bar\lambda}{1-2\bar\lambda}\,,
                                    \label{eq:Rp-beta0}
  \\
    R'_{\betaobs \neq 0}(\bar\lambda) &= \frac{2C_i}{\pi b_0\beta_\text{obs}} \left[\ln\left(1-\frac{2\bar\lambda}{1+\beta_\text{obs}}\right) - \ln(1-2\bar\lambda) \right].
\end{align}
\end{subequations}
In Sec.~\ref{sec:global-obs} we also considered standard DIS event
shapes, that have the property of being continuously global, and can be
defined considering only
partons in the current hemisphere: the thrust
with respect to the photon axis and normalised to
$Q/2$, $\tau_{zQ}$ (eq.~\eqref{eq:tauQ}), and the broadening with respect
to the photon axis, normalised either to the energy in the current
hemisphere $B_{zE}$ (eq.~\eqref{eq:BzE}) or to $Q$, $B_{zQ}$ (eq.~\eqref{eq:BzQ}).
The NLL analytic predictions for $\tau_{zQ}$ were computed for the
first time in Ref.~\cite{Antonelli:1999kx}, and they are identical to
those for $S_{\betaobs =1}$. 
The two definitions of broadening $B_{zE}$ and $B_{zQ}$ are equivalent at NLL, and the analytic prediction can be
read from Ref.~\cite{Dasgupta:2001eq} to read
\begin{align}
	\ln \Sigma(B_z<e^L) =  -g_1^{\betaobs = 0}(\bar{\lambda})\,L 
	+ \bar{g}^{\beta_\text{obs}=0}_2(\bar{\lambda}) + \ln \mathcal{F}_B(\bar{\lambda}) + \ln \frac{f_i\left(x_{\dis},  Q^{2}
		e^{2L}\right) }{f_i\left(x_{\dis},  Q^{2}
		\right)}\,,
\end{align}
where $B_z$ can be either $B_{zE}$ or $B_{zQ}$.
The observable-dependent factor  $\ln \mathcal{F}_B$
had an analytic form that reads
\begin{subequations}
\begin{align}
	\ln \mathcal{F}_B =& -\frac{1}{2} (\ln2 +2\gamma_E)R' + \ln \Lambda(R')-\ln\Gamma(1+R')\,,\\
	\Lambda(R') =& \int_0^\infty \frac{y dy}{(1+y^2)^{3/2}} \left(\frac{y(1+\sqrt{1+y^2})}{8}\right)^{-R'/2} \nonumber \\
	=& 8^{R'/2} \Gamma(1-R'/4)\Gamma(1+R') \frac{{}_2{F}_1(3R'/4, 1+R', 2+3R'/4;-1)}{\Gamma(2+3R'/4)},
\end{align}
\end{subequations}
where
$R'= R'_{\betaobs = 0}(\bar\lambda)$ defined in eq.~\eqref{eq:Rp-beta0} and 
${}_2{F}_1(a,b,c;z)$ is
the hypergeometric function.

\section{The $\alpha_s \to 0$ extrapolation and size of subleading contributions}
\label{sec:asto0}

In this appendix, we show the $\alpha_s \to 0$ extrapolation for $\tau_{zQ}$ as specified in Eq.~\eqref{eq:tauQ}.
\begin{figure}[t]
	\centering
	\mbox{
        \includegraphics[page=1, width=0.48\textwidth]{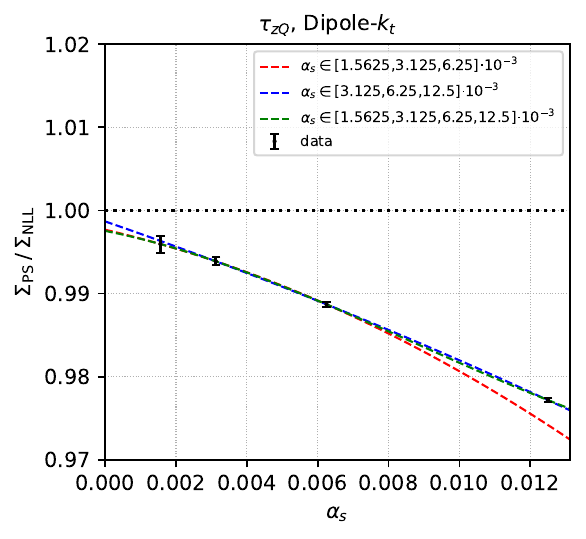}
	\includegraphics[page=2, width=0.48\textwidth]{plots/plot-asto0-sil-25-10-23.pdf}
	}
        \mbox{
        \includegraphics[page=3, width=0.48\textwidth]{plots/plot-asto0-sil-25-10-23.pdf}
	\includegraphics[page=4, width=0.48\textwidth]{plots/plot-asto0-sil-25-10-23.pdf}
	} 
	\caption{$\Sigma_{\rm PS}/\Sigma_{\rm NLL}$ for $\tau_{zQ}$ of
          eq.~\eqref{eq:tauQ} for Dipole-$k_t$ (upper left),
          PanLocal($\betaps=0.5$) (upper right), PanGlobal($\betaps =
          0$) (lower left) and PanGlobal($\betaps = 0.5$) (lower
          right) showers.
	  The black dots represent the data (with the Monte Carlo
          error bars) obtained at $\alpha_s$ values of $0.0015625,
          0.003125, 0.00625$ and $0.0125$ with $\lambda = \alpha_s \ln \tau_{zQ} = -0.5$ held constant.
	  The dashed curves show the interpolation of the data points,
          where red, blue and green represent the interpolation of the
          data at $\alpha_s \in [0.0015625, 0.003125, 0.00625]$,
          $\alpha_s \in [0.003125, 0.00625, 0.0125]$ and $\alpha_s \in
          [0.0015625, 0.003125, 0.00625, 0.0125]$ respectively.
	}
	\label{fig:extrap-as-to-0}
\end{figure}
In Fig.~\ref{fig:extrap-as-to-0}, we show the results obtained for
ratio between the shower result and the analytic NLL expectation for
the cumulative distributions for the Dipole-$k_t$,
PanLocal($\betaps=0.5$), PanGlobal($\betaps=0$) and
PanGlobal($\betaps=0.5$) showers for $\lambda=\alpha_s \ln \tau_{zQ} = -0.5$.
These results are obtained with four different values
$\alpha_s \in [0.0015625, 0.003125, 0.00625, 0.0125]$.
The size of subleading contributions and the stability of the extrapolation 
depends on both the shower (as can be seen in Fig.~\ref{fig:extrap-as-to-0}) and the observable. 

We then fit the results using a polynomial $f(\alpha_s) =
\sum_{i=0}^n c_i \alpha^i$, with $n$ the (variable) number of data points. 
The $\alpha_s \to 0$ extrapolation is obtained through the coefficient $c_0$, i.e.~the point where the curves cross the $y$-axis. 
The slope of the curves near $\alpha_s = 0$ indicates the size
of NNLL corrections generated by the showers. 
Our current numerical precison prevents us from getting their precise estimate, but in all cases we find a value of order 1, as is
expected since $\lambda=\mathcal{O}(1)$.


\section{Spin correlations}
\label{sec:spin}
Spin correlations must be included to reproduce the correct
azimuthal structure of strongly angular-ordered collinear splittings.
This can be achieved by for example using the Collins-Knowles
algorithm~\cite{Collins:1987cp,Knowles:1987cu,Knowles:1988vs,Knowles:1988hu},
which was also recently applied to the PanScales final-state and initial-state
showers~\cite{Karlberg:2021kwr, vanBeekveld:2022zhl}.
In Ref.~\cite{Hamilton:2021dyz} the algorithm was further improved to
include for the first time the treatment of the dominant
leading-colour soft azimuthal correlations.
In this appendix, we show that the algorithm introduced
in Ref.~\cite{vanBeekveld:2022zhl} can be  applied straightforwardly
 to DIS.

We consider the shower effective matrix element for the production of
two extra partons $i$ and $j$, and compare the result against the
analytic expectation as a function of the azimuthal angle
$\Delta\psi_{ij}$ between the planes spanned by two emissions $i$ and
$j$.
At $\mathcal{O}(\alpha_s^2)$, the differential cross section can be written as
\begin{align}
    \label{eq:spin-azimuthal-angle}
    \frac{d\sigma}{d\Delta \psi_{ij}} \propto a_0 \left( 1 + \frac{a_2}{a_0}
    \cos(2\Delta \psi_{ij}) \right) = a_0 \left( 1 + A(z_i) B(z_{j})
    \cos(2\Delta \psi_{ij}) \right)\,,
\end{align}
 where $\psi_{ij}$ is the azimuthal difference between the plane
 defined by the primary and secondary splittings with light-cone momentum
 fraction $z_i$ and $z_j$, respectively.
The values of $a_0$ and $a_2$ depend on the type of branching,
and are a function of $z_{i}$ and $z_j$.
In the absence of spin correlations, the ratio $a_2/a_0$ would be 0.

\begin{figure}[t]
    \centering
    \includegraphics[width=\textwidth]{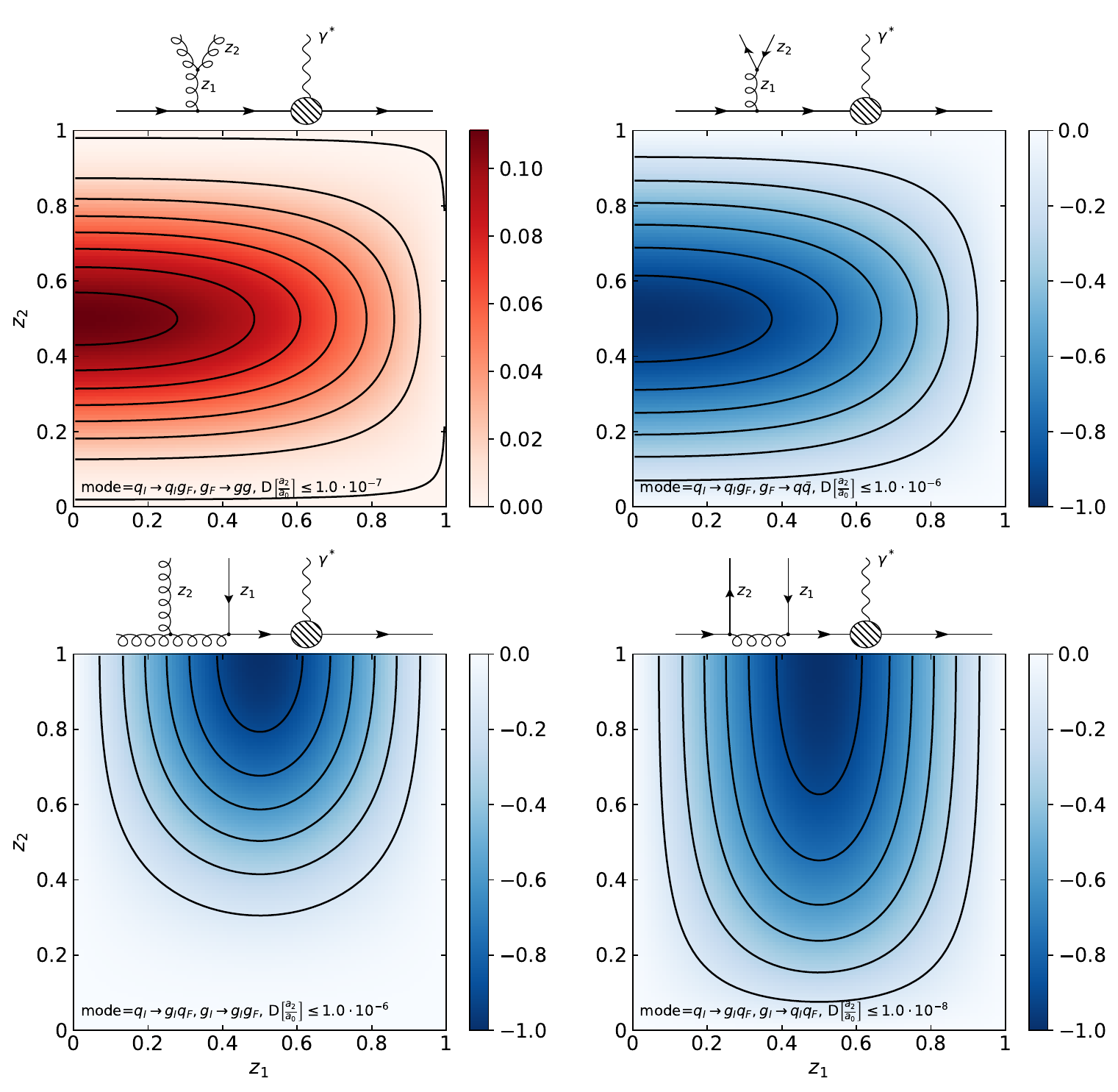}
    \caption{Size of the ratio between the two Fourier coefficients
      $a_2/a_0$, defined in eq.~\eqref{eq:spin-azimuthal-angle} at
      $\mathcal{O}(\as^2)$ for collinear splittings, for the
      PanGlobal($\betaobs=0$) shower.
      The Feynman diagrams indicate the sequence of splittings under
      consideration.
      The  black lines indicate constant values for this ratio, and
      are obtained using the analytic predictions.
      The (maximum) deviation of the analytic prediction and the shower
      is given by ${\rm D}\big[\frac{a_2}{a_0}\big] \leq {\rm
        max}\big|\big(\frac{a_2}{a_0}\big)_{\rm PS} -
      \big(\frac{a_2}{a_0}\big)_{\rm ME} \big|$, with
      $\big(\frac{a_2}{a_0}\big)_{\rm PS}$ the shower prediction and
      $\big(\frac{a_2}{a_0}\big)_{\rm ME}$ the analytic prediction for
      the matrix element.
    }
    \label{fig:plot-spin-collinear}
\end{figure}

In Fig.~\ref{fig:plot-spin-collinear} we illustrate the ratio
$a_2/a_0$ across several values of $z_1$ and $z_2$ for
collinear-splittings involving an intermediate gluon.     
In particular, in the left panel we show the case in which the first
emission is a final-state gluon, that further branches into a $q\bar{q}$
pair.
In the right panel we instead consider the backward evolution of the
initial-state quark into an initial-state gluon~($g_I$), and then an
emission of a final-state gluon very collinear to the incoming beam.
We only show the predictions obtained with the
PanGlobal($\betaobs=0$) shower, but identical
results can be obtained by considering
the other PanScales showers. 
The differences between the shower predictions and the analytic
expectation were always smaller that $10^{-6}$,  confirming our implementation at fixed-order accuracy.

\bibliographystyle{JHEP}
\bibliography{MC}

\providecommand{\href}[2]{#2}\begingroup\raggedright\begin{thebibliography}{100}

\bibitem{Bahr:2008pv}
M.~Bahr et~al., \emph{{Herwig++ Physics and Manual}},
  \href{http://dx.doi.org/10.1140/epjc/s10052-008-0798-9}{\emph{Eur. Phys. J.}
  {\bf C58} (2008) 639--707}, [\href{http://arxiv.org/abs/0803.0883}{{\tt
  0803.0883}}].

\bibitem{Bellm:2019zci}
J.~Bellm et~al., \emph{{Herwig 7.2 release note}},
  \href{http://dx.doi.org/10.1140/epjc/s10052-020-8011-x}{\emph{Eur. Phys. J.
  C} {\bf 80} (2020) 452}, [\href{http://arxiv.org/abs/1912.06509}{{\tt
  1912.06509}}].

\bibitem{Gieseke:2003rz}
S.~Gieseke, P.~Stephens and B.~Webber, \emph{{New formalism for QCD parton
  showers}}, \href{http://dx.doi.org/10.1088/1126-6708/2003/12/045}{\emph{JHEP}
  {\bf 12} (2003) 045}, [\href{http://arxiv.org/abs/hep-ph/0310083}{{\tt
  hep-ph/0310083}}].

\bibitem{Bewick:2019rbu}
G.~Bewick, S.~Ferrario~Ravasio, P.~Richardson and M.~H. Seymour,
  \emph{{Logarithmic accuracy of angular-ordered parton showers}},
  \href{http://dx.doi.org/10.1007/JHEP04(2020)019}{\emph{JHEP} {\bf 04} (2020)
  019}, [\href{http://arxiv.org/abs/1904.11866}{{\tt 1904.11866}}].

\bibitem{Bewick:2021nhc}
G.~Bewick, S.~Ferrario~Ravasio, P.~Richardson and M.~H. Seymour, \emph{{Initial
  state radiation in the Herwig 7 angular-ordered parton shower}},
  \href{http://dx.doi.org/10.1007/JHEP01(2022)026}{\emph{JHEP} {\bf 01} (2022)
  026}, [\href{http://arxiv.org/abs/2107.04051}{{\tt 2107.04051}}].

\bibitem{Banfi:2006gy}
A.~Banfi, G.~Corcella and M.~Dasgupta, \emph{{Angular ordering and parton
  showers for non-global QCD observables}},
  \href{http://dx.doi.org/10.1088/1126-6708/2007/03/050}{\emph{JHEP} {\bf 03}
  (2007) 050}, [\href{http://arxiv.org/abs/hep-ph/0612282}{{\tt
  hep-ph/0612282}}].

\bibitem{Gustafson:1987rq}
G.~Gustafson and U.~Pettersson, \emph{{Dipole Formulation of QCD Cascades}},
  \href{http://dx.doi.org/10.1016/0550-3213(88)90441-5}{\emph{Nucl. Phys. B}
  {\bf 306} (1988) 746--758}.

\bibitem{Catani:2001cc}
S.~Catani, F.~Krauss, R.~Kuhn and B.~R. Webber, \emph{{QCD matrix elements +
  parton showers}},
  \href{http://dx.doi.org/10.1088/1126-6708/2001/11/063}{\emph{JHEP} {\bf 11}
  (2001) 063}, [\href{http://arxiv.org/abs/hep-ph/0109231}{{\tt
  hep-ph/0109231}}].

\bibitem{Krauss:2002up}
F.~Krauss, \emph{{Matrix elements and parton showers in hadronic
  interactions}},
  \href{http://dx.doi.org/10.1088/1126-6708/2002/08/015}{\emph{JHEP} {\bf 08}
  (2002) 015}, [\href{http://arxiv.org/abs/hep-ph/0205283}{{\tt
  hep-ph/0205283}}].

\bibitem{Lavesson:2008ah}
N.~Lavesson and L.~Lonnblad, \emph{{Extending CKKW-merging to One-Loop Matrix
  Elements}},
  \href{http://dx.doi.org/10.1088/1126-6708/2008/12/070}{\emph{JHEP} {\bf 12}
  (2008) 070}, [\href{http://arxiv.org/abs/0811.2912}{{\tt 0811.2912}}].

\bibitem{Hoeche:2009rj}
S.~Hoeche, F.~Krauss, S.~Schumann and F.~Siegert, \emph{{QCD matrix elements
  and truncated showers}},
  \href{http://dx.doi.org/10.1088/1126-6708/2009/05/053}{\emph{JHEP} {\bf 05}
  (2009) 053}, [\href{http://arxiv.org/abs/0903.1219}{{\tt 0903.1219}}].

\bibitem{Giele:2011cb}
W.~T. Giele, D.~A. Kosower and P.~Z. Skands, \emph{{Higher-Order Corrections to
  Timelike Jets}},
  \href{http://dx.doi.org/10.1103/PhysRevD.84.054003}{\emph{Phys. Rev. D} {\bf
  84} (2011) 054003}, [\href{http://arxiv.org/abs/1102.2126}{{\tt 1102.2126}}].

\bibitem{Platzer:2012bs}
S.~Pl\"atzer, \emph{{Controlling inclusive cross sections in parton shower +
  matrix element merging}},
  \href{http://dx.doi.org/10.1007/JHEP08(2013)114}{\emph{JHEP} {\bf 08} (2013)
  114}, [\href{http://arxiv.org/abs/1211.5467}{{\tt 1211.5467}}].

\bibitem{Lonnblad:2012ix}
L.~L\"onnblad and S.~Prestel, \emph{{Merging Multi-leg NLO Matrix Elements with
  Parton Showers}},
  \href{http://dx.doi.org/10.1007/JHEP03(2013)166}{\emph{JHEP} {\bf 03} (2013)
  166}, [\href{http://arxiv.org/abs/1211.7278}{{\tt 1211.7278}}].

\bibitem{Frederix:2012ps}
R.~Frederix and S.~Frixione, \emph{{Merging meets matching in MC@NLO}},
  \href{http://dx.doi.org/10.1007/JHEP12(2012)061}{\emph{JHEP} {\bf 12} (2012)
  061}, [\href{http://arxiv.org/abs/1209.6215}{{\tt 1209.6215}}].

\bibitem{Lonnblad:2012ng}
L.~Lonnblad and S.~Prestel, \emph{{Unitarising Matrix Element + Parton Shower
  merging}}, \href{http://dx.doi.org/10.1007/JHEP02(2013)094}{\emph{JHEP} {\bf
  02} (2013) 094}, [\href{http://arxiv.org/abs/1211.4827}{{\tt 1211.4827}}].

\bibitem{Bellm:2017ktr}
J.~Bellm, S.~Gieseke and S.~Pl\"atzer, \emph{{Merging NLO Multi-jet
  Calculations with Improved Unitarization}},
  \href{http://dx.doi.org/10.1140/epjc/s10052-018-5723-2}{\emph{Eur. Phys. J.
  C} {\bf 78} (2018) 244}, [\href{http://arxiv.org/abs/1705.06700}{{\tt
  1705.06700}}].

\bibitem{Brooks:2020mab}
H.~Brooks and C.~T. Preuss, \emph{{Efficient multi-jet merging with the Vincia
  sector shower}},
  \href{http://dx.doi.org/10.1016/j.cpc.2021.107985}{\emph{Comput. Phys.
  Commun.} {\bf 264} (2021) 107985},
  [\href{http://arxiv.org/abs/2008.09468}{{\tt 2008.09468}}].

\bibitem{Hamilton:2012rf}
K.~Hamilton, P.~Nason, C.~Oleari and G.~Zanderighi, \emph{{Merging H/W/Z + 0
  and 1 jet at NLO with no merging scale: a path to parton shower + NNLO
  matching}}, \href{http://dx.doi.org/10.1007/JHEP05(2013)082}{\emph{JHEP} {\bf
  05} (2013) 082}, [\href{http://arxiv.org/abs/1212.4504}{{\tt 1212.4504}}].

\bibitem{Alioli:2013hqa}
S.~Alioli, C.~W. Bauer, C.~Berggren, F.~J. Tackmann, J.~R. Walsh and S.~Zuberi,
  \emph{{Matching Fully Differential NNLO Calculations and Parton Showers}},
  \href{http://dx.doi.org/10.1007/JHEP06(2014)089}{\emph{JHEP} {\bf 06} (2014)
  089}, [\href{http://arxiv.org/abs/1311.0286}{{\tt 1311.0286}}].

\bibitem{Hoche:2014dla}
S.~Hoeche, Y.~Li and S.~Prestel, \emph{{Higgs-boson production through gluon
  fusion at NNLO QCD with parton showers}},
  \href{http://dx.doi.org/10.1103/PhysRevD.90.054011}{\emph{Phys. Rev.} {\bf
  D90} (2014) 054011}, [\href{http://arxiv.org/abs/1407.3773}{{\tt
  1407.3773}}].

\bibitem{Monni:2019whf}
P.~F. Monni, P.~Nason, E.~Re, M.~Wiesemann and G.~Zanderighi,
  \emph{{MiNNLO$_{\text{PS}}$: A new method to match NNLO QCD to parton
  showers}},  \href{http://arxiv.org/abs/1908.06987}{{\tt 1908.06987}}.

\bibitem{Campbell:2021svd}
J.~M. Campbell, S.~H\"oche, H.~T. Li, C.~T. Preuss and P.~Skands,
  \emph{{Towards NNLO+PS matching with sector showers}},
  \href{http://dx.doi.org/10.1016/j.physletb.2022.137614}{\emph{Phys. Lett. B}
  {\bf 836} (2023) 137614}, [\href{http://arxiv.org/abs/2108.07133}{{\tt
  2108.07133}}].

\bibitem{Prestel:2021vww}
S.~Prestel, \emph{{Matching N3LO QCD calculations to parton showers}},
  \href{http://dx.doi.org/10.1007/JHEP11(2021)041}{\emph{JHEP} {\bf 11} (2021)
  041}, [\href{http://arxiv.org/abs/2106.03206}{{\tt 2106.03206}}].

\bibitem{Mangano:2001xp}
M.~L. Mangano, M.~Moretti and R.~Pittau, \emph{{Multijet matrix elements and
  shower evolution in hadronic collisions: $W b \bar{b}$ + $n$ jets as a case
  study}}, \href{http://dx.doi.org/10.1016/S0550-3213(02)00249-3}{\emph{Nucl.
  Phys.} {\bf B632} (2002) 343--362},
  [\href{http://arxiv.org/abs/hep-ph/0108069}{{\tt hep-ph/0108069}}].

\bibitem{Hamilton:2009ne}
K.~Hamilton, P.~Richardson and J.~Tully, \emph{{A Modified CKKW matrix element
  merging approach to angular-ordered parton showers}},
  \href{http://dx.doi.org/10.1088/1126-6708/2009/11/038}{\emph{JHEP} {\bf 11}
  (2009) 038}, [\href{http://arxiv.org/abs/0905.3072}{{\tt 0905.3072}}].

\bibitem{Martinez:2021chk}
A.~B. Martinez, F.~Hautmann and M.~L. Mangano, \emph{{TMD evolution and
  multi-jet merging}},
  \href{http://dx.doi.org/10.1016/j.physletb.2021.136700}{\emph{Phys. Lett. B}
  {\bf 822} (2021) 136700}, [\href{http://arxiv.org/abs/2107.01224}{{\tt
  2107.01224}}].

\bibitem{Schumann:2007mg}
S.~Schumann and F.~Krauss, \emph{{A Parton shower algorithm based on
  Catani-Seymour dipole factorisation}},
  \href{http://dx.doi.org/10.1088/1126-6708/2008/03/038}{\emph{JHEP} {\bf 03}
  (2008) 038}, [\href{http://arxiv.org/abs/0709.1027}{{\tt 0709.1027}}].

\bibitem{Platzer:2009jq}
S.~Platzer and S.~Gieseke, \emph{{Coherent Parton Showers with Local Recoils}},
  \href{http://dx.doi.org/10.1007/JHEP01(2011)024}{\emph{JHEP} {\bf 01} (2011)
  024}, [\href{http://arxiv.org/abs/0909.5593}{{\tt 0909.5593}}].

\bibitem{Hoche:2015sya}
S.~Hoeche and S.~Prestel, \emph{{The midpoint between dipole and parton
  showers}}, \href{http://dx.doi.org/10.1140/epjc/s10052-015-3684-2}{\emph{Eur.
  Phys. J.} {\bf C75} (2015) 461}, [\href{http://arxiv.org/abs/1506.05057}{{\tt
  1506.05057}}].

\bibitem{Cabouat:2017rzi}
B.~Cabouat and T.~Sjöstrand, \emph{{Some Dipole Shower Studies}},
  \href{http://dx.doi.org/10.1140/epjc/s10052-018-5645-z,
  10.1140/s10052-018-5645-z}{\emph{Eur. Phys. J.} {\bf C78} (2018) 226},
  [\href{http://arxiv.org/abs/1710.00391}{{\tt 1710.00391}}].

\bibitem{Brooks:2020upa}
H.~Brooks, C.~T. Preuss and P.~Skands, \emph{{Sector Showers for Hadron
  Collisions}}, \href{http://dx.doi.org/10.1007/JHEP07(2020)032}{\emph{JHEP}
  {\bf 07} (2020) 032}, [\href{http://arxiv.org/abs/2003.00702}{{\tt
  2003.00702}}].

\bibitem{Sjostrand:2006za}
T.~Sjostrand, S.~Mrenna and P.~Z. Skands, \emph{{PYTHIA 6.4 Physics and
  Manual}}, \href{http://dx.doi.org/10.1088/1126-6708/2006/05/026}{\emph{JHEP}
  {\bf 05} (2006) 026}, [\href{http://arxiv.org/abs/hep-ph/0603175}{{\tt
  hep-ph/0603175}}].

\bibitem{Sjostrand:2014zea}
T.~Sjöstrand, S.~Ask, J.~R. Christiansen, R.~Corke, N.~Desai, P.~Ilten et~al.,
  \emph{{An Introduction to PYTHIA 8.2}},
  \href{http://dx.doi.org/10.1016/j.cpc.2015.01.024}{\emph{Comput. Phys.
  Commun.} {\bf 191} (2015) 159--177},
  [\href{http://arxiv.org/abs/1410.3012}{{\tt 1410.3012}}].

\bibitem{Gleisberg:2008ta}
T.~Gleisberg, S.~Hoeche, F.~Krauss, M.~Schonherr, S.~Schumann, F.~Siegert
  et~al., \emph{{Event generation with SHERPA 1.1}},
  \href{http://dx.doi.org/10.1088/1126-6708/2009/02/007}{\emph{JHEP} {\bf 02}
  (2009) 007}, [\href{http://arxiv.org/abs/0811.4622}{{\tt 0811.4622}}].

\bibitem{Sherpa:2019gpd}
{\scshape Sherpa} collaboration, E.~Bothmann et~al., \emph{{Event Generation
  with Sherpa 2.2}},
  \href{http://dx.doi.org/10.21468/SciPostPhys.7.3.034}{\emph{SciPost Phys.}
  {\bf 7} (2019) 034}, [\href{http://arxiv.org/abs/1905.09127}{{\tt
  1905.09127}}].

\bibitem{Bierlich:2022pfr}
C.~Bierlich et~al., \emph{{A comprehensive guide to the physics and usage of
  PYTHIA 8.3}},  \href{http://arxiv.org/abs/2203.11601}{{\tt 2203.11601}}.

\bibitem{Dasgupta:2020fwr}
M.~Dasgupta, F.~A. Dreyer, K.~Hamilton, P.~F. Monni, G.~P. Salam and G.~Soyez,
  \emph{{Parton showers beyond leading logarithmic accuracy}},
  \href{http://dx.doi.org/10.1103/PhysRevLett.125.052002}{\emph{Phys. Rev.
  Lett.} {\bf 125} (2020) 052002}, [\href{http://arxiv.org/abs/2002.11114}{{\tt
  2002.11114}}].

\bibitem{Dasgupta:2018nvj}
M.~Dasgupta, F.~A. Dreyer, K.~Hamilton, P.~F. Monni and G.~P. Salam,
  \emph{{Logarithmic accuracy of parton showers: a fixed-order study}},
  \href{http://dx.doi.org/10.1007/JHEP09(2018)033}{\emph{JHEP} {\bf 09} (2018)
  033}, [\href{http://arxiv.org/abs/1805.09327}{{\tt 1805.09327}}].

\bibitem{Forshaw:2020wrq}
J.~R. Forshaw, J.~Holguin and S.~Pl\"atzer, \emph{{Building a consistent parton
  shower}}, \href{http://dx.doi.org/10.1007/JHEP09(2020)014}{\emph{JHEP} {\bf
  09} (2020) 014}, [\href{http://arxiv.org/abs/2003.06400}{{\tt 2003.06400}}].

\bibitem{Nagy:2020dvz}
Z.~Nagy and D.~E. Soper, \emph{{Summations by parton showers of large
  logarithms in electron-positron annihilation}},
  \href{http://arxiv.org/abs/2011.04777}{{\tt 2011.04777}}.

\bibitem{Herren:2022jej}
F.~Herren, S.~H\"oche, F.~Krauss, D.~Reichelt and M.~Schoenherr, \emph{{A new
  approach to color-coherent parton evolution}},
  \href{http://arxiv.org/abs/2208.06057}{{\tt 2208.06057}}.

\bibitem{vanBeekveld:2022zhl}
M.~van Beekveld, S.~Ferrario~Ravasio, G.~P. Salam, A.~Soto-Ontoso, G.~Soyez and
  R.~Verheyen, \emph{{PanScales parton showers for hadron collisions:
  formulation and fixed-order studies}},
  \href{http://dx.doi.org/10.1007/JHEP11(2022)019}{\emph{JHEP} {\bf 11} (2022)
  019}, [\href{http://arxiv.org/abs/2205.02237}{{\tt 2205.02237}}].

\bibitem{vanBeekveld:2022ukn}
M.~van Beekveld, S.~Ferrario~Ravasio, K.~Hamilton, G.~P. Salam, A.~Soto-Ontoso,
  G.~Soyez et~al., \emph{{PanScales showers for hadron collisions: all-order
  validation}}, \href{http://dx.doi.org/10.1007/JHEP11(2022)020}{\emph{JHEP}
  {\bf 11} (2022) 020}, [\href{http://arxiv.org/abs/2207.09467}{{\tt
  2207.09467}}].

\bibitem{Nagy:2009vg}
Z.~Nagy and D.~E. Soper, \emph{{On the transverse momentum in Z-boson
  production in a virtuality ordered parton shower}},
  \href{http://dx.doi.org/10.1007/JHEP03(2010)097}{\emph{JHEP} {\bf 03} (2010)
  097}, [\href{http://arxiv.org/abs/0912.4534}{{\tt 0912.4534}}].

\bibitem{Covarelli:2021gyz}
R.~Covarelli, M.~Pellen and M.~Zaro, \emph{{Vector-Boson scattering at the LHC:
  Unraveling the electroweak sector}},
  \href{http://dx.doi.org/10.1142/S0217751X2130009X}{\emph{Int. J. Mod. Phys.
  A} {\bf 36} (2021) 2130009}, [\href{http://arxiv.org/abs/2102.10991}{{\tt
  2102.10991}}].

\bibitem{BuarqueFranzosi:2021wrv}
D.~Buarque~Franzosi et~al., \emph{{Vector boson scattering processes: Status
  and prospects}},
  \href{http://dx.doi.org/10.1016/j.revip.2022.100071}{\emph{Rev. Phys.} {\bf
  8} (2022) 100071}, [\href{http://arxiv.org/abs/2106.01393}{{\tt
  2106.01393}}].

\bibitem{ATLAS:2020fzp}
{\scshape ATLAS} collaboration, G.~Aad et~al., \emph{{A search for the dimuon
  decay of the Standard Model Higgs boson with the ATLAS detector}},
  \href{http://dx.doi.org/10.1016/j.physletb.2020.135980}{\emph{Phys. Lett. B}
  {\bf 812} (2021) 135980}, [\href{http://arxiv.org/abs/2007.07830}{{\tt
  2007.07830}}].

\bibitem{CMS:2020xwi}
{\scshape CMS} collaboration, A.~M. Sirunyan et~al., \emph{{Evidence for Higgs
  boson decay to a pair of muons}},
  \href{http://dx.doi.org/10.1007/JHEP01(2021)148}{\emph{JHEP} {\bf 01} (2021)
  148}, [\href{http://arxiv.org/abs/2009.04363}{{\tt 2009.04363}}].

\bibitem{CMS:2022kdi}
{\scshape CMS} collaboration, \emph{{Measurements of Higgs boson production in
  the decay channel with a pair of $\tau$ leptons in proton-proton collisions
  at $\sqrt{s}$ = 13 TeV}},  \href{http://arxiv.org/abs/2204.12957}{{\tt
  2204.12957}}.

\bibitem{ATLAS:2022yrq}
{\scshape ATLAS} collaboration, G.~Aad et~al., \emph{{Measurements of Higgs
  boson production cross-sections in the~$H\to\tau^{+}\tau^{-}$ decay channel
  in pp collisions at $ \sqrt{s} $ = 13 TeV with the ATLAS detector}},
  \href{http://dx.doi.org/10.1007/JHEP08(2022)175}{\emph{JHEP} {\bf 08} (2022)
  175}, [\href{http://arxiv.org/abs/2201.08269}{{\tt 2201.08269}}].

\bibitem{ATLAS:2022yvh}
{\scshape ATLAS} collaboration, G.~Aad et~al., \emph{{Search for invisible
  Higgs-boson decays in events with vector-boson fusion signatures using 139
  fb$^{-1}$ of proton-proton data recorded by the ATLAS experiment}},
  \href{http://dx.doi.org/10.1007/JHEP08(2022)104}{\emph{JHEP} {\bf 08} (2022)
  104}, [\href{http://arxiv.org/abs/2202.07953}{{\tt 2202.07953}}].

\bibitem{CMS:2022qva}
{\scshape CMS} collaboration, A.~Tumasyan et~al., \emph{{Search for invisible
  decays of the Higgs boson produced via vector boson fusion in proton-proton
  collisions at s=13\,\,TeV}},
  \href{http://dx.doi.org/10.1103/PhysRevD.105.092007}{\emph{Phys. Rev. D} {\bf
  105} (2022) 092007}, [\href{http://arxiv.org/abs/2201.11585}{{\tt
  2201.11585}}].

\bibitem{Ballestrero:2018anz}
A.~Ballestrero et~al., \emph{{Precise predictions for same-sign W-boson
  scattering at the LHC}},
  \href{http://dx.doi.org/10.1140/epjc/s10052-018-6136-y}{\emph{Eur. Phys. J.
  C} {\bf 78} (2018) 671}, [\href{http://arxiv.org/abs/1803.07943}{{\tt
  1803.07943}}].

\bibitem{Jager:2020hkz}
B.~J\"ager, A.~Karlberg, S.~Pl\"atzer, J.~Scheller and M.~Zaro,
  \emph{{Parton-shower effects in Higgs production via Vector-Boson Fusion}},
  \href{http://dx.doi.org/10.1140/epjc/s10052-020-8326-7}{\emph{Eur. Phys. J.
  C} {\bf 80} (2020) 756}, [\href{http://arxiv.org/abs/2003.12435}{{\tt
  2003.12435}}].

\bibitem{Hoche:2021mkv}
S.~H\"oche, S.~Mrenna, S.~Payne, C.~T. Preuss and P.~Skands, \emph{{A Study of
  QCD Radiation in VBF Higgs Production with Vincia and Pythia}},
  \href{http://dx.doi.org/10.21468/SciPostPhys.12.1.010}{\emph{SciPost Phys.}
  {\bf 12} (2022) 010}, [\href{http://arxiv.org/abs/2106.10987}{{\tt
  2106.10987}}].

\bibitem{ATLAS:2022tnm}
{\scshape ATLAS} collaboration, \emph{{Measurement of the properties of Higgs
  boson production at $\sqrt{s} = 13$ TeV in the $H\to\gamma\gamma$ channel
  using $139$ fb$^{-1}$ of $pp$ collision data with the ATLAS experiment}},
  \href{http://arxiv.org/abs/2207.00348}{{\tt 2207.00348}}.

\bibitem{ATLAS:2022ooq}
{\scshape ATLAS} collaboration, \emph{{Measurements of Higgs boson production
  by gluon$-$gluon fusion and vector-boson fusion using $H\rightarrow W W^*
  \rightarrow e\nu \mu\nu$ decays in $pp$ collisions at $\sqrt{s}=13$ TeV with
  the ATLAS detector}},  \href{http://arxiv.org/abs/2207.00338}{{\tt
  2207.00338}}.

\bibitem{ATLAS:2020rej}
{\scshape ATLAS} collaboration, G.~Aad et~al., \emph{{Higgs boson production
  cross-section measurements and their EFT interpretation in the $4\ell $ decay
  channel at $\sqrt{s}=$13 TeV with the ATLAS detector}},
  \href{http://dx.doi.org/10.1140/epjc/s10052-020-8227-9}{\emph{Eur. Phys. J.
  C} {\bf 80} (2020) 957}, [\href{http://arxiv.org/abs/2004.03447}{{\tt
  2004.03447}}].

\bibitem{CMS:2021kom}
{\scshape CMS} collaboration, A.~M. Sirunyan et~al., \emph{{Measurements of
  Higgs boson production cross sections and couplings in the diphoton decay
  channel at $ \sqrt{\mathrm{s}} $ = 13 TeV}},
  \href{http://dx.doi.org/10.1007/JHEP07(2021)027}{\emph{JHEP} {\bf 07} (2021)
  027}, [\href{http://arxiv.org/abs/2103.06956}{{\tt 2103.06956}}].

\bibitem{CMS:2022uhn}
{\scshape CMS} collaboration, \emph{{Measurements of the Higgs boson production
  cross section and couplings in the W boson pair decay channel in
  proton-proton collisions at $\sqrt{s}$ = 13 TeV}},
  \href{http://arxiv.org/abs/2206.09466}{{\tt 2206.09466}}.

\bibitem{Buckley:2021gfw}
A.~Buckley et~al., \emph{{A comparative study of Higgs boson production from
  vector-boson fusion}},
  \href{http://dx.doi.org/10.1007/JHEP11(2021)108}{\emph{JHEP} {\bf 11} (2021)
  108}, [\href{http://arxiv.org/abs/2105.11399}{{\tt 2105.11399}}].

\bibitem{Han:1992hr}
T.~Han, G.~Valencia and S.~Willenbrock, \emph{{Structure function approach to
  vector boson scattering in p p collisions}},
  \href{http://dx.doi.org/10.1103/PhysRevLett.69.3274}{\emph{Phys. Rev. Lett.}
  {\bf 69} (1992) 3274--3277}, [\href{http://arxiv.org/abs/hep-ph/9206246}{{\tt
  hep-ph/9206246}}].

\bibitem{Bolzoni:2010xr}
P.~Bolzoni, F.~Maltoni, S.-O. Moch and M.~Zaro, \emph{{Higgs production via
  vector-boson fusion at NNLO in QCD}},
  \href{http://dx.doi.org/10.1103/PhysRevLett.105.011801}{\emph{Phys. Rev.
  Lett.} {\bf 105} (2010) 011801}, [\href{http://arxiv.org/abs/1003.4451}{{\tt
  1003.4451}}].

\bibitem{Bolzoni:2011cu}
P.~Bolzoni, F.~Maltoni, S.-O. Moch and M.~Zaro, \emph{{Vector boson fusion at
  NNLO in QCD: SM Higgs and beyond}},
  \href{http://dx.doi.org/10.1103/PhysRevD.85.035002}{\emph{Phys. Rev. D} {\bf
  85} (2012) 035002}, [\href{http://arxiv.org/abs/1109.3717}{{\tt 1109.3717}}].

\bibitem{Liu:2019tuy}
T.~Liu, K.~Melnikov and A.~A. Penin, \emph{{Nonfactorizable QCD Effects in
  Higgs Boson Production via Vector Boson Fusion}},
  \href{http://dx.doi.org/10.1103/PhysRevLett.123.122002}{\emph{Phys. Rev.
  Lett.} {\bf 123} (2019) 122002}, [\href{http://arxiv.org/abs/1906.10899}{{\tt
  1906.10899}}].

\bibitem{Dreyer:2020urf}
F.~A. Dreyer, A.~Karlberg and L.~Tancredi, \emph{{On the impact of
  non-factorisable corrections in VBF single and double Higgs production}},
  \href{http://dx.doi.org/10.1007/JHEP10(2020)131}{\emph{JHEP} {\bf 10} (2020)
  131}, [\href{http://arxiv.org/abs/2005.11334}{{\tt 2005.11334}}].

\bibitem{Webber:1993bm}
B.~R. Webber, \emph{{Factorization and jet clustering algorithms for deep
  inelastic scattering}},
  \href{http://dx.doi.org/10.1088/0954-3899/19/10/012}{\emph{J. Phys. G} {\bf
  19} (1993) 1567--1575}.

\bibitem{Catani:1990rr}
S.~Catani, B.~R. Webber and G.~Marchesini, \emph{{QCD coherent branching and
  semiinclusive processes at large x}},
  \href{http://dx.doi.org/10.1016/0550-3213(91)90390-J}{\emph{Nucl. Phys.} {\bf
  B349} (1991) 635--654}.

\bibitem{Catani:1996vz}
S.~Catani and M.~H. Seymour, \emph{{A General algorithm for calculating jet
  cross-sections in NLO QCD}},
  \href{http://dx.doi.org/10.1016/S0550-3213(96)00589-5,
  10.1016/S0550-3213(98)81022-5}{\emph{Nucl. Phys.} {\bf B485} (1997)
  291--419}, [\href{http://arxiv.org/abs/hep-ph/9605323}{{\tt
  hep-ph/9605323}}].

\bibitem{Ritzmann:2012ca}
M.~Ritzmann, D.~A. Kosower and P.~Skands, \emph{{Antenna Showers with Hadronic
  Initial States}},
  \href{http://dx.doi.org/10.1016/j.physletb.2012.12.003}{\emph{Phys. Lett. B}
  {\bf 718} (2013) 1345--1350}, [\href{http://arxiv.org/abs/1210.6345}{{\tt
  1210.6345}}].

\bibitem{DOUBLESOFT}
S.~Ferrario~Ravasio, K.~Hamilton, A.~Karlberg, G.~P. Salam, L.~Scyboz and
  G.~Soyez, \emph{{A parton shower with higher-logarithmic accuracy for soft
  emissions}},  \href{http://arxiv.org/abs/2307.11142}{{\tt 2307.11142}}.

\bibitem{Antonelli:1999kx}
V.~Antonelli, M.~Dasgupta and G.~P. Salam, \emph{{Resummation of thrust
  distributions in DIS}},
  \href{http://dx.doi.org/10.1088/1126-6708/2000/02/001}{\emph{JHEP} {\bf 02}
  (2000) 001}, [\href{http://arxiv.org/abs/hep-ph/9912488}{{\tt
  hep-ph/9912488}}].

\bibitem{Hamilton:2023dwb}
K.~Hamilton, A.~Karlberg, G.~P. Salam, L.~Scyboz and R.~Verheyen,
  \emph{{Matching and event-shape NNDL accuracy in parton showers}},
  \href{http://dx.doi.org/10.1007/JHEP03(2023)224}{\emph{JHEP} {\bf 03} (2023)
  224}, [\href{http://arxiv.org/abs/2301.09645}{{\tt 2301.09645}}].

\bibitem{Cruz-Martinez:2018rod}
J.~Cruz-Martinez, T.~Gehrmann, E.~W.~N. Glover and A.~Huss, \emph{{Second-order
  QCD effects in Higgs boson production through vector boson fusion}},
  \href{http://dx.doi.org/10.1016/j.physletb.2018.04.046}{\emph{Phys. Lett. B}
  {\bf 781} (2018) 672--677}, [\href{http://arxiv.org/abs/1802.02445}{{\tt
  1802.02445}}].

\bibitem{Dreyer:2018rfu}
F.~A. Dreyer and A.~Karlberg, \emph{{Fully differential Vector-Boson Fusion
  Higgs Pair Production at Next-to-Next-to-Leading Order}},
  \href{http://dx.doi.org/10.1103/PhysRevD.99.074028}{\emph{Phys. Rev. D} {\bf
  99} (2019) 074028}, [\href{http://arxiv.org/abs/1811.07918}{{\tt
  1811.07918}}].

\bibitem{Asteriadis:2021gpd}
K.~Asteriadis, F.~Caola, K.~Melnikov and R.~R\"ontsch, \emph{{NNLO QCD
  corrections to weak boson fusion Higgs boson production in the H
  \textrightarrow{} b$ \overline{b} $ and H \textrightarrow{} WW$^{*}$
  \textrightarrow{} 4l decay channels}},
  \href{http://dx.doi.org/10.1007/JHEP02(2022)046}{\emph{JHEP} {\bf 02} (2022)
  046}, [\href{http://arxiv.org/abs/2110.02818}{{\tt 2110.02818}}].

\bibitem{Andersson:1988gp}
B.~Andersson, G.~Gustafson, L.~Lonnblad and U.~Pettersson, \emph{{Coherence
  Effects in Deep Inelastic Scattering}},
  \href{http://dx.doi.org/10.1007/BF01550942}{\emph{Z. Phys.} {\bf C43} (1989)
  625}.

\bibitem{Hamilton:2020rcu}
K.~Hamilton, R.~Medves, G.~P. Salam, L.~Scyboz and G.~Soyez, \emph{{Colour and
  logarithmic accuracy in final-state parton showers}},
  \href{http://dx.doi.org/10.1007/JHEP03(2021)041}{\emph{JHEP} {\bf 03} (2021)
  041}, [\href{http://arxiv.org/abs/2011.10054}{{\tt 2011.10054}}].

\bibitem{Gribov:1972ri}
V.~N. Gribov and L.~N. Lipatov, \emph{{Deep inelastic e p scattering in
  perturbation theory}}, {\emph{Sov. J. Nucl. Phys.} {\bf 15} (1972) 438--450}.

\bibitem{Dokshitzer:1977sg}
Y.~L. Dokshitzer, \emph{{Calculation of the Structure Functions for Deep
  Inelastic Scattering and e+ e- Annihilation by Perturbation Theory in Quantum
  Chromodynamics.}}, {\emph{Sov. Phys. JETP} {\bf 46} (1977) 641--653}.

\bibitem{Altarelli:1977zs}
G.~Altarelli and G.~Parisi, \emph{{Asymptotic Freedom in Parton Language}},
  \href{http://dx.doi.org/10.1016/0550-3213(77)90384-4}{\emph{Nucl. Phys.} {\bf
  B126} (1977) 298--318}.

\bibitem{Salam:2008qg}
G.~P. Salam and J.~Rojo, \emph{{A Higher Order Perturbative Parton Evolution
  Toolkit (HOPPET)}},
  \href{http://dx.doi.org/10.1016/j.cpc.2008.08.010}{\emph{Comput. Phys.
  Commun.} {\bf 180} (2009) 120--156},
  [\href{http://arxiv.org/abs/0804.3755}{{\tt 0804.3755}}].

\bibitem{Catani:1991pm}
S.~Catani, Y.~L. Dokshitzer, F.~Fiorani and B.~R. Webber, \emph{{Average number
  of jets in e+ e- annihilation}},
  \href{http://dx.doi.org/10.1016/0550-3213(92)90296-N}{\emph{Nucl. Phys.} {\bf
  B377} (1992) 445--460}.

\bibitem{Catani:1993yx}
S.~Catani, Y.~L. Dokshitzer and B.~R. Webber, \emph{{Average number of jets in
  deep inelastic scattering}},
  \href{http://dx.doi.org/10.1016/0370-2693(94)91118-5}{\emph{Phys. Lett. B}
  {\bf 322} (1994) 263--269}.

\bibitem{Medves:2022ccw}
R.~Medves, A.~Soto-Ontoso and G.~Soyez, \emph{{Lund and Cambridge
  multiplicities for precision physics}},
  \href{http://dx.doi.org/10.1007/JHEP10(2022)156}{\emph{JHEP} {\bf 10} (2022)
  156}, [\href{http://arxiv.org/abs/2205.02861}{{\tt 2205.02861}}].

\bibitem{Banfi:2004yd}
A.~Banfi, G.~P. Salam and G.~Zanderighi, \emph{{Principles of general
  final-state resummation and automated implementation}},
  \href{http://dx.doi.org/10.1088/1126-6708/2005/03/073}{\emph{JHEP} {\bf 03}
  (2005) 073}, [\href{http://arxiv.org/abs/hep-ph/0407286}{{\tt
  hep-ph/0407286}}].

\bibitem{Catani:2007vq}
S.~Catani and M.~Grazzini, \emph{{An NNLO subtraction formalism in hadron
  collisions and its application to Higgs boson production at the LHC}},
  \href{http://dx.doi.org/10.1103/PhysRevLett.98.222002}{\emph{Phys. Rev.
  Lett.} {\bf 98} (2007) 222002},
  [\href{http://arxiv.org/abs/hep-ph/0703012}{{\tt hep-ph/0703012}}].

\bibitem{Stewart:2010tn}
I.~W. Stewart, F.~J. Tackmann and W.~J. Waalewijn, \emph{{N-Jettiness: An
  Inclusive Event Shape to Veto Jets}},
  \href{http://dx.doi.org/10.1103/PhysRevLett.105.092002}{\emph{Phys. Rev.
  Lett.} {\bf 105} (2010) 092002}, [\href{http://arxiv.org/abs/1004.2489}{{\tt
  1004.2489}}].

\bibitem{Dasgupta:2001eq}
M.~Dasgupta and G.~P. Salam, \emph{{Resummation of the jet broadening in DIS}},
  \href{http://dx.doi.org/10.1007/s100520200915}{\emph{Eur. Phys. J. C} {\bf
  24} (2002) 213--236}, [\href{http://arxiv.org/abs/hep-ph/0110213}{{\tt
  hep-ph/0110213}}].

\bibitem{Dasgupta:2002dc}
M.~Dasgupta and G.~P. Salam, \emph{{Resummed event shape variables in DIS}},
  \href{http://dx.doi.org/10.1088/1126-6708/2002/08/032}{\emph{JHEP} {\bf 08}
  (2002) 032}, [\href{http://arxiv.org/abs/hep-ph/0208073}{{\tt
  hep-ph/0208073}}].

\bibitem{Dasgupta:2006ru}
M.~Dasgupta and Y.~Delenda, \emph{{The Q(t) distribution of the Breit current
  hemisphere in DIS as a probe of small-x broadening effects}},
  \href{http://dx.doi.org/10.1088/1126-6708/2006/08/080}{\emph{JHEP} {\bf 08}
  (2006) 080}, [\href{http://arxiv.org/abs/hep-ph/0606285}{{\tt
  hep-ph/0606285}}].

\bibitem{Dasgupta:2001sh}
M.~Dasgupta and G.~Salam, \emph{{Resummation of nonglobal QCD observables}},
  \href{http://dx.doi.org/10.1016/S0370-2693(01)00725-0}{\emph{Phys. Lett. B}
  {\bf 512} (2001) 323--330}, [\href{http://arxiv.org/abs/hep-ph/0104277}{{\tt
  hep-ph/0104277}}].

\bibitem{Dasgupta:2002bw}
M.~Dasgupta and G.~P. Salam, \emph{{Accounting for coherence in interjet E(t)
  flow: A Case study}},
  \href{http://dx.doi.org/10.1088/1126-6708/2002/03/017}{\emph{JHEP} {\bf 03}
  (2002) 017}, [\href{http://arxiv.org/abs/hep-ph/0203009}{{\tt
  hep-ph/0203009}}].

\bibitem{Caletti:2021oor}
S.~Caletti, O.~Fedkevych, S.~Marzani, D.~Reichelt, S.~Schumann, G.~Soyez
  et~al., \emph{{Jet angularities in Z+jet production at the LHC}},
  \href{http://dx.doi.org/10.1007/JHEP07(2021)076}{\emph{JHEP} {\bf 07} (2021)
  076}, [\href{http://arxiv.org/abs/2104.06920}{{\tt 2104.06920}}].

\bibitem{ATLAS:2020jgy}
{\scshape ATLAS} collaboration, G.~Aad et~al., \emph{{Search for the $HH
  \rightarrow b \bar{b} b \bar{b}$ process via vector-boson fusion production
  using proton-proton collisions at $\sqrt{s} = 13$ TeV with the ATLAS
  detector}}, \href{http://dx.doi.org/10.1007/JHEP07(2020)108}{\emph{JHEP} {\bf
  07} (2020) 108}, [\href{http://arxiv.org/abs/2001.05178}{{\tt 2001.05178}}].

\bibitem{NNPDF:2021njg}
{\scshape NNPDF} collaboration, R.~D. Ball et~al., \emph{{The path to proton
  structure at 1\% accuracy}},
  \href{http://dx.doi.org/10.1140/epjc/s10052-022-10328-7}{\emph{Eur. Phys. J.
  C} {\bf 82} (2022) 428}, [\href{http://arxiv.org/abs/2109.02653}{{\tt
  2109.02653}}].

\bibitem{Buckley:2014ana}
A.~Buckley, J.~Ferrando, S.~Lloyd, K.~Nordstr\"om, B.~Page, M.~R\"ufenacht
  et~al., \emph{{LHAPDF6: parton density access in the LHC precision era}},
  \href{http://dx.doi.org/10.1140/epjc/s10052-015-3318-8}{\emph{Eur. Phys. J.
  C} {\bf 75} (2015) 132}, [\href{http://arxiv.org/abs/1412.7420}{{\tt
  1412.7420}}].

\bibitem{Mrenna:2016sih}
S.~Mrenna and P.~Skands, \emph{{Automated Parton-Shower Variations in Pythia
  8}}, \href{http://dx.doi.org/10.1103/PhysRevD.94.074005}{\emph{Phys. Rev. D}
  {\bf 94} (2016) 074005}, [\href{http://arxiv.org/abs/1605.08352}{{\tt
  1605.08352}}].

\bibitem{Cacciari:2008gp}
M.~Cacciari, G.~P. Salam and G.~Soyez, \emph{{The anti-$k_t$ jet clustering
  algorithm}},
  \href{http://dx.doi.org/10.1088/1126-6708/2008/04/063}{\emph{JHEP} {\bf 04}
  (2008) 063}, [\href{http://arxiv.org/abs/0802.1189}{{\tt 0802.1189}}].

\bibitem{Cacciari:2011ma}
M.~Cacciari, G.~P. Salam and G.~Soyez, \emph{{FastJet User Manual}},
  \href{http://dx.doi.org/10.1140/epjc/s10052-012-1896-2}{\emph{Eur. Phys. J.}
  {\bf C72} (2012) 1896}, [\href{http://arxiv.org/abs/1111.6097}{{\tt
  1111.6097}}].

\bibitem{Dokshitzer:1997in}
Y.~L. Dokshitzer, G.~D. Leder, S.~Moretti and B.~R. Webber, \emph{{Better jet
  clustering algorithms}},
  \href{http://dx.doi.org/10.1088/1126-6708/1997/08/001}{\emph{JHEP} {\bf 08}
  (1997) 001}, [\href{http://arxiv.org/abs/hep-ph/9707323}{{\tt
  hep-ph/9707323}}].

\bibitem{Wobisch:1998wt}
M.~Wobisch and T.~Wengler, \emph{{Hadronization corrections to jet
  cross-sections in deep inelastic scattering}},  in \emph{{Workshop on Monte
  Carlo Generators for HERA Physics (Plenary Starting Meeting)}}, pp.~270--279,
  4, 1998.
\newblock \href{http://arxiv.org/abs/hep-ph/9907280}{{\tt hep-ph/9907280}}.

\bibitem{Collins:1987cp}
J.~C. Collins, \emph{{Spin Correlations in Monte Carlo Event Generators}},
  \href{http://dx.doi.org/10.1016/0550-3213(88)90654-2}{\emph{Nucl. Phys.} {\bf
  B304} (1988) 794--804}.

\bibitem{Knowles:1987cu}
I.~Knowles, \emph{{Angular Correlations in QCD}},
  \href{http://dx.doi.org/10.1016/0550-3213(88)90653-0}{\emph{Nucl. Phys. B}
  {\bf 304} (1988) 767--793}.

\bibitem{Knowles:1988vs}
I.~Knowles, \emph{{Spin Correlations in Parton - Parton Scattering}},
  \href{http://dx.doi.org/10.1016/0550-3213(88)90092-2}{\emph{Nucl. Phys. B}
  {\bf 310} (1988) 571--588}.

\bibitem{Knowles:1988hu}
I.~G. Knowles, \emph{{A Linear Algorithm for Calculating Spin Correlations in
  Hadronic Collisions}},
  \href{http://dx.doi.org/10.1016/0010-4655(90)90063-7}{\emph{Comput. Phys.
  Commun.} {\bf 58} (1990) 271--284}.

\bibitem{Karlberg:2021kwr}
A.~Karlberg, G.~P. Salam, L.~Scyboz and R.~Verheyen, \emph{{Spin correlations
  in final-state parton showers and jet observables}},
  \href{http://dx.doi.org/10.1140/epjc/s10052-021-09378-0}{\emph{Eur. Phys. J.
  C} {\bf 81} (2021) 681}, [\href{http://arxiv.org/abs/2103.16526}{{\tt
  2103.16526}}].

\bibitem{Hamilton:2021dyz}
K.~Hamilton, A.~Karlberg, G.~P. Salam, L.~Scyboz and R.~Verheyen, \emph{{Soft
  spin correlations in final-state parton showers}},
  \href{http://dx.doi.org/10.1007/JHEP03(2022)193}{\emph{JHEP} {\bf 03} (2022)
  193}, [\href{http://arxiv.org/abs/2111.01161}{{\tt 2111.01161}}].

\end{thebibliography}\endgroup

\end{document}